\documentclass[12pt]{article}
\usepackage{amsmath}
\usepackage{graphicx,psfrag,epsf}
\usepackage{caption}
\usepackage{subcaption}
\usepackage{enumerate}
\usepackage{natbib}
\usepackage{rotating}
\usepackage[table]{xcolor}
\usepackage{xr-hyper}
\usepackage{hyperref}
\usepackage{multirow}
\usepackage{natbib}
\usepackage{amsthm}
\usepackage{amssymb}
\usepackage{mathtools}
\usepackage{booktabs}
\usepackage{float}
\usepackage{adjustbox}
\usepackage[linesnumbered,ruled, lined, longend,boxed,plain]{algorithm2e}
\usepackage{ulem}
\usepackage{blkarray}

\makeatletter
\newcommand*{\addFileDependency}[1]{
  \typeout{(#1)}
  \@addtofilelist{#1}
  \IfFileExists{#1}{}{\typeout{No file #1.}}
}
\makeatother

\newcommand*{\myexternaldocument}[1]{%
\externaldocument{#1}%
\addFileDependency{#1.tex}%
\addFileDependency{#1.aux}%
}
\myexternaldocument{mGHS_suppl_mat}

\newcommand{\blind}{0}

\newtheorem{proposition}{\bf Proposition}[section]

\graphicspath{{Figures/}}

\begin{document}

\def\spacingset#1{\renewcommand{\baselinestretch}%
{#1}\small\normalsize} \spacingset{1}


\if0\blind
{
  \title{\bf Inference of multiple high-dimensional networks with the Graphical Horseshoe prior}
  \author{Claudio Busatto \\
    Department of Statistics, Computer Science, Applications "G. Parenti", \\ University of Florence, Florence, Italy\\
    and \\
    Francesco Claudio Stingo \\
    Department of Statistics, Computer Science, Applications "G. Parenti", \\ University of Florence, Florence, Italy}
  \maketitle
} \fi

\if1\blind
{
  \bigskip
  \bigskip
  \bigskip
  \begin{center}
    {\LARGE\bf Multiple Graphical Horseshoe estimator for modeling correlated precision matrices}
\end{center}
  \medskip
} \fi

\bigskip
\begin{abstract}
\noindent We develop a novel full-Bayesian approach for multiple correlated precision matrices, called multiple Graphical Horseshoe (mGHS). The proposed approach relies on a novel multivariate shrinkage prior based on the Horseshoe prior that borrows strength and shares sparsity patterns across groups, improving posterior edge selection when the precision matrices are similar. On the other hand, there is no loss of performance when the groups are independent. Moreover, mGHS provides a similarity matrix estimate, useful for understanding network similarities across groups. We implement an efficient Metropolis-within-Gibbs for posterior inference; specifically, local variance parameters are updated via a novel and efficient modified rejection sampling algorithm that samples from a three-parameter Gamma distribution. The method scales well with respect to the number of variables and provides one of the fastest full-Bayesian approaches for the estimation of multiple precision matrices. Finally, edge selection is performed with a novel approach based on model cuts. 
We empirically demonstrate that mGHS outperforms competing approaches through both simulation studies and an application to a bike-sharing dataset.
\end{abstract}

\noindent%

\spacingset{1.45}


\section{Introduction}
\label{sec:introduction}

Graphical models are a popular tool used in many scientific fields to analyze and infer networks. 
In the Gaussian setting, the main challenges in graph estimation are the positive-definiteness constraint on precision matrices (inverse-covariance matrices) and the quadratic growth, with respect to the number of variables included in the analysis, of the number of free parameters. Traditional methods, such as the ones based on pairwise model comparisons, become computationally infeasible as the number of considered variables increases.
For exchangeable observations, a collection of the existing methods for high-dimensional covariance matrix estimation is available in \cite{Pourahmadi-2011}, in which the author proposes to reduce the problem to multiple independent (penalized) least-squares regressions. Other common approaches, such as the Graphical LASSO of \cite{Friedman-2008} and the Graphical SCAD of \cite{Fan-2009}, are based on a penalized likelihood optimization and provide a sparse solution for the precision matrix in high-dimensional settings. A few approaches for the estimation of high-dimensional sparse networks have also been proposed within the Bayesian framework. In particular, the Bayesian version of the Graphical LASSO \citep{Wang-2012}, the spike and slab stochastic search method \citep{Wang-2015}, and the more recent Graphical Horseshoe presented in \cite{Bahdra-2019}; all Bayesian methods implemented a block Gibbs sampler that has shown good computational performances up to a few hundred variables. 

We are interested in settings where observations can be considered exchangeable only within groups; in these settings, a separate group-specific estimation will reduce the statistical power, while an analysis of data pooled across groups will lead to spurious findings \citep{Peterson2015}. Generalizations of the graphical models, called multiple graphical models, have been proposed with the aim of jointly estimating multiple correlated networks. Among the penalized likelihood approaches, the fused Graphical LASSO and the group Graphical LASSO of \cite{Danaher-2014} rely on convex optimization problems and force similar edge values and similar graph structures, respectively. Bayesian approaches have been first proposed to encourage similar network structures across related subgroups \citep{Peterson2015,Shaddox}. More recent attempts, such as the generalization of the Bayesian spike and slab stochastic method of \cite{Peterson-2020} and the GemBAG of \cite{Yang-2021}, focus on shared sparsity structures and precision matrix elements. See \cite{Ni2022} for a recent review of Bayesian approaches for complex graphical models, including methods for multiple groups.

Here we propose a generalization of the Graphical Horseshoe of \cite{Bahdra-2019} in the presence of multiple correlated sample groups, which we refer to as the \textit{multiple Graphical Horseshoe} (mGHS). This model works under the multivariate Gaussianity assumption with multiple dependent precision matrices. The proposed model is based on a novel prior on multiple covariance matrices that builds upon the Horseshoe prior proposed in \cite{Carvalho-2010} and lets the data decide whether borrowing strength across groups and then encouraging similar precision matrices is appropriate. The properties of the Horseshoe prior are well-studied and include the improved Kullback-Leibler risk bound \citep{Carvalho-2010}, minimaxity in estimation under the $l_2$ loss \citep{vdPas-2014} and improved risk properties in linear regression \citep{Bhadra-2016}. Through simulation studies, we empirically show that the model benefits from the similar structures of the groups and provides better statistical performances than the Graphical Horseshoe applied separately to each group. The model relies on a Metropolis-within-Gibbs sampler where the parameters are updated by sampling from their full-conditional distributions and, in particular, a novel method is introduced in order to sample the local variance parameters. 
This method scales well with respect to the number of variables and is the first full Bayesian approach (to our knowledge) able to analyze multiple graphs of hundreds of nodes. Finally, we discuss a novel idea for posterior edge selection based on model cuts. The main novelties can be summarized as follow: 1) a novel shrinkage prior for multiple precision matrices, 2) an efficient algorithm that scales exceptionally well, and 3) a novel approach for edge selection based on model cuts.

The paper is organized as follows. In Section \ref{sec:model} the proposed sampling model is introduced. Section \ref{sec:G3p} illustrates how to sample from a three-parameters Gamma distribution ($\mathcal{G}_{3p}$) with a modified rejection sampling approach. Section \ref{sec:posterior} outlines the proposed algorithm in detail. In Section \ref{sec:edgesselection} we present a novel proposal for model selection. Section \ref{sec:simulation} illustrates comparative simulation studies, whereas in Section \ref{sec:application} we present an application to a benchmark bike-sharing dataset. Discussions and comments are presented in Section \ref{sec:conclusion}.

%

\section{The model}
\label{sec:model}

In this section, we introduce the sampling model used to infer relationships among variables within each of $K$ possibly related sample groups, each represented by a graph $G_k = (V, E_k)$, where $V$ corresponds to a set of vertices and $E_k$ to a set of group-specific edges. Let $\mathbf{y}_{sk}$ be the $p$-dimensional random vector related to the observation $s$ in group $k$, where $s = 1, \dots, n_k$ and $k = 1, \dots, K$. Under the multivariate normal distribution, the corresponding sampling model is
\begin{equation*}
	\mathbf{y}_{sk} \sim \mathcal{N}_p \left(\mathbf{0}, \boldsymbol{\Sigma}_{k}\right),
\end{equation*}
where $\boldsymbol{\Omega}_{k} \equiv (\omega_{ij}^{k})_{p\times p} = \boldsymbol{\Sigma}_k^{-1}$ is the precision matrix of group $k$. There is a one-to-one correspondence between the zero patterns in a precision matrix and an undirected graph $G_k$ that, in turn, can be used to learn conditional independencies. Specifically, it can be shown that $\omega_{ij}^k = 0$ if and only if variables $i$ and $j$ are conditionally independent conditioning on the other variables \citep{Dempster-1972}; in this case, the undirected graph $G_k$ will have a missing edge between nodes $i$ and $j$. Therefore, the goal is the joint estimation of non-zero entries in precision matrices with the aim of capturing significant connections among variables. In high-dimensional settings, the number of parameters to be estimated in $\boldsymbol{\Omega}_k$ is of order $O\left(p^2\right)$. This task is particularly challenging since these precision matrices, in addition to being very large, are constrained to the cone of symmetric positive definite matrices. 
Building upon the Graphical Horseshoe proposed by \cite{Bahdra-2019}, we propose in Sections \ref{prior} and \ref{sec:posterior} model and algorithm, respectively, that use shrinkage priors to perform full Bayesian inference of multiple related high-dimensional undirected graphical models.

\subsection{An horseshoe prior for multiple related precision matrices} 
\label{prior}
\noindent \cite{Bahdra-2019} have successfully developed the Graphical Horseshoe prior, a shrinkage prior for (single) precision matrices. In this section, we describe how to extend the Graphical Horseshoe prior to multiple related precision matrices. The proposed approach will both achieve shrinkage and borrowing strength across related subgroups; as a key modeling feature, our approach will learn from the data which pairs of groups are related and which ones can be considered independent. With respect to the model proposed by \cite{Peterson-2020}, the only alternative full Bayesian approach that uses a joint prior on related multiple precision matrices, the proposed approach will result in a much more scalable algorithm, as detailed in Section \ref{sec:posterior}.

Let $\boldsymbol{\omega}_{ij} = \left( \omega_{ij}^1, \dots, \omega_{ij}^K \right)^\intercal$ be the vector of precision matrix entries corresponding to edge $\left(i, j\right)$ across $K$ groups.  Our approach builds upon the Graphical Horseshoe prior \citep{Bahdra-2019}, as we shrink non-informative edges $\omega_{ij}^k$ with a novel multivariate Horseshoe prior \citep{Carvalho-2010}; we assume a non-informative prior for diagonal entries $\omega_{jj}^k$. The joint prior distribution for precision matrices $\boldsymbol{\Omega}_1, \dots, \boldsymbol{\Omega}_K$ can be written as
\begin{align*}
	\pi\left(\omega_{jj}^k\right) \propto & \; 1, \quad k = 1, \dots, K, \quad j = 1, \dots, p \\
	\pi\left(\boldsymbol{\Omega}_1, \dots, \boldsymbol{\Omega}_K \vert \boldsymbol{\Psi}_{ij} : i < j \right) \propto & \; \prod_{i < j} \mathcal{N}_K \left(\boldsymbol{\omega}_{ij} \vert \mathbf{0}, \boldsymbol{\Psi}_{ij} \right) \cdot \mathbb{I}_{\left(\boldsymbol{\Omega}_1, \dots, \boldsymbol{\Omega}_K \in \mathbb{M}_+^p\right)}
\end{align*}
where $\mathbb{M}_+^p$ denotes the space of $p \times p$ positive-definite symmetric matrices. 
The proposed prior jointly models multiple precision matrices and, specifically, accounts for similarity between groups by imposing a $K$-variate normal prior distribution for $\boldsymbol{\omega}_{ij}$ with prior covariance matrix specific for each pair $ij$. As in \cite{Peterson-2020}, the proposed prior jointly learns both the within-group and across-group associations from the data in a single step, but it is computationally more efficient because it is based on continuous mixtures of multivariate normal distributions. Indeed, there is no need to sample the binary edge inclusion indicators as in \cite{Peterson-2020}.

Following the \textit{separation strategy} introduced by \cite{Barnard-2000}, the across-group covariance matrices $ \boldsymbol{\Psi}_{ij}$ can be decomposed as $\boldsymbol{\Psi}_{ij} = \boldsymbol{\Delta}_{ij} \mathbf{R} \boldsymbol{\Delta}_{ij}$, where $\boldsymbol{\Delta}_{ij} = \text{diag}\{\delta_{ij,1}, \dots, \delta_{ij,K}\}$ contains the standard deviations of edge $(i,j)$ and $\mathbf{R} = \{r_{k'k}: k' < k\} \in \mathbb{M}_+^K$ is a valid correlation matrix with diagonal entries equal to one. As suggested by \cite{Barnard-2000}, we model variances  $\delta_{ij,k}$ and correlations $r_{k'k}$ separately since it is generally not clear how these elements interact with each other. We apply the Horseshoe prior from \cite{Carvalho-2010} by decomposing $\delta_{ij, k} = \tau_k \lambda_{ij, k}$ and imposing the following priors:
\begin{align}
	\lambda_{ij, k} \sim & \; \mathcal{C} ^+ \left(0, 1 \right), \label{eq:HS1}\\
	\tau_k \sim & \; \mathcal{C}^+ \left(0, 1\right), \label{eq:HS2}
\end{align}
where $\mathcal{C}^+$ denotes the positive half-Cauchy distribution. In \eqref{eq:HS1} and \eqref{eq:HS2}, parameters $\tau_k$ and $\lambda_{ij, k}$ control the global and local shrinkage of $\omega_{ij}^k$, respectively. The heavy-tail distribution of $\lambda_{ij, k}$ allows $\omega_{ij}^k$ to avoid overshrinkage and lets the coefficients free to reach larger values. The amount of common shrinkage shared by the entries $\omega_{ij}^k$ is then controlled by the global scale parameter $\tau_k$. When $K = 1$, the proposed model reduces to the Graphical Horseshoe of \cite{Bahdra-2019}. 

The selection of the prior distribution for correlation matrix $\mathbf{R}$ is often more complicated. \cite{Barnard-2000} give an overview of the most common prior for a correlation matrix. Here we follow \cite{Peterson-2020} and choose the prior distribution
\begin{equation*}
	\label{eq:R}
	\pi\left(\mathbf{R}\right) \propto 1 \cdot \mathbb{I}_{\left(\mathbf{R}\in \mathbb{C}_+^K\right)},
\end{equation*}
where $\mathbb{C}_+^K$ denotes the space of $K \times K$ definite-positive correlation matrices with diagonal entries equal to $1$. The matrix $\mathbf{R}$ allows the local variances $\lambda_{ij}^k$ to share information between each other when the correlations between groups are large. On the contrary, the model reduces to the Graphical Horseshoe of \cite{Bahdra-2019} applied separately to each group when $\mathbf{R} = \mathbf{I}_K$ is the identity matrix. In Section \ref{sec:G3p} we introduce a new sampling algorithm for the three-parameter Gamma distribution that will be used within the algorithm for posterior inference detailed in Section \ref{sec:posterior}. 

\section{The three-parameter Gamma $\mathcal{G}_{3p} \left(\gamma, \alpha, \beta\right)$ distribution and a modified rejection sampling algorithm}
\label{sec:G3p}
%

In this section, we introduce a modified acceptance-rejection method designed to generate samples from the three-parameter Gamma ($\mathcal{G}_{3p}$) distribution. \cite{Ahrens-1982} and  \cite{Stadlober-1982} demonstrated how to apply a rejection sampling for a target distribution when no valid proposal distribution is available. In particular, they proposed a modified rejection sampling to sample from a Gamma distribution and a $t$-Student distribution, respectively. Here the same situation applies since no trivial distribution, such as Gaussian or Gamma distributions, can be used as a valid proposal distribution. Indeed, it can be shown that these densities do not cover the target function on the latter's support, as required by the standard rejection sampling method. Therefore, we propose to overcome this problem by applying a modified rejection sampling with a Gaussian proposal distribution. The technical and theoretical aspects of this approach are detailed in Appendix \ref{app:a}, where we also provide a proof that the method proposed in this section draws samples from the target distribution \eqref{eq:f}. For the sake of clarity, the notation used in this section does not relate to the notation used in the other sections.

Let $X \sim \mathcal{G}_{3p} \left(\gamma, \alpha, \beta \right)$, $\alpha, \beta \ne 0$, $\gamma \in \mathbb{N}^+$, a random variable with density
\begin{equation}
	f_X\left(x\mid \gamma, \alpha, \beta \right) =  \frac{e^{-\frac{\beta^2}{8 \alpha^2}} \left(2 \alpha^2\right)^{\frac{\gamma+1}{2}}}{\Gamma\left(\gamma+1\right) D_{-\gamma-1}\left(-\frac{\beta}{\alpha \sqrt{2}}\right)} x^\gamma e^{-\alpha^2 x^2 + \beta x} \cdot \mathbb{I}_{\left(x > 0\right)},
	\label{eq:f}
\end{equation}
where $D_{a}\left(b\right)$ is the Parabolic Cylinder function with parameters $a$ and $b$. The mean and variance of variable $X$ are
\begin{align*}
	E\left(X\right) \equiv \mu = &\; \frac{\gamma + 1}{\alpha \sqrt{2}} \frac{D_{-\gamma-2}\left(-\frac{\beta}{\alpha \sqrt{2}}\right)}{ D_{-\gamma-1}\left(-\frac{\beta}{\alpha \sqrt{2}}\right)} \\
	Var\left(X\right) \equiv \sigma^2 = &\; \frac{\left(\gamma + 1\right) \left(\gamma+2\right)}{2\alpha^2} \frac{D_{-\gamma-3}\left(-\frac{\beta}{\alpha \sqrt{2}}\right)}{D_{-\gamma-1}\left(-\frac{\beta}{\alpha \sqrt{2}}\right)} - \frac{\left(\gamma + 1\right)^2}{2\alpha^2}\frac{D_{-\gamma-2}\left(-\frac{\beta}{\alpha \sqrt{2}}\right)^2}{D_{-\gamma-1}\left(-\frac{\beta}{\alpha \sqrt{2}}\right)^2}. 
\end{align*}

\begin{figure}[t]
	\centering
	\includegraphics[scale = 0.5]{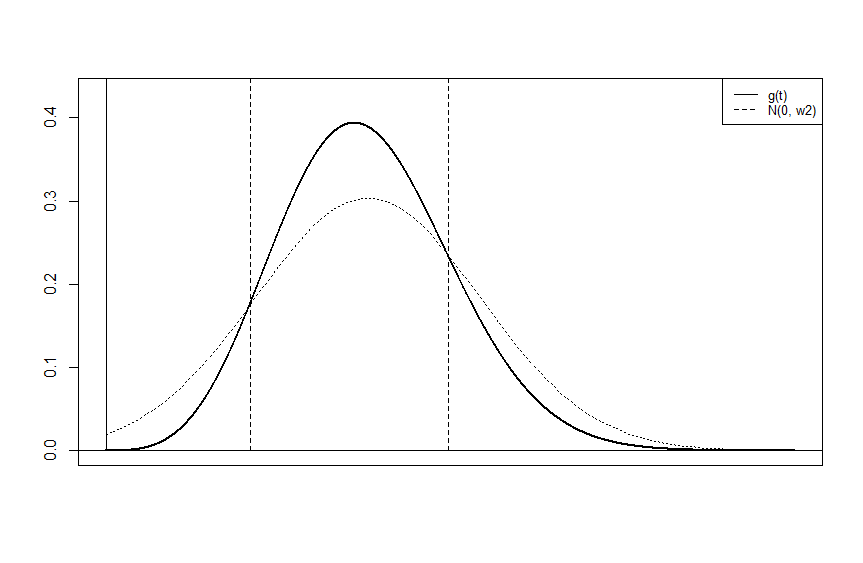}
	\caption{density $g$ and $h$ with $\gamma = 4$, $\alpha = 2.75$, $\beta = 3.3$; dotted lines represent $t_1$ and $t_2$, whereas vertical line is the minimum $-\mu/\sigma$.}
	\label{fig:1}
\end{figure}

\noindent The density $f(x) \sim \mathcal{G}_{3p}\left(\gamma, \alpha, \beta\right)$ is transformed into a standardized distribution $g(t) = \sigma f \left(\sigma t  + \mu\right)$ by the transformation $t = (x - \mu)/\sigma$, with support on the interval $\left(-\frac{\mu}{\sigma}, \infty\right)$. A new value $t_\star$ can be drawn from $g(t)$ using the modified rejection sampling described below. Finally, the value $x_\star = \sigma t_\star + \mu$ is returned. 

\noindent Consider the proposal distribution $h(t) \sim \mathcal{N}\left(0, \omega^2\right)$ and the ratio
\begin{align}
	r(t_\star) = &\; \frac{g(t_\star)}{h(t_\star)} = \frac{\sigma f \left(\sigma t_\star  + \mu\right)}{h(t_\star)} \nonumber\\
	= &\; \omega \sigma C_f \sqrt{2 \pi} \left(\sigma t_\star  + \mu\right)^\gamma e^{-\alpha^2 \left(\sigma t_\star  + \mu\right)^2 + \beta \left(\sigma t_\star  + \mu\right) - \frac{t_\star^2}{2\omega^2}} \cdot \mathbb{I}_{\left(t_\star > -\frac{\mu}{\sigma}\right)} \nonumber\\
	= &\; \omega \sigma C_f \sqrt{2 \pi} \left(\sigma t_\star  + \mu\right)^\gamma e^{\left(\frac{1}{2\omega^2} - \alpha^2 \sigma^2\right) t_\star^2 + \left(\beta  - 2 \mu \alpha^2\right)\sigma t_\star + \beta\mu - \alpha^2\mu^2},
	\label{eq:rt}
\end{align}
where $C_f$ is the normalizing constant of $f(x)$ and $ \left(\beta - 2 \mu \alpha^2\right) < 0$. The analysis of $r(t)$ gives insights on how to correctly choose the variance $\omega^2$ of the proposal distribution $h(t)$, as $r(t)$ needs to be bounded and should go to zero as $t$ increases. For this reason we set the variance to $w^2 = \frac{1}{2 \alpha^2 \sigma^2}$ and the ratio in \eqref{eq:rt} evaluated at $t_\star$ can be re-written as
\begin{equation*}
	r(t_\star) =  \omega \sigma C_f \sqrt{2 \pi} \left(\sigma t_\star  + \mu\right)^\gamma e^{\left(\beta  - 2  \mu \alpha^2\right) \left(\sigma t_\star + \mu\right) + \alpha^2\mu^2},
	\label{eq:ratio}
\end{equation*}
which is analytically tractable. In order to apply a standard rejection sampling, the method requires that $r\left(t_\star\right) \le 1$. However, as shown in Figure \ref{fig:1}, the proposal density $h(t)$ lays below the target density $g(t)$ in the interval $\left[ t_1, t_2\right]$, with 
\begin{align}
	t_1 = & \; \frac{\gamma}{\sigma  \left(\beta  - 2  \mu \alpha^2\right)} W_{0}\left(\frac{\left(\beta  - 2  \mu \alpha^2\right)}{\gamma} \left(\frac{e^{-\alpha^2\mu^2}}{ \omega \sigma C_f \sqrt{2 \pi}}\right)^{\frac{1}{\gamma}}\right) - \frac{\mu}{\sigma}, \nonumber \\
	t_2 = & \; \frac{\gamma}{\sigma  \left(\beta  - 2  \mu \alpha^2\right)} W_{-1}\left(\frac{\left(\beta  - 2  \mu \alpha^2\right)}{\gamma} \left(\frac{e^{-\alpha^2\mu^2}}{ \omega \sigma C_f \sqrt{2 \pi}}\right)^{\frac{1}{\gamma}}\right) - \frac{\mu}{\sigma}, \nonumber 
\end{align} 
where $W$ denotes the Lambert function. It can be analytically shown that $r(t_{max}) \ge 1$, where $t_{max} =  -\frac{\gamma}{\sigma \left(\beta - 2 \mu \alpha^2\right) } -\frac{\mu}{\sigma}$ is the global maximum of the ratio. Therefore, a standard rejection sampling cannot be applied. Noting that in the intervals $\left(-\frac{\mu}{\sigma}, t_1 \right)$ and $\left(t_2, \infty\right)$ it yields $h(t) > g(t)$, the rejection sampling algorithm can be modified as follows:
\begin{itemize}
	\item Step 1: generate a sample $t_\star$ from $h(t)$ and immediately accept $x_\star = \sigma t_\star + \mu$ if $t_1 \le t_\star \le t_2$;
	\item Step 2: if $t_\star < t_1$ or $t_\star > t_2$, generate a sample $u$ from a $\mathcal{U}\left(0, 1\right)$ density and compute $r\left(t_\star\right)$. Accept $x_\star = \sigma t_\star + \mu$ if $u \le r(t_\star)$. The computation of $r(t_\star)$ can often be avoided if an accurate lower bound for the tails of the ratio is available;
	\item Step 3: if Step 2 leads to rejection, take a new sample $t_\star'$ from the distribution $d(t) = g(t) - h(t)$, in the interval $\left[t_1, t_2\right]$ and return $x_\star' = \sigma t_\star' + \mu$. Sampling from $d(t)$ can be achieved by means of a standard rejection sampling, as in \cite{Ahrens-1982}, \cite{Stadlober-1982}. More details about this step can be found in Appendix \ref{app:a2}.
\end{itemize}
The acceptance probability of each step is discussed in Appendix \ref{app:a1}.

\begin{proposition}\label{prep_1}
    The modified rejection sampling defined by steps 1, 2, and 3 draws a sample from a $\mathcal{G}_{3p}$ distribution with probability 1.
\end{proposition}
\noindent \textit{Proof} See Appendix \ref{app:a3}.

\vspace{0.3cm}
\noindent The main computational bottleneck of the method is the evaluation of the Parabolic Cylinder function $D$. This issue can be alleviated by exploiting the following proposition and by the application of sharp approximations.
\begin{proposition}\label{prep_KL}
The Kullback-Leibler divergence (KL) between a distribution $q_x \sim \mathcal{G}_{3p} \left(\gamma, \alpha, \beta\right)$ and a distribution $p_x \sim \mathcal{G} \left(d, c\right)$, where $d = \gamma + 1$ and $c = -\beta$, goes to zero when $\beta / \alpha \to - \infty$.
\end{proposition}
\noindent \textit{Proof} See Appendix \ref{app:a4}.

\vspace{0.3cm}
\noindent Furthermore, when $\beta / \alpha \to \infty$ or $\gamma \to \infty$, the three-parameter Gamma distribution can be conveniently 
approximated by a Normal distribution. We empirically show that, in these cases, the KL divergence between a distribution $q_x \sim \mathcal{G}_{3p} \left(\gamma, \alpha, \beta\right)$ and a distribution $p_x \sim \mathcal{N} \left(m, s^2\right)$ asymptotically goes to $0$, where estimates of $m$ and $s^2$ are given in Appendix \ref{app:a4}. These empirical results, along with proposition \ref{prep_KL}, can be used to efficiently evaluate the mean and variance of the target distribution without the need to compute the function $D$ for some combinations of the parameters' value.


\section{Posterior sampling}
\label{sec:posterior}
%
We develop an efficient MCMC algorithm to sample from the posterior distribution of the parameters. The algorithm can be divided into three main steps: 1. a Gibbs step for the update of parameters $\boldsymbol{\Omega}_1, \dots, \boldsymbol{\Omega}_K$; 2. a Gibbs step for the update of shrinkage parameters $\boldsymbol{\Lambda}_1^2, \dots, \boldsymbol{\Lambda}_K^2$ and $\boldsymbol{\tau}^2$; 3. a Metropolis-Hastings (MH) step for the update of correlation matrix $\mathbf{R}$. In step 2 we make use of the modified rejection sampler introduced in Section \ref{sec:G3p}. The complete algorithm is shown in Appendix A of Supplementary Materials.

\vspace{0.4cm}
\noindent \textbf{\textit{1. Sampling $\boldsymbol{\Omega}_1, \dots, \boldsymbol{\Omega}_K$.}} The full conditional distribution of $\boldsymbol{\Omega}_1, \dots, \boldsymbol{\Omega}_K$ is
\begin{align*}
\label{eq:jdomega}
	\pi \left(\boldsymbol{\Omega}_1, \dots, \boldsymbol{\Omega}_K \vert \cdot\right) \propto & \;  \prod_{k = 1}^K \big|\boldsymbol{\Omega}_k\big|^{\frac{n_k}{2}} \exp\left\{-\frac{1}{2} tr\left(\mathbf{S}_k \boldsymbol{\Omega}_k\right)\right\} \cdot \nonumber \\
	& \; \quad \prod_{i < j} \exp \left\{- \frac{1}{2} \boldsymbol{\omega}_{ij}^\intercal   \boldsymbol{\Delta}_{ij}^{-1} \mathbf{R}^{-1} \boldsymbol{\Delta}_{ij}^{-1} \boldsymbol{\omega}_{ij} \right\} \cdot \mathbb{I}_{\left(\boldsymbol{\Omega}_1, \dots, \boldsymbol{\Omega}_K \in \mathbb{M}_+^K\right)}
\end{align*}
where $\mathbf{S}_k = \sum_{s = 1}^{n_k} \mathbf{y}_{sk} \mathbf{y}_{sk}^\intercal$ and $tr \left( \cdot \right)$ denotes the trace. Precision matrices $\boldsymbol{\Omega}_1, \dots, \boldsymbol{\Omega}_K$ can be updated by adapting the block Gibbs sampler proposed in \cite{Wang-2015} for the estimation of a single precision matrix. Following \cite{Peterson-2020}, for each sample group $k = 1, \dots, K$ precision matrix $\boldsymbol{\Omega}_k$ is updated column-wise by sampling from the full-conditional distribution of each column $j = 1, \dots, p$ conditionally on both the rest of the columns of group $k$ and on the $j$-th column of the reaming $k-1$ sample groups. Consider the following partition of vector $\boldsymbol{\omega}_{ij}$ and matrices $\boldsymbol{\Delta}_{ij}$ and $\mathbf{R}$:
\begin{equation}
	\boldsymbol{\omega}_{ij} = \begin{bmatrix} \boldsymbol{\omega}_{ij}^{ -k} \\ \omega_{ij}^{k} \end{bmatrix}, \quad \boldsymbol{\Delta}_{ij} =  \begin{bmatrix} \boldsymbol{\Delta}_{ij, -k} & \mathbf{0} \\ \mathbf{0}^\intercal & \delta_{ij, k}  \end{bmatrix} \quad \text{and} \quad \mathbf{R} = \begin{bmatrix} \mathbf{R}_{-k} & \mathbf{r}_{k} \\ \mathbf{r}_{k}^\intercal & 1 \end{bmatrix}.
	\label{eq:partition}
\end{equation}
The full conditional of $\boldsymbol{\Omega}_k$ is:
\begin{equation}
\label{eq:omegaK}
	\pi\left(\boldsymbol{\Omega}_k \vert \cdot \right) \propto \vert\boldsymbol{\Omega}_k\vert^{\frac{n_k}{2}} \exp\left\{-\frac{1}{2} tr\left(\mathbf{S}_k \boldsymbol{\Omega}_k\right)\right\} \prod_{i < j} \exp \left\{- \frac{1}{2 d_{ij}^k }\left( \omega_{ij}^k - \delta_{ij, k} \mathbf{r}_k^\intercal \mathbf{R}_{-k}^{-1} \boldsymbol{\Delta}_{ij, -k}^{-1} \boldsymbol{\omega}_{ij}^{-k}\right)^2 \right\},
\end{equation}
where $d_{ij}^k =  \delta_{ij, k}^2 \left( 1 - \mathbf{r}_k^\intercal \mathbf{R}_{-k}^{-1} \mathbf{r}_{k}\right)$. As proposed in \cite{Wang-2015}, sampling from \eqref{eq:omegaK} can be achieved by updating one column of $\boldsymbol{\Omega}_k$ at the time. Without loss of generality, consider the permutation of the columns such that the $j$-th column becomes the last one. This permutation leads to the following partition:
\begin{equation*}
	 \mathbf{S}_k = \begin{bmatrix} \mathbf{S}_{-j}^k & \mathbf{s}_j^k \\ \left(\mathbf{s}_j^k\right)^\intercal & s_{jj}^k  \end{bmatrix} \quad \text{and} \quad \boldsymbol{\Omega}_k = \begin{bmatrix} \boldsymbol{\Omega}_{-j}^k & \boldsymbol{\omega}_j^k \\ \left(\boldsymbol{\omega}_j^k\right)^\intercal & \omega_{jj}^k  \end{bmatrix}.
\end{equation*}
The full-conditional distribution of parameters $\left(\omega_{jj}^k, \boldsymbol{\omega}_j^k\right)$ is
\begin{align}
\label{eq:block}
	\pi\left(\omega_{jj}^k, \boldsymbol{\omega}_j^k \vert \cdot\right) \propto &\; \left(\omega_{jj}^k - \left(\boldsymbol{\omega}_j^k\right)^\intercal \left(\boldsymbol{\Omega}_{-j}^k\right)^{-1}  \boldsymbol{\omega}_j^k\right)^{\frac{n_K}{2}} \cdot  \nonumber\\
		& \qquad e^{-\frac{1}{2} \left(\left( \boldsymbol{\omega}_j^k - \mathbf{m}_{j,k}\right)^\intercal \mathbf{D}_{j,k}^{-1} \left( \boldsymbol{\omega}_j^k - \mathbf{m}_{j,k}\right) + 2 \left(\boldsymbol{\omega}_j^k\right)^\intercal \mathbf{s}_j^k + s_{jj}^k \omega_{jj}^k\right)},
\end{align}
where $\mathbf{m}_{j,k}$ is the $(p-1)$-dimensional vector with entries $m_{j, k}^i =\delta_{ij, k} \mathbf{r}_{k}^\intercal \mathbf{R}_{-k}^{-1} \boldsymbol{\Delta}_{ij, -k}^{-1} \boldsymbol{\omega}_{ij}^{-k}$ and $\mathbf{D}_{j,k}$ is diagonal with entries $d_{ij}^k$, $i = 0, \dots, p, i \ne j$. A closed form for sampling from \eqref{eq:block} can be obtained with the transformation $\left(\mathbf{v}_{j,k}, \gamma_{jj}^k \right) \to \left(\boldsymbol{\omega}_j^k, \omega_{jj}^k - \left(\boldsymbol{\omega}_j^k\right)^\intercal \left(\boldsymbol{\Omega}_{-j}^k\right)^{-1}  \boldsymbol{\omega}_j^k\right)$, which yields
\begin{align*}
	\gamma_{jj}^k \vert \cdot \sim &\; \mathcal{G} \left(\frac{n_k}{2} + 1, \frac{s_{jj}^k}{2}\right),  \\
	\mathbf{v}_{j,k} \vert \cdot \sim &\; \mathcal{N}_{p-1} \left(\boldsymbol{\Sigma}_{j, k}^{-1}\left(\mathbf{D}_{j,k}^{-1}\mathbf{m}_{j,k} - \mathbf{s}_{jj}^k\right), \boldsymbol{\Sigma}_{j, k}^{-1}\right)
\end{align*}
where $\mathcal{G}$ denotes the Gamma distribution and $\boldsymbol{\Sigma}_{j, k} = \mathbf{D}_{j,k}^{-1} + s_{jj}^k\left(\boldsymbol{\Omega}_{-j}^k\right)^{-1} $. Therefore, values $\omega_{jj}^k$ and $\boldsymbol{\omega}_j^k$ can be updated by first sampling $\gamma_{jj}^k$ and $\mathbf{v}_{j,k}$ and then applying the inverse transformation.

Computationally, this is the most expensive step of the algorithm due to the need to invert the matrices $\boldsymbol{\Omega}_{-j}^k$ and $\boldsymbol{\Sigma}_{j,k}$. 
In our implementation of the Gibbs steps for $\gamma_{jj}^k$ and $\mathbf{v}_{j,k}$, we make use of Shermann-Morrison formula to update $\left(\boldsymbol{\Omega}_{-j}^k \right)^{-1}$ with $O(p^2)$ operations, instead of $O(p^3)$.


\vspace{0.4cm}
\noindent \textbf{\textit{2. Sampling $\boldsymbol{\Lambda}_1^2, \dots, \boldsymbol{\Lambda}_K^2$ and $\boldsymbol{\tau}^2$.}} Samplers commonly used in conjunction with Horseshoe prior cannot be implemented for the proposed model. Indeed, the positive half-Cauchy distribution is not conjugated to the variance in a multivariate normal means model. Our approach builds upon the data-augmentation scheme proposed \cite{Makalic-2016}. We introduce the auxiliary variables $\eta_{ij, k}$ and $\zeta_k$ such that
\begin{itemize}
    \item if $\lambda_{ij, k}^2 \mid \eta_{ij, k} \sim \; \mathcal{IG} \left(\frac{1}{2}, \frac{1}{\eta_{ij, k}}\right)$ and $\eta_{ij, k}  \sim \mathcal{IG} \left(\frac{1}{2},1\right)$, then $\lambda_{ij, k} \sim \mathcal{C}^+ \left(0, 1\right)$;
    \item if $\tau_k^2 \mid \zeta_k \sim \; \mathcal{IG} \left(\frac{1}{2}, \frac{1}{\zeta_k}\right)$ and $\zeta_k  \sim \mathcal{IG} \left(\frac{1}{2},1\right)$, then $\tau_k \sim \mathcal{C}^+ \left(0, 1\right)$.
\end{itemize}
After conditioning on the auxiliary variables  $\eta_{ij,k}$ and $\zeta_k$, the full conditional distribution of parameters $\boldsymbol{\Lambda}$ and $\boldsymbol{\tau}$ can be written as
\begin{align*}
	\pi\left(\boldsymbol{\Lambda}, \boldsymbol{\tau} \vert \cdot \right) \propto & \; \prod_{i<j} \vert \boldsymbol{\Delta}_{ij} \vert^{-1} \exp\left\{ -\frac{1}{2} \boldsymbol{\omega}_{ij}^\intercal \left( \boldsymbol{\Delta}_{ij} \mathbf{R} \boldsymbol{\Delta}_{ij}\right)^{-1} \boldsymbol{\omega}_{ij}\right\} \cdot \nonumber \\
	& \; \quad \prod_{k = 1}^K \tau_k^{-3} \exp\left\{-\frac{1}{\zeta_k \tau_k^2}\right\} \cdot\prod_{i<j} \lambda_{ij, k}^{-3} \exp\left\{-\frac{1}{\eta_{ij, k} \lambda_{ij, k}^2}\right\} . 
\end{align*}
Local shrinkage matrix $\boldsymbol{\Lambda}_k$ is updated column-wise alongside precision matrix $\boldsymbol{\Omega}_k$. Considering the partition of $\boldsymbol{\omega}_{ij}$, $\boldsymbol{\Delta}_{ij}$ and $\mathbf{R}$ in \eqref{eq:partition}, the full-conditionals of parameters $\lambda_{ij, k}^2$ and $\tau_k^2$ related to group $k$ are
\begin{align}
	\pi\left(\lambda_{ij, k}^2 \mid \cdot \right) \propto & \; \lambda_{ij, k}^{-4} \exp\left\{-\frac{\alpha_{\lambda_{ij,k}}}{\lambda_{ij, k}^2} + \frac{\beta_{\lambda_{ij, k}}}{\lambda_{ij, k} }  \right\} \cdot \mathbb{I}_{\left(\lambda_{ij,k}^2 > 0\right)}, \nonumber \\
	& \; \alpha_{\lambda_{ij,k}} = \frac{1}{\eta_{ij, k}} + \frac{\left(\omega_{ij}^k\right)^2}{2 \tau_k^2 \mu_k} \quad \text{and} \quad \beta_{\lambda_{ij, k}} = \frac{\omega_{ij}^k}{ \tau_k \mu_k}  \mathbf{r}_{k}^\intercal \mathbf{R}_{-k}^{-1} \boldsymbol{\Delta}_{ij, -k}^{-1} \boldsymbol{\omega}_{ij}^{-k}, \label{eq:Glambda}\\
	\pi\left(\tau_k^2 \mid \cdot \right) \propto & \; \tau_k^{-\frac{p(p-1)}{2}-3} \exp\left\{- \frac{\alpha_{\tau_k}}{\tau_k^2} + \frac{\beta_{\tau_k}}{\tau_k}  \right\} \cdot  \mathbb{I}_{\left(\tau_k^2 > 0\right)} , \nonumber\\
	& \; \alpha_{\tau_k} =\frac{1}{\zeta_k} + \frac{1}{2}\sum_{i<j}\frac{\left(\omega_{ij}^k\right)^2}{\lambda_{ij, k}^2  \mu_k} \quad \text{and} \quad \beta_{\tau_k} = \sum_{i<j} \frac{\omega_{ij}^k}{ \lambda_{ij, k} \mu_k}  \mathbf{r}_{k}^\intercal \mathbf{R}_{-k}^{-1} \boldsymbol{\Delta}_{ij, -k}^{-1} \boldsymbol{\omega}_{ij}^{-k} \label{eq:Gtau}
\end{align}
where $\mu_k =  1 - \mathbf{r}_{k}^\intercal \mathbf{R}_{-k}^{-1} \mathbf{r}_{k}$. Note that the full conditional distributions show a shared amount of global and local shrinkage, as the model exploits the similarity among groups and learns from the structures of the other graphs. Densities \eqref{eq:Glambda} and \eqref{eq:Gtau} are a transformation of $\mathcal{G}_{3p}$ random variables introduced in Section \ref{sec:G3p}. Specifically,
\begin{align}
	& \text{if} \quad u \sim \mathcal{G}_{3p} \left(1, \alpha_{\lambda_{ij,k}}, \beta_{\lambda_{ij,k}}\right), \quad \text{then} \quad \lambda_{ij,k}^2 = 1 / u^2, \nonumber\\
	& \text{if} \quad u \sim \mathcal{G}_{3p} \left(p(p-1)/2, \alpha_{\tau_k}, \beta_{\tau_k}\right), \quad \text{then} \quad \tau^2_k = 1 / u^2. \nonumber
\end{align}
We use the sampling algorithm introduced in Section \ref{sec:G3p} to efficiently obtain samples from these distributions. Finally, hyper-parameters $\eta_{ij, k}$ and $\zeta_k$ are updated by sampling from the inverse-Gamma distributions $\eta_{ij, k} \sim \mathcal{IG} \left(1, 1 + 1/ \lambda_{ij, k}^2\right)$ and $\zeta_k \sim \mathcal{IG} \left(1, 1 + 1/\tau_k^2\right)$.


\vspace{0.4cm}
\noindent \textbf{\textit{3. Sampling $\mathbf{R}$.}} The similarity among groups is captured through correlation matrix $\mathbf{R} \in \mathbb{C}_+^K$. Following \cite{Peterson-2020}, 
we implement a modified version of the Metropolis-Hastings sampler proposed by \cite{Liu-2006}, which relies on a candidate prior distribution $\pi^\star\left(\mathbf{R}\right)$ that is used to define a proposal distribution for correlation matrices. In the first step of this data-augmentation approach a $K \times K$ covariance matrix $\boldsymbol{\Theta}$ is sampled from an Inverse-Wishart distribution; in the second step, a reduction function is applied to map the covariance matrix to a valid correlation matrix, that is eventually accepted with an MH step. 

\indent We introduce a diagonal matrix $\mathbf{V}$ such that $\boldsymbol{\Theta} = \mathbf{V} \mathbf{R} \mathbf{V}$; the matrix $\mathbf{V}$ maps the correlation matrix $\mathbf{R}$ to the covariance matrix $\boldsymbol{\Theta}$. 
Following \cite{Peterson-2020}, the transformation from the standard parameter space to the expanded space is achieved as
\begin{equation}
	\boldsymbol{\omega}_{ij} = \mathbf{V}^{-1} \boldsymbol{\epsilon}_{ij}, \quad
	\mathbf{R} = \mathbf{V}^{-1} \boldsymbol{\Theta} \mathbf{V}^{-1},
	 \label{eq:a1}
\end{equation}
where $\sum_{i < j} \epsilon_{ijk}^2 = 1$, for $k = 1, \dots, K$ and $\mathbf{V} = \text{diag} \left\{\sum_{i<j} \left(\omega_{ij}^{1}\right)^2, \dots, \sum_{i<j} \left(\omega_{ij}^K\right)^2\right\}$. Let the candidate prior distribution be
\begin{equation}
	\label{eq:propR}
	\pi^\star\left(\mathbf{R}\right) \propto \big| \mathbf{R}\big|^{-\frac{K+1}{2}} \cdot \mathbb{I}_{\left(\mathbf{R}\in \mathbb{C}_+^K\right)},
\end{equation}
then the proposal density for matrix $\mathbf{R}$ is
\begin{align*}
	q\left(\mathbf{R} \mid \cdot\right) \propto & \; \pi^\star\left(\mathbf{R}\right) \cdot \pi\left(\boldsymbol{\Omega}_1, \dots, \boldsymbol{\Omega}_K \mid \mathbf{R}\right) \nonumber \\
	\propto & \; \left| \mathbf{R} \right|^{-\frac{K+1 + p(p-1)/2}{2}} \cdot \prod_{i<j} e^{-\frac{1}{2} \boldsymbol{\omega}_{ij}^\intercal \boldsymbol{\Delta}_{ij}^{-1} \mathbf{R}^{-1} \boldsymbol{\Delta}_{ij}^{-1} \boldsymbol{\omega}_{ij}},
\end{align*}
which is conditioned on the current state of the algorithm and accounts for the dependency with parameters $\boldsymbol{\Omega}_1, \dots, \boldsymbol{\Omega}_K$, $\boldsymbol{\Lambda}_1, \dots, \boldsymbol{\Lambda}_K$ and $\boldsymbol{\tau}^2$. Note that \eqref{eq:propR} concentrates its mass around zero when $K$ increases; for this reason, a reasonably small number of sample groups $K$ is required. The Jacobian of the transformation defined in \eqref{eq:a1} is $J = \left| \mathbf{V}^{-1}\right|^{\frac{p(p-1)}{2} + K + 1}$, thus the proposal distribution for the MH sampler is
\begin{align}
	q\left(\boldsymbol{\Theta} \mid \cdot\right) \propto &\; \pi^\star\left(\boldsymbol{\Theta}\right) \cdot \pi\left(\boldsymbol{\Omega}_1, \dots, \boldsymbol{\Omega}_K \mid \boldsymbol{\Theta}\right) \nonumber \\
        \propto & \; \left|\boldsymbol{\Theta} \right|^{-\frac{K+1 + p(p-1)/2}{2}} e^{-\frac{1}{2}\sum_{i < j} \boldsymbol{\epsilon}_{ij}^\intercal \boldsymbol{\Delta}_{ij}^{-1} \boldsymbol{\Theta}^{-1} \boldsymbol{\Delta}_{ij}^{-1} \boldsymbol{\epsilon}_{ij}}
	\label{eq:postTheta}
\end{align}
which is a $\mathcal{IW} \left(\frac{p(p-1)}{2}, \mathbf{H}\right)$, where $\mathbf{H} = \sum_{i < j}  \boldsymbol{\Delta}_{ij}^{-1} \boldsymbol{\epsilon}_{ij} \boldsymbol{\epsilon}_{ij}^\intercal \boldsymbol{\Delta}_{ij}^{-1} $. Therefore, a candidate $\boldsymbol{\Theta}^\star$ is sampled from \eqref{eq:postTheta} and then mapped to $\mathbf{R}^\star$ via the inverse transformation $\mathbf{R}^\star = \mathbf{V}^{-1} \boldsymbol{\Theta}^\star \mathbf{V}^{-1}$. New correlation matrix $\mathbf{R}^\star$ is accepted with probability
\begin{align*}
	\alpha = &\; \text{min} \left\{1, \frac{\pi\left(\mathbf{R}^\star \mid \cdot\right) \cdot q\left(\mathbf{R} \mid \cdot \right)}{\pi\left(\mathbf{R} \mid \cdot\right) \cdot q\left(\mathbf{R}^\star \mid \cdot \right)}\right\} \nonumber \\
	= & \; \text{min} \left\{1, e^{\frac{K+1}{2} \left(\log{\left|\mathbf{R}^\star\right|} - \log{\left|\mathbf{R}\right|}\right)} \right\},
\end{align*}
where $p\left(\mathbf{R} \mid \cdot \right) \propto \pi\left(\mathbf{R}\right) \cdot \pi\left(\boldsymbol{\Omega}_1, \dots, \boldsymbol{\Omega}_K \mid \mathbf{R}\right)$ denotes the full-conditional distribution of $\mathbf{R}$.

\section{Posterior edge selection}
\label{sec:edgesselection}
%
A practical problem with continuous shrinkage priors is model selection since the parameters are shrunk toward zero but never exactly zero. A common method relies on posterior marginal credible intervals. However, \cite{vdPas-2017} have shown that under the Horseshoe prior in a Normal means problem, this method leads to a conservative variables selection procedure where some of the zero parameters are falsely selected, whereas some signal is not, due to wide intervals for non-zero parameters. To avoid such a problem, \cite{Bahdra-2019} used $50\%$ credible intervals to control the number of false negatives. This choice is in line with the \textit{median probability model} (MPM) of \cite{Barbieri-2004}. The MPM model is defined as the model that includes only those edges with marginal posterior probability greater (or equal) than $1/2$. In the context of linear regression models, \cite{Barbieri-2004} have shown that this method represents the predictive optimal model under some common but strict hypothesis, such as orthogonality of the covariates. The result is extended to $g$-type spike and slab priors in \cite{Barbieri-2021}. This approach is used, among many others, in \cite{Wang-2015} and \cite{Peterson-2020}. A practical example of an MPM-like strategy can be found in \cite{Carvalho-2010}. The authors show that the Horseshoe estimator is $\beta_j^{\text{HS}} = \lambda_{j}^2 / \left(1 + \lambda_{j}^2\right) \beta_j^{\text{OLS}}$, where $\lambda_j^2$ and $\beta_j$ denote the local shrinkage parameter and the regression parameter of variable $j$, respectively, and propose to set to zero those variables for which $\lambda_{j}^2 / \left(1 + \lambda_{j}^2\right) < 1/2$. 

The cited methods present two main drawbacks. First, the optimality results in \cite{Barbieri-2004} only hold for fixed design $\tilde{\mathbf{X}}$ of prediction point or for stochastic predictors with $\mathbb{E}\left(\tilde{\mathbf{X}}^\intercal \tilde{\mathbf{X}}\right)$, which are often unrealistic assumptions; therefore, the threshold $1/2$ does not ensure the optimality of the selected model under the considered framework, where the goal is to analyze the connections between variables. Secondly, the considered selection procedures rely on marginal values and do not account for any posterior correlation among the parameters.

To overcome these problems, we propose a ``quasi-bayesian'' approach for edge selection that accounts for the posterior dependencies among the parameters. The method relies on a \textit{cut} function that ``cuts'' the relationship between the parameters to prevent model feedback which could negatively affect the performances of the model \citep{Zigler-2013,plummer-2015}. Cuts have been used in different contexts \citep{Lunn-2009,Liu-2009,McCandless-2010,Blangiardo-2011,Zigler-2016} either to control the flow of information or to gain a computational advantage. \cite{Liu-2009} consider the cut function as a ``\textit{modularization}'' of the model. This approach breaks a bigger model into smaller parts called modules, modifying the magnitude of the interactions between the parameters in different modules.

\subsection{An extended model and algorithm for edge selection}
\label{subsec:es}
In this section, we extend the model presented in the previous sections introducing two parameters $t_\alpha$ and $\mathbf{z}$, and an algorithm that updates these parameters with a Metropolis-within-Gibbs step. Notation refers to a single graph and can be easily extended to the case of multiple graphs.

\begin{figure}
	\centering
	\includegraphics[scale = 0.12]{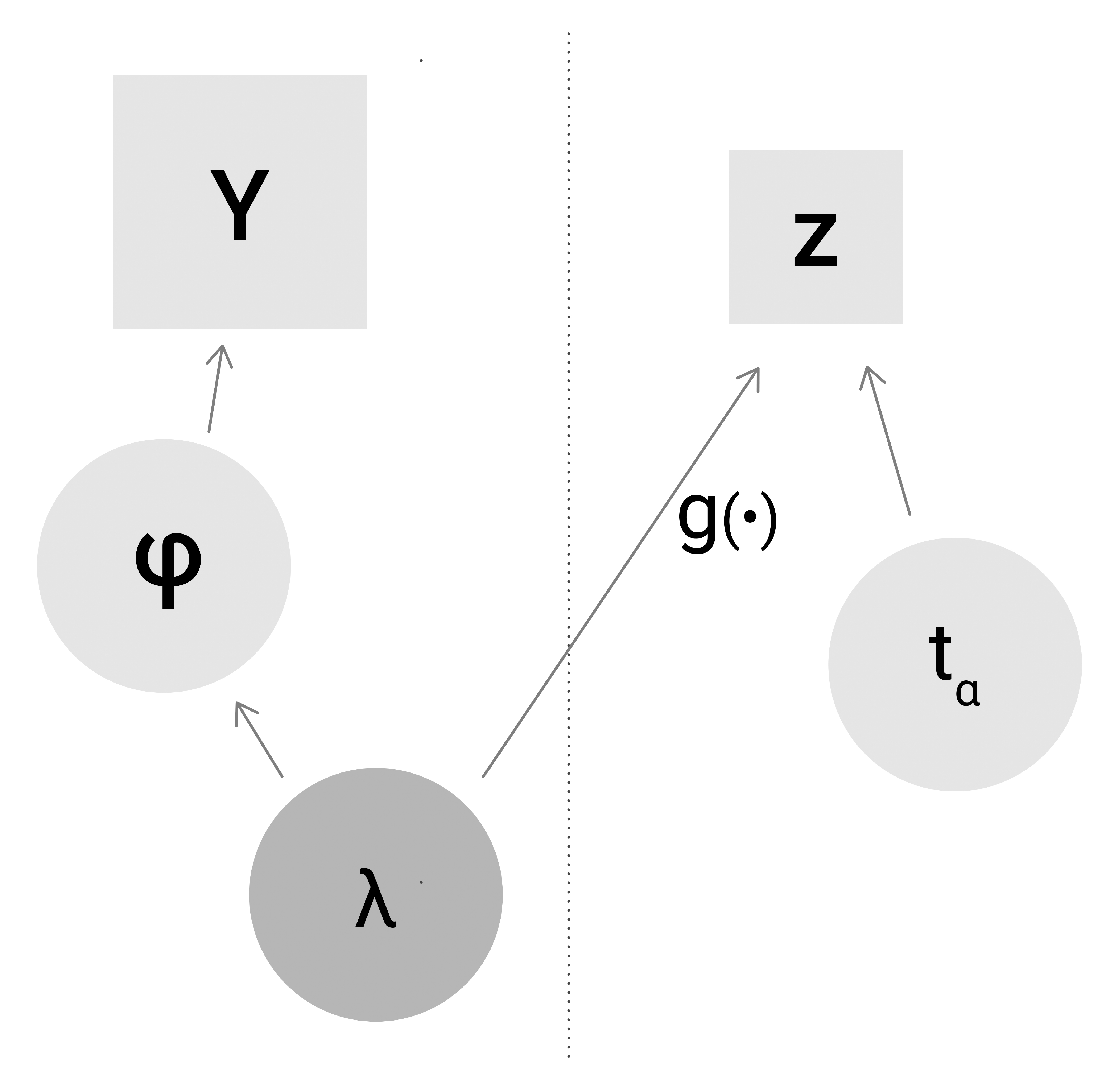}
	\caption{Graphical representation of the model. The dotted line denotes the cut function, stopping the flows of information from $\mathbf{z}$ to $\boldsymbol{\lambda}$.}
	\label{fig:cut}
\end{figure}

The parameter $t_\alpha \in \left(0, 1\right)$ can be interpreted as a threshold for edge selection, and the latent variable $\mathbf{z}$ is a $p(p-1)/2$-binary vector with generic element $z_{ij} = 1$ if the corresponding edge $\omega_{ij}$, $i<j$, is included in the model, $z_{ij} = 0$ otherwise. Formally, the model is defined as
\begin{equation*}
	z_{ij} = 1 \text{ if } \kappa_{ij} \ge t_\alpha, \text{ and} \quad z_{ij} = 0 \text{ otherwise},
\end{equation*}	
where $\kappa_{ij} = \lambda_{ij}^2 / \left(1 + \lambda_{ij}^2\right)$. Here the goal is to estimate parameter $t_\alpha$ based on the posterior values of $\boldsymbol{\lambda}$. At the same time, we want to prevent the flow of information from $t_\alpha$ and $\mathbf{z}$ to $\boldsymbol{\lambda}$. The cut function comes in handy to avoid such issues. The modularization of the proposed model is shown in Figure \ref{fig:cut}, where $\boldsymbol{\varphi} = \left(\boldsymbol{\Omega}, \boldsymbol{\tau}, \mathbf{R}\right)$ and parameters $\mathbf{z}$ and $\boldsymbol{\lambda}$ are connected through the reparametrization $\kappa_{ij}$. 

Different prior distributions can be assumed for $t_\alpha$; a natural choice is $t^\alpha \sim \mathsf{Beta}(a, b)$. Parameters $z_{ij}$ can be seen as the realization of $p(p-1)/2$ Bernoulli distributions $z_{ij} \mid \kappa_{ij}, \boldsymbol{\varphi}, t_\alpha \sim \mathsf{Ber}\left(q_{ij}^\alpha\right)$, where $q_{ij}^\alpha = 1 - \mathbb{P}\left(\kappa_{ij} \le t^\alpha \mid \boldsymbol{\varphi}\right)$. The joint likelihood of the model can be factorized as
\begin{align*}
	\pi\left(\mathbf{Y}, \boldsymbol{\lambda}, \boldsymbol{\varphi}, \mathbf{z}, t_\alpha  \right) \propto & \; \pi\left( \boldsymbol{\varphi}\mid \mathbf{Y}, \boldsymbol{\lambda}\right) \pi\left(\mathbf{z}, t_\alpha\mid \boldsymbol{\kappa}, \boldsymbol{\varphi}\right) \pi\left(\boldsymbol{\lambda}\right), \nonumber \\
	\propto & \; \pi\left(\boldsymbol{\lambda}, \boldsymbol{\varphi} \mid \mathbf{Y}\right) \pi\left(\mathbf{z}, t_\alpha\mid \boldsymbol{\kappa}, \boldsymbol{\varphi} \right).
\end{align*}
The modularization of the model allows us to sample directly from the conditional distributions $\pi\left(\boldsymbol{\lambda}, \boldsymbol{\varphi} \mid \mathbf{Y}\right)$ and $\pi\left(\mathbf{z}, t_\alpha\mid \boldsymbol{\kappa}, \boldsymbol{\varphi}\right)$, thus evaluating parameters $\boldsymbol{\lambda}$ and $\boldsymbol{\varphi}$ without the influence of the unknown quantity $\mathbf{z}$. The joint posterior distribution of parameters $t_\alpha$ and $\mathbf{z}$ is
\begin{equation}
	\pi\left(\mathbf{z}, t^\alpha \mid \boldsymbol{\kappa}, \boldsymbol{\varphi} \right) \propto \left(t^\alpha\right)^{a-1} \left(1 - t^\alpha\right)^{b-1} \cdot \prod_{j = 1}^p \prod_{i < j} \left(q_{ij}^\alpha\right)^{z_{ij}} \left(1 - q_{ij}^\alpha\right)^{1-z_{ij}} .
	\label{eq:taa}
\end{equation}
We propose a Metropolis-within-Gibbs algorithm in order to sample from \eqref{eq:taa}. Parameters $z_{ij}$ are sampled from the full-conditional distribution 
\begin{equation*}
    z_{ij} \mid \kappa_{ij}, \boldsymbol{\varphi}, t^\alpha \sim  \mathsf{Ber}\left(q_{ij}^\alpha\right).
\end{equation*}
Under the framework introduced in Section \ref{sec:posterior}, the transformation $\kappa_{ij} = \lambda_{ij}^2 / \left(1 + \lambda_{ij}^2\right)$ with Jacobian $J_{\kappa_{ij}} = \left(1 + \kappa_{ij}\right)^{-2}$ yields
\begin{equation*}
	\pi\left(\kappa_{ij} \mid \mathbf{Y}, \boldsymbol{\varphi}\right) \propto \kappa_{ij}^{-2} \exp\left\{-\alpha_{\lambda_{ij}}\frac{1-\kappa_{ij}}{\kappa_{ij}} + \beta_{\lambda_{ij}}\sqrt{\frac{1-\kappa_{ij}}{\kappa_{ij}}}\right\} \cdot \mathbb{I}_{\left(\kappa_{ij} \in (0, 1)\right)},
\end{equation*}
where the cumulative density function $F_{\kappa_{ij} \mid \mathbf{Y}, \boldsymbol{\varphi}} \left(t^\alpha\right)$ is available in closed form. Therefore, the quantity $q_{ij}^\alpha$ can be analytically and efficiently computed conditionally on the current state of $\boldsymbol{\varphi}$. 

The threshold $t^\alpha$ is then updated with a MH step, where the new values $t^\alpha_\star$ are sampled from the prior distribution. The acceptance probability of this step is
\begin{equation*}
	\alpha_{\text{MH}} = \min\left\{1, \frac{\pi\left(\mathbf{z}, t^\alpha_\star \mid \boldsymbol{\kappa}, \boldsymbol{\varphi}\right)}{\pi\left(\mathbf{z}, t^\alpha \mid \boldsymbol{\kappa}, \boldsymbol{\varphi}\right)}\right\}.
\end{equation*}
The sampled values of $t^\alpha$ can be used to perform graph selection; specifically, we include in the graph all edges such that $P(\kappa_{ij} \mid \mathbf{Y}, \boldsymbol{\varphi})>t^\alpha$. Hereafter, we consider both this approach and the MPM method \citep[$t^\alpha = 1/2$;][]{Carvalho-2010} as two alternative approaches to posterior edge selection.

\section{Simulation studies}
\label{sec:simulation}
%
We perform simulation studies that cover several scenarios of interest. The performances of the proposed model and competing approaches are tested in four scenarios all comprising $K = 4$ groups:
\begin{itemize}
	\item[$\bullet$] \textbf{Independence set-up}: the groups are simulated from multivariate Gaussian distributions with a different precision matrix for each group;
	\item[$\bullet$] \textbf{Coupled set-up}: each pair of groups is simulated from a multivariate Gaussian distribution with the same precision matrix;
	\item[$\bullet$] \textbf{P2020 set-up}: the groups are simulated following the scheme of \cite{Peterson-2020}, where each precision matrix is created by adding (deleting) new edges to (from) the other precision matrices;
	\item[$\bullet$] \textbf{Full-dependence set-up}: the groups are simulated from multivariate Gaussian distributions with equal precision matrices.
\end{itemize}
The precision matrices are simulated following the approach of \cite{Peterson-2020}, which relies on a generalization of the method proposed by \cite{Danaher-2014}. Edges are divided into independent subgroups with size either equal to $5$ or $10$. 
Diagonal entries of the precision matrices are set to $1$. We test our model against the fused and grouped Graphical LASSO (fJGL and gJGL, respectively) of \cite{Danaher-2014}, the ordinary Graphical Horseshoe (GHS) of \cite{Bahdra-2019} estimated for each group independently, and the group estimation of multiple Bayesian graphical models (GemBAG) from \cite{Yang-2021}. Among all competing approaches, the proposed approach is the only one that provides uncertainty quantification through posterior inference on all model parameters. 

Different combinations of $n$ and $p$ are evaluated, and the results are reported in Tables \ref{tab:simn50p50}-\ref{tab:simn100p500}, where $p_0$ refers to the mean number of true significant edges across groups. Edge selection is assessed based on accuracy, the Matthews correlation coefficient (MCC), true and false positive rate (TPR and FPR, respectively) and the AUC criterion. We take the mean Frobenius loss among groups matrices to evaluate the goodness of the precision matrices estimates. Subscripts MPM and $t_\alpha$ indicate whether the posterior edge selection is performed based on the MPM method or with the cut-model proposed in Section \ref{sec:edgesselection}, respectively. Hyperparameters $a$ and $b$ of the Beta prior on $t_\alpha$ should reflect prior beliefs in graphs' sparsity; 
to control the number of false positives, we set $a = 30$ and $b = 25$. 
For the fused and grouped Graphical LASSO, regulation parameters $\lambda_1$ and $\lambda_2$ are selected by performing a grid search to find the combination of values that minimizes the AIC \citep{Danaher-2014,Peterson-2020}. For GemBAG, hyperparameters related to the two levels of sparseness are set to $p_1 = 0.4$ and $p_2 = 0.8$ for all the considered cases. Prior variances $v_0$ and $v_1$ are estimated by minimizing the BIC criterion over a grid of values, as done in \cite{Yang-2021}.

In all scenarios, see tables \ref{tab:simn50p50}-\ref{tab:simn100p500}, mGHS performs better than GHS applied to each group separately when the groups are actually similar, as it provides better selection performances in all the coupled, P2020 and full-dependence settings. Moreover, our model is the only competitor able to approach the performances of the GHS in the independent set-up. Indeed, in this case the latter shows better performances than all the other competitors for all the considered values of $n$ and $p$, whereas the Graphical LASSO and GemBAG behave poorly and their selection results worsen as $p$ increases.

The P2020 set-up provides the most realistic scheme, where the groups have similar but different precision matrices. Under these circumstances, the best model is GemBAG, which gives higher values of MCC and AUC for $p \ge 100$. The only competitive model is mGHS, which has the highest AUC when $p = 50$ and it is the only competitor able to approach GemBAG's performances in the other considered cases.


In this simulation study, edge selection based on the cut model completely overtakes the selection procedure based on the MPM model. Indeed, the approach based on cuts strongly reduces the number of false discoveries, resulting in a higher value of the MCC index. Note that the value of the estimated threshold is affected by the choice of the prior distribution of $t^{\alpha}$. We used $t^\alpha \sim \mathsf{Beta}(30, 25)$ across all simulation scenarios and data analyses; in our experience, this is a viable option that leads to control of the FPR
even though different choices may lead to a different level of sparsity in the estimated graphs.

Finally, the GemBAG and fJGL provide the lowest values of the Frobenius loss. Except for the independent setting, none of the other methods gives better performances in terms of precision matrices estimation. GemBAG is the most efficient method, as it takes an average of only a few hours for the estimation of a network with $p = 500$. On the contrary, the mGHS provides a fully Bayesian inference at the cost of a 10-fold increase in computational time. GHS and Graphical LASSO have not been included in this case, as the computational time increases dramatically.

\begin{table}[H]
\begin{adjustbox}{width = \textwidth}
\begin{tabular}{|c|cccccc|c|cccccc|}
\hline
\multirow{2}{*}{$n = 50, p = 50$} & \multicolumn{6}{c|}{$Independent$ ($p_0 = 82.5$)}                                                                                                                                                                                                                                                                                                                                                                                                                                                                                                   &                   & \multicolumn{6}{c|}{$Coupled$ ($p_0 = 77.5$)}                                                                                                                                                                                                                                                                                                                                                                                                                                                                                                       \\ \cline{2-14} 
                                  & \multicolumn{1}{c|}{Acc}                                                                & \multicolumn{1}{c|}{MCC}                                                                & \multicolumn{1}{c|}{TPR}                                                                & \multicolumn{1}{c|}{FPR}                                                                & \multicolumn{1}{c|}{AUC}                                                                & Fr Loss                                                            & \multirow{7}{*}{} & \multicolumn{1}{c|}{Acc}                                                                & \multicolumn{1}{c|}{MCC}                                                                & \multicolumn{1}{c|}{TPR}                                                                & \multicolumn{1}{c|}{FPR}                                                                & \multicolumn{1}{c|}{AUC}                                                                & Fr Loss                                                            \\ \cline{1-14} \cline{1-14} \cline{1-14}
$\mathsf{mGHS}_{\mathsf{MPM}}$        & \multicolumn{1}{c|}{\begin{tabular}[c]{@{}c@{}}0.775\\ (0.018)\end{tabular}}            & \multicolumn{1}{c|}{\begin{tabular}[c]{@{}c@{}}0.299\\ (0.030)\end{tabular}}            & \multicolumn{1}{c|}{\begin{tabular}[c]{@{}c@{}}0.744\\ (0.040)\end{tabular}}            & \multicolumn{1}{c|}{\begin{tabular}[c]{@{}c@{}}0.223\\ (0.019)\end{tabular}}            & \multicolumn{1}{c|}{\begin{tabular}[c]{@{}c@{}}0.824\\ (0.027)\end{tabular}}            & \begin{tabular}[c]{@{}c@{}}10.231\\ (1.224)\end{tabular}           &                   & \multicolumn{1}{c|}{\begin{tabular}[c]{@{}c@{}}0.715\\ (0.039)\end{tabular}}            & \multicolumn{1}{c|}{\begin{tabular}[c]{@{}c@{}}0.230\\ (0.039)\end{tabular}}            & \multicolumn{1}{c|}{\begin{tabular}[c]{@{}c@{}}$\mathbf{0.723}$\\ (0.048)\end{tabular}} & \multicolumn{1}{c|}{\begin{tabular}[c]{@{}c@{}}0.286\\ (0.041)\end{tabular}}            & \multicolumn{1}{c|}{\begin{tabular}[c]{@{}c@{}}$\mathbf{0.789}$\\ (0.037)\end{tabular}} & \begin{tabular}[c]{@{}c@{}}8.624\\ (1.029)\end{tabular}            \\ \cline{1-7} \cline{9-14} 
$\mathsf{mGHS}_{t_\alpha}$          & \multicolumn{1}{c|}{\begin{tabular}[c]{@{}c@{}}0.926\\ (0.008)\end{tabular}}            & \multicolumn{1}{c|}{\begin{tabular}[c]{@{}c@{}}$\mathbf{0.459}$\\ (0.038)\end{tabular}} & \multicolumn{1}{c|}{\begin{tabular}[c]{@{}c@{}}0.544\\ (0.052)\end{tabular}}            & \multicolumn{1}{c|}{\begin{tabular}[c]{@{}c@{}}0.046\\ (0.009)\end{tabular}}            & \multicolumn{1}{c|}{\begin{tabular}[c]{@{}c@{}}0.824\\ (0.027)\end{tabular}}            & \begin{tabular}[c]{@{}c@{}}10.231\\ (1.224)\end{tabular}           &                   & \multicolumn{1}{c|}{\begin{tabular}[c]{@{}c@{}}0.930\\ (0.009)\end{tabular}}            & \multicolumn{1}{c|}{\begin{tabular}[c]{@{}c@{}}$\mathbf{0.392}$\\ (0.055)\end{tabular}} & \multicolumn{1}{c|}{\begin{tabular}[c]{@{}c@{}}0.421\\ (0.086)\end{tabular}}            & \multicolumn{1}{c|}{\begin{tabular}[c]{@{}c@{}}0.035\\ (0.012)\end{tabular}}            & \multicolumn{1}{c|}{\begin{tabular}[c]{@{}c@{}}$\mathbf{0.789}$\\ (0.037)\end{tabular}} & \begin{tabular}[c]{@{}c@{}}8.624\\ (1.209)\end{tabular}            \\ \cline{1-7} \cline{9-14} 
$\mathsf{GHS}_{\mathsf{MPM}}$         & \multicolumn{1}{c|}{\begin{tabular}[c]{@{}c@{}}0.786\\ (0.015)\end{tabular}}            & \multicolumn{1}{c|}{\begin{tabular}[c]{@{}c@{}}0.315\\ (0.029)\end{tabular}}            & \multicolumn{1}{c|}{\begin{tabular}[c]{@{}c@{}}$\mathbf{0.754}$\\ (0.037)\end{tabular}} & \multicolumn{1}{c|}{\begin{tabular}[c]{@{}c@{}}0.211\\ (0.015)\end{tabular}}            & \multicolumn{1}{c|}{\begin{tabular}[c]{@{}c@{}}$\mathbf{0.840}$\\ (0.024)\end{tabular}} & \begin{tabular}[c]{@{}c@{}}10.199\\ (1.246)\end{tabular}           &                   & \multicolumn{1}{c|}{\begin{tabular}[c]{@{}c@{}}0.702\\ (0.047)\end{tabular}}            & \multicolumn{1}{c|}{\begin{tabular}[c]{@{}c@{}}0.204\\ (0.044)\end{tabular}}            & \multicolumn{1}{c|}{\begin{tabular}[c]{@{}c@{}}0.684\\ (0.048)\end{tabular}}            & \multicolumn{1}{c|}{\begin{tabular}[c]{@{}c@{}}0.297\\ (0.049)\end{tabular}}            & \multicolumn{1}{c|}{\begin{tabular}[c]{@{}c@{}}0.760\\ (0.040)\end{tabular}}            & \begin{tabular}[c]{@{}c@{}}8.745\\ (0.940)\end{tabular}            \\ \cline{1-7} \cline{9-14} 
$\mathsf{fJGL}$                             & \multicolumn{1}{c|}{\begin{tabular}[c]{@{}c@{}}0.873\\ (0.024)\end{tabular}}            & \multicolumn{1}{c|}{\begin{tabular}[c]{@{}c@{}}0.384\\ (0.037)\end{tabular}}            & \multicolumn{1}{c|}{\begin{tabular}[c]{@{}c@{}}0.648\\ (0.063)\end{tabular}}            & \multicolumn{1}{c|}{\begin{tabular}[c]{@{}c@{}}0.110\\ (0.029)\end{tabular}}            & \multicolumn{1}{c|}{\begin{tabular}[c]{@{}c@{}}0.769\\ (0.024)\end{tabular}}            & \begin{tabular}[c]{@{}c@{}}$\mathbf{9.186}$\\ (0.709)\end{tabular} &                   & \multicolumn{1}{c|}{\begin{tabular}[c]{@{}c@{}}0.907\\ (0.021)\end{tabular}}            & \multicolumn{1}{c|}{\begin{tabular}[c]{@{}c@{}}0.333\\ (0.044)\end{tabular}}            & \multicolumn{1}{c|}{\begin{tabular}[c]{@{}c@{}}0.437\\ (0.088)\end{tabular}}            & \multicolumn{1}{c|}{\begin{tabular}[c]{@{}c@{}}0.061\\ (0.026)\end{tabular}}            & \multicolumn{1}{c|}{\begin{tabular}[c]{@{}c@{}}0.688\\ (0.034)\end{tabular}}            & \begin{tabular}[c]{@{}c@{}}$\mathbf{7.863}$\\ (0.535)\end{tabular} \\ \cline{1-7} \cline{9-14} 
$\mathsf{gJGL}$                              & \multicolumn{1}{c|}{\begin{tabular}[c]{@{}c@{}}0.874\\ (0.024)\end{tabular}}            & \multicolumn{1}{c|}{\begin{tabular}[c]{@{}c@{}}0.383\\ (0.036)\end{tabular}}            & \multicolumn{1}{c|}{\begin{tabular}[c]{@{}c@{}}0.645\\ (0.062)\end{tabular}}            & \multicolumn{1}{c|}{\begin{tabular}[c]{@{}c@{}}0.109\\ (0.028)\end{tabular}}            & \multicolumn{1}{c|}{\begin{tabular}[c]{@{}c@{}}0.768\\ (0.024)\end{tabular}}            & \begin{tabular}[c]{@{}c@{}}9.232\\ (0.720)\end{tabular}            &                   & \multicolumn{1}{c|}{\begin{tabular}[c]{@{}c@{}}0.906\\ (0.021)\end{tabular}}            & \multicolumn{1}{c|}{\begin{tabular}[c]{@{}c@{}}0.328\\ (0.043)\end{tabular}}            & \multicolumn{1}{c|}{\begin{tabular}[c]{@{}c@{}}0.436\\ (0.091)\end{tabular}}            & \multicolumn{1}{c|}{\begin{tabular}[c]{@{}c@{}}0.062\\ (0.027)\end{tabular}}            & \multicolumn{1}{c|}{\begin{tabular}[c]{@{}c@{}}0.687\\ (0.036)\end{tabular}}            & \begin{tabular}[c]{@{}c@{}}7.998\\ (0.557)\end{tabular}            \\ \cline{1-7} \cline{9-14} 
$\mathsf{GemBAG}_{\mathsf{MPM}}$      & \multicolumn{1}{c|}{\begin{tabular}[c]{@{}c@{}}$\mathbf{0.940}$\\ (0.002)\end{tabular}} & \multicolumn{1}{c|}{\begin{tabular}[c]{@{}c@{}}0.311\\ (0.052)\end{tabular}}            & \multicolumn{1}{c|}{\begin{tabular}[c]{@{}c@{}}0.124\\ (0.041)\end{tabular}}            & \multicolumn{1}{c|}{\begin{tabular}[c]{@{}c@{}}$\mathbf{0.001}$\\ (0.002)\end{tabular}} & \multicolumn{1}{c|}{\begin{tabular}[c]{@{}c@{}}0.791\\ (0.057)\end{tabular}}            & \begin{tabular}[c]{@{}c@{}}11.835\\ (1.150)\end{tabular}           &                   & \multicolumn{1}{c|}{\begin{tabular}[c]{@{}c@{}}$\mathbf{0.940}$\\ (0.002)\end{tabular}} & \multicolumn{1}{c|}{\begin{tabular}[c]{@{}c@{}}0.238\\ (0.064)\end{tabular}}            & \multicolumn{1}{c|}{\begin{tabular}[c]{@{}c@{}}0.081\\ (0.036)\end{tabular}}            & \multicolumn{1}{c|}{\begin{tabular}[c]{@{}c@{}}$\mathbf{0.001}$\\ (0.002)\end{tabular}} & \multicolumn{1}{c|}{\begin{tabular}[c]{@{}c@{}}0.786\\ (0.050)\end{tabular}}            & \begin{tabular}[c]{@{}c@{}}8.580\\ (0.775)\end{tabular}            \\ \hline \hline
\multirow{2}{*}{}                 & \multicolumn{6}{c|}{$P2020$ ($p_0 = 82.5$)}                                                                                                                                                                                                                                                                                                                                                                                                                                                                                                         &                   & \multicolumn{6}{c|}{$Full \; dependence$ ($p_0 = 85$)}                                                                                                                                                                                                                                                                                                                                                                                                                                                                                               \\ \cline{2-14} 
                                  & \multicolumn{1}{c|}{Acc}                                                                & \multicolumn{1}{c|}{MCC}                                                                & \multicolumn{1}{c|}{TPR}                                                                & \multicolumn{1}{c|}{FPR}                                                                & \multicolumn{1}{c|}{AUC}                                                                & Fr Loss                                                            & \multirow{7}{*}{} & \multicolumn{1}{c|}{Acc}                                                                & \multicolumn{1}{c|}{MCC}                                                                & \multicolumn{1}{c|}{TPR}                                                                & \multicolumn{1}{c|}{FPR}                                                                & \multicolumn{1}{c|}{AUC}                                                                & Fr Loss                                                            \\ \cline{1-14} \cline{1-14} \cline{1-14}
$\mathsf{mGHS}_{\mathsf{MPM}}$        & \multicolumn{1}{c|}{\begin{tabular}[c]{@{}c@{}}0.796\\ (0.011)\end{tabular}}            & \multicolumn{1}{c|}{\begin{tabular}[c]{@{}c@{}}0.358\\ (0.021)\end{tabular}}            & \multicolumn{1}{c|}{\begin{tabular}[c]{@{}c@{}}$\mathbf{0.822}$\\ (0.030)\end{tabular}} & \multicolumn{1}{c|}{\begin{tabular}[c]{@{}c@{}}0.206\\ (0.011)\end{tabular}}            & \multicolumn{1}{c|}{\begin{tabular}[c]{@{}c@{}}$\mathbf{0.875}$\\ (0.020)\end{tabular}} & \begin{tabular}[c]{@{}c@{}}8.498\\ (1.323)\end{tabular}            &                   & \multicolumn{1}{c|}{\begin{tabular}[c]{@{}c@{}}0.716\\ (0.034)\end{tabular}}            & \multicolumn{1}{c|}{\begin{tabular}[c]{@{}c@{}}0.247\\ (0.036)\end{tabular}}            & \multicolumn{1}{c|}{\begin{tabular}[c]{@{}c@{}}$\mathbf{0.735}$\\ (0.046)\end{tabular}} & \multicolumn{1}{c|}{\begin{tabular}[c]{@{}c@{}}0.285\\ (0.036)\end{tabular}}            & \multicolumn{1}{c|}{\begin{tabular}[c]{@{}c@{}}0.792\\ (0.032)\end{tabular}}            & \begin{tabular}[c]{@{}c@{}}8.349\\ (1.184)\end{tabular}            \\ \cline{1-7} \cline{9-14} 
$\mathsf{mGHS}_{t_\alpha}$          & \multicolumn{1}{c|}{\begin{tabular}[c]{@{}c@{}}0.925\\ (0.008)\end{tabular}}            & \multicolumn{1}{c|}{\begin{tabular}[c]{@{}c@{}}0.532\\ (0.034)\end{tabular}}            & \multicolumn{1}{c|}{\begin{tabular}[c]{@{}c@{}}0.698\\ (0.037)\end{tabular}}            & \multicolumn{1}{c|}{\begin{tabular}[c]{@{}c@{}}0.059\\ (0.009)\end{tabular}}            & \multicolumn{1}{c|}{\begin{tabular}[c]{@{}c@{}}$\mathbf{0.875}$\\ (0.020)\end{tabular}} & \begin{tabular}[c]{@{}c@{}}8.498\\ (1.323)\end{tabular}            &                   & \multicolumn{1}{c|}{\begin{tabular}[c]{@{}c@{}}0.923\\ (0.009)\end{tabular}}            & \multicolumn{1}{c|}{\begin{tabular}[c]{@{}c@{}}$\mathbf{0.408}$\\ (0.046)\end{tabular}} & \multicolumn{1}{c|}{\begin{tabular}[c]{@{}c@{}}0.446\\ (0.074)\end{tabular}}            & \multicolumn{1}{c|}{\begin{tabular}[c]{@{}c@{}}0.041\\ (0.012)\end{tabular}}            & \multicolumn{1}{c|}{\begin{tabular}[c]{@{}c@{}}0.792\\ (0.032)\end{tabular}}            & \begin{tabular}[c]{@{}c@{}}8.349\\ (1.184\end{tabular}             \\ \cline{1-7} \cline{9-14} 
$\mathsf{GHS}_{\mathsf{MPM}}$         & \multicolumn{1}{c|}{\begin{tabular}[c]{@{}c@{}}0.795\\ (0.013)\end{tabular}}            & \multicolumn{1}{c|}{\begin{tabular}[c]{@{}c@{}}0.321\\ (0.027)\end{tabular}}            & \multicolumn{1}{c|}{\begin{tabular}[c]{@{}c@{}}0.748\\ (0.037)\end{tabular}}            & \multicolumn{1}{c|}{\begin{tabular}[c]{@{}c@{}}0.202\\ (0.013)\end{tabular}}            & \multicolumn{1}{c|}{\begin{tabular}[c]{@{}c@{}}0.840\\ (0.022)\end{tabular}}            & \begin{tabular}[c]{@{}c@{}}9.371\\ (1.216)\end{tabular}            &                   & \multicolumn{1}{c|}{\begin{tabular}[c]{@{}c@{}}0.670\\ (0.055)\end{tabular}}            & \multicolumn{1}{c|}{\begin{tabular}[c]{@{}c@{}}0.165\\ (0.046)\end{tabular}}            & \multicolumn{1}{c|}{\begin{tabular}[c]{@{}c@{}}0.631\\ (0.046)\end{tabular}}            & \multicolumn{1}{c|}{\begin{tabular}[c]{@{}c@{}}0.327\\ (0.059)\end{tabular}}            & \multicolumn{1}{c|}{\begin{tabular}[c]{@{}c@{}}0.710\\ (0.043)\end{tabular}}            & \begin{tabular}[c]{@{}c@{}}8.616\\ (0.954)\end{tabular}            \\ \cline{1-7} \cline{9-14} 
$\mathsf{fJGL}$                              & \multicolumn{1}{c|}{\begin{tabular}[c]{@{}c@{}}0.874\\ (0.023)\end{tabular}}            & \multicolumn{1}{c|}{\begin{tabular}[c]{@{}c@{}}0.412\\ (0.046)\end{tabular}}            & \multicolumn{1}{c|}{\begin{tabular}[c]{@{}c@{}}0.697\\ (0.050)\end{tabular}}            & \multicolumn{1}{c|}{\begin{tabular}[c]{@{}c@{}}0.113\\ (0.025)\end{tabular}}            & \multicolumn{1}{c|}{\begin{tabular}[c]{@{}c@{}}0.792\\ (0.025)\end{tabular}}            & \begin{tabular}[c]{@{}c@{}}8.205\\ (0.702)\end{tabular}            &                   & \multicolumn{1}{c|}{\begin{tabular}[c]{@{}c@{}}0.905\\ (0.021)\end{tabular}}            & \multicolumn{1}{c|}{\begin{tabular}[c]{@{}c@{}}0.309\\ (0.056)\end{tabular}}            & \multicolumn{1}{c|}{\begin{tabular}[c]{@{}c@{}}0.373\\ (0.100)\end{tabular}}            & \multicolumn{1}{c|}{\begin{tabular}[c]{@{}c@{}}0.055\\ (0.028)\end{tabular}}            & \multicolumn{1}{c|}{\begin{tabular}[c]{@{}c@{}}0.659\\ (0.040)\end{tabular}}            & \begin{tabular}[c]{@{}c@{}}$\mathbf{7.711}$\\ (0.611)\end{tabular}            \\ \cline{1-7} \cline{9-14} 
$\mathsf{gJGL}$                              & \multicolumn{1}{c|}{\begin{tabular}[c]{@{}c@{}}0.864\\ (0.025)\end{tabular}}            & \multicolumn{1}{c|}{\begin{tabular}[c]{@{}c@{}}0.376\\ (0.036)\end{tabular}}            & \multicolumn{1}{c|}{\begin{tabular}[c]{@{}c@{}}0.660\\ (0.054)\end{tabular}}            & \multicolumn{1}{c|}{\begin{tabular}[c]{@{}c@{}}0.121\\ (0.029)\end{tabular}}            & \multicolumn{1}{c|}{\begin{tabular}[c]{@{}c@{}}0.770\\ (0.022)\end{tabular}}            & \begin{tabular}[c]{@{}c@{}}8.851\\ (0.714)\end{tabular}            &                   & \multicolumn{1}{c|}{\begin{tabular}[c]{@{}c@{}}0.902\\ (0.023)\end{tabular}}            & \multicolumn{1}{c|}{\begin{tabular}[c]{@{}c@{}}0.293\\ (0.048)\end{tabular}}            & \multicolumn{1}{c|}{\begin{tabular}[c]{@{}c@{}}0.358\\ (0.100)\end{tabular}}            & \multicolumn{1}{c|}{\begin{tabular}[c]{@{}c@{}}0.057\\ (0.030)\end{tabular}}            & \multicolumn{1}{c|}{\begin{tabular}[c]{@{}c@{}}0.650\\ (0.039)\end{tabular}}            & \begin{tabular}[c]{@{}c@{}}7.989\\ (0.579)\end{tabular}            \\ \cline{1-7} \cline{9-14} 
$\mathsf{GemBAG}_{\mathsf{MPM}}$      & \multicolumn{1}{c|}{\begin{tabular}[c]{@{}c@{}}$\mathbf{0.956}$\\ (0.004)\end{tabular}} & \multicolumn{1}{c|}{\begin{tabular}[c]{@{}c@{}}$\mathbf{0.580}$\\ (0.049)\end{tabular}} & \multicolumn{1}{c|}{\begin{tabular}[c]{@{}c@{}}0.367\\ (0.065)\end{tabular}}            & \multicolumn{1}{c|}{\begin{tabular}[c]{@{}c@{}}$\mathbf{0.001}$\\ (0.001)\end{tabular}} & \multicolumn{1}{c|}{\begin{tabular}[c]{@{}c@{}}0.871\\ (0.035)\end{tabular}}            & \begin{tabular}[c]{@{}c@{}}$\mathbf{7.835}$\\ (1.043)\end{tabular} &                   & \multicolumn{1}{c|}{\begin{tabular}[c]{@{}c@{}}$\mathbf{0.938}$\\ (0.002)\end{tabular}} & \multicolumn{1}{c|}{\begin{tabular}[c]{@{}c@{}}0.318\\ (0.049)\end{tabular}}            & \multicolumn{1}{c|}{\begin{tabular}[c]{@{}c@{}}0.112\\ (0.032)\end{tabular}}            & \multicolumn{1}{c|}{\begin{tabular}[c]{@{}c@{}}$\mathbf{0.000}$\\ (0.000)\end{tabular}} & \multicolumn{1}{c|}{\begin{tabular}[c]{@{}c@{}}$\mathbf{0.838}$\\ (0.031)\end{tabular}} & \begin{tabular}[c]{@{}c@{}}$7.984$\\ (0.651)\end{tabular} \\ \hline
\end{tabular}
\end{adjustbox}
\caption{Simulation results for $n = 50$ and $p = 50$ ($50$ replicates). Methods mGHS and GHS are evaluated over $B = 10000$ post burn-in samples.}
\label{tab:simn50p50}
\end{table}

\begin{table}[H]
\begin{adjustbox}{width = \textwidth}
\begin{tabular}{|c|cccccc|c|cccccc|}
\hline
\multirow{2}{*}{$n = 50, p = 100$} & \multicolumn{6}{c|}{$Independent$ ($p_0 = 195$)}                                                                                                                                                                         &                   & \multicolumn{6}{c|}{$Coupled$ ($p_0 = 177.5$)}                                                                                                                                                                                                                                                                                                                                                                                                                                                                                                        \\ \cline{2-14} 
                                   & \multicolumn{1}{c|}{Acc}                                                                & \multicolumn{1}{c|}{MCC}                                                                & \multicolumn{1}{c|}{TPR}                                                                & \multicolumn{1}{c|}{FPR}                                                                & \multicolumn{1}{c|}{AUC}                                                                & Fr Loss                                                             & \multirow{7}{*}{} & \multicolumn{1}{c|}{Acc}                                                                & \multicolumn{1}{c|}{MCC}                                                                & \multicolumn{1}{c|}{TPR}                                                                & \multicolumn{1}{c|}{FPR}                                                                & \multicolumn{1}{c|}{AUC}                                                                & Fr Loss                                                             \\ \cline{1-14} \cline{1-14} \cline{1-14} 
$\mathsf{mGHS}_{\mathsf{MPM}}$     & \multicolumn{1}{c|}{\begin{tabular}[c]{@{}c@{}}0.655\\ (0.024)\end{tabular}}            & \multicolumn{1}{c|}{\begin{tabular}[c]{@{}c@{}}0.146\\ (0.018)\end{tabular}}            & \multicolumn{1}{c|}{\begin{tabular}[c]{@{}c@{}}0.712\\ (0.030)\end{tabular}}            & \multicolumn{1}{c|}{\begin{tabular}[c]{@{}c@{}}0.348\\ (0.025)\end{tabular}}            & \multicolumn{1}{c|}{\begin{tabular}[c]{@{}c@{}}0.759\\ (0.023)\end{tabular}}            & \begin{tabular}[c]{@{}c@{}}20.607\\ (1.444)\end{tabular}            &                   & \multicolumn{1}{c|}{\begin{tabular}[c]{@{}c@{}}0.568\\ (0.028)\end{tabular}}            & \multicolumn{1}{c|}{\begin{tabular}[c]{@{}c@{}}0.082\\ (0.020)\end{tabular}}            & \multicolumn{1}{c|}{\begin{tabular}[c]{@{}c@{}}$\mathbf{0.653}$\\ (0.036)\end{tabular}} & \multicolumn{1}{c|}{\begin{tabular}[c]{@{}c@{}}0.436\\ (0.028)\end{tabular}}            & \multicolumn{1}{c|}{\begin{tabular}[c]{@{}c@{}}0.671\\ (0.037)\end{tabular}}            & \begin{tabular}[c]{@{}c@{}}17.547\\ (1.283)\end{tabular}            \\ \cline{1-7} \cline{9-14} 
$\mathsf{mGHS}_{t_\alpha}$         & \multicolumn{1}{c|}{\begin{tabular}[c]{@{}c@{}}0.953\\ (0.004)\end{tabular}}            & \multicolumn{1}{c|}{\begin{tabular}[c]{@{}c@{}}$\mathbf{0.348}$\\ (0.032)\end{tabular}} & \multicolumn{1}{c|}{\begin{tabular}[c]{@{}c@{}}0.361\\ (0.044)\end{tabular}}            & \multicolumn{1}{c|}{\begin{tabular}[c]{@{}c@{}}0.022\\ (0.005)\end{tabular}}            & \multicolumn{1}{c|}{\begin{tabular}[c]{@{}c@{}}0.759\\ (0.023)\end{tabular}}            & \begin{tabular}[c]{@{}c@{}}20.607\\ (1.444)\end{tabular}            &                   & \multicolumn{1}{c|}{\begin{tabular}[c]{@{}c@{}}0.961\\ (0.003)\end{tabular}}            & \multicolumn{1}{c|}{\begin{tabular}[c]{@{}c@{}}0.228\\ (0.050)\end{tabular}}            & \multicolumn{1}{c|}{\begin{tabular}[c]{@{}c@{}}0.152\\ (0.060)\end{tabular}}            & \multicolumn{1}{c|}{\begin{tabular}[c]{@{}c@{}}0.009\\ (0.005)\end{tabular}}            & \multicolumn{1}{c|}{\begin{tabular}[c]{@{}c@{}}0.671\\ (0.037)\end{tabular}}            & \begin{tabular}[c]{@{}c@{}}17.547\\ (1.283)\end{tabular}            \\ \cline{1-7} \cline{9-14} 
$\mathsf{GHS}_{\mathsf{MPM}}$      & \multicolumn{1}{c|}{\begin{tabular}[c]{@{}c@{}}0.669\\ (0.023)\end{tabular}}            & \multicolumn{1}{c|}{\begin{tabular}[c]{@{}c@{}}0.155\\ (0.019)\end{tabular}}            & \multicolumn{1}{c|}{\begin{tabular}[c]{@{}c@{}}$\mathbf{0.715}$\\ (0.030)\end{tabular}} & \multicolumn{1}{c|}{\begin{tabular}[c]{@{}c@{}}0.333\\ (0.024)\end{tabular}}            & \multicolumn{1}{c|}{\begin{tabular}[c]{@{}c@{}}$\mathbf{0.769}$\\ (0.024)\end{tabular}} & \begin{tabular}[c]{@{}c@{}}20.594\\ (1.453)\end{tabular}            &                   & \multicolumn{1}{c|}{\begin{tabular}[c]{@{}c@{}}0.563\\ (0.029)\end{tabular}}            & \multicolumn{1}{c|}{\begin{tabular}[c]{@{}c@{}}0.074\\ (0.020)\end{tabular}}            & \multicolumn{1}{c|}{\begin{tabular}[c]{@{}c@{}}0.638\\ (0.036)\end{tabular}}            & \multicolumn{1}{c|}{\begin{tabular}[c]{@{}c@{}}0.439\\ (0.029)\end{tabular}}            & \multicolumn{1}{c|}{\begin{tabular}[c]{@{}c@{}}0.655\\ (0.036)\end{tabular}}            & \begin{tabular}[c]{@{}c@{}}17.545\\ (1.283)\end{tabular}            \\ \cline{1-7} \cline{9-14} 
$\mathsf{fJGL}$                    & \multicolumn{1}{c|}{\begin{tabular}[c]{@{}c@{}}0.931\\ (0.012)\end{tabular}}            & \multicolumn{1}{c|}{\begin{tabular}[c]{@{}c@{}}0.315\\ (0.028)\end{tabular}}            & \multicolumn{1}{c|}{\begin{tabular}[c]{@{}c@{}}0.451\\ (0.054)\end{tabular}}            & \multicolumn{1}{c|}{\begin{tabular}[c]{@{}c@{}}0.049\\ (0.014)\end{tabular}}            & \multicolumn{1}{c|}{\begin{tabular}[c]{@{}c@{}}0.701\\ (0.022)\end{tabular}}            & \begin{tabular}[c]{@{}c@{}}$\mathbf{19.892}$\\ (0.988)\end{tabular} &                   & \multicolumn{1}{c|}{\begin{tabular}[c]{@{}c@{}}0.952\\ (0.009)\end{tabular}}            & \multicolumn{1}{c|}{\begin{tabular}[c]{@{}c@{}}0.234\\ (0.033)\end{tabular}}            & \multicolumn{1}{c|}{\begin{tabular}[c]{@{}c@{}}0.226\\ (0.074)\end{tabular}}            & \multicolumn{1}{c|}{\begin{tabular}[c]{@{}c@{}}0.021\\ (0.012)\end{tabular}}            & \multicolumn{1}{c|}{\begin{tabular}[c]{@{}c@{}}0.603\\ (0.032)\end{tabular}}            & \begin{tabular}[c]{@{}c@{}}$\mathbf{16.296}$\\ (0.694)\end{tabular} \\ \cline{1-7} \cline{9-14} 
$\mathsf{gJGL}$                    & \multicolumn{1}{c|}{\begin{tabular}[c]{@{}c@{}}0.929\\ (0.013)\end{tabular}}            & \multicolumn{1}{c|}{\begin{tabular}[c]{@{}c@{}}0.312\\ (0.029)\end{tabular}}            & \multicolumn{1}{c|}{\begin{tabular}[c]{@{}c@{}}0.456\\ (0.058)\end{tabular}}            & \multicolumn{1}{c|}{\begin{tabular}[c]{@{}c@{}}0.051\\ (0.015)\end{tabular}}            & \multicolumn{1}{c|}{\begin{tabular}[c]{@{}c@{}}0.702\\ (0.023)\end{tabular}}            & \begin{tabular}[c]{@{}c@{}}19.921\\ (1.055)\end{tabular}            &                   & \multicolumn{1}{c|}{\begin{tabular}[c]{@{}c@{}}0.952\\ (0.009)\end{tabular}}            & \multicolumn{1}{c|}{\begin{tabular}[c]{@{}c@{}}$\mathbf{0.229}$\\ (0.034)\end{tabular}} & \multicolumn{1}{c|}{\begin{tabular}[c]{@{}c@{}}0.219\\ (0.074)\end{tabular}}            & \multicolumn{1}{c|}{\begin{tabular}[c]{@{}c@{}}0.021\\ (0.011)\end{tabular}}            & \multicolumn{1}{c|}{\begin{tabular}[c]{@{}c@{}}0.599\\ (0.032)\end{tabular}}            & \begin{tabular}[c]{@{}c@{}}16.689\\ (0.722)\end{tabular}            \\ \cline{1-7} \cline{9-14} 
$\mathsf{GemBAG}_{\mathsf{MPM}}$   & \multicolumn{1}{c|}{\begin{tabular}[c]{@{}c@{}}$\mathbf{0.962}$\\ (0.001)\end{tabular}} & \multicolumn{1}{c|}{\begin{tabular}[c]{@{}c@{}}0.179\\ (0.043)\end{tabular}}            & \multicolumn{1}{c|}{\begin{tabular}[c]{@{}c@{}}0.052\\ (0.026)\end{tabular}}            & \multicolumn{1}{c|}{\begin{tabular}[c]{@{}c@{}}$\mathbf{0.001}$\\ (0.001)\end{tabular}} & \multicolumn{1}{c|}{\begin{tabular}[c]{@{}c@{}}0.698\\ (0.069)\end{tabular}}            & \begin{tabular}[c]{@{}c@{}}23.012\\ (2.174)\end{tabular}            &                   & \multicolumn{1}{c|}{\begin{tabular}[c]{@{}c@{}}$\mathbf{0.965}$\\ (0.001)\end{tabular}} & \multicolumn{1}{c|}{\begin{tabular}[c]{@{}c@{}}0.143\\ (0.046)\end{tabular}}            & \multicolumn{1}{c|}{\begin{tabular}[c]{@{}c@{}}0.034\\ (0.016)\end{tabular}}            & \multicolumn{1}{c|}{\begin{tabular}[c]{@{}c@{}}$\mathbf{0.001}$\\ (0.000)\end{tabular}} & \multicolumn{1}{c|}{\begin{tabular}[c]{@{}c@{}}$\mathbf{0.708}$\\ (0.044)\end{tabular}} & \begin{tabular}[c]{@{}c@{}}16.986\\ (0.894)\end{tabular}            \\ \hline \hline
\multirow{2}{*}{}                  & \multicolumn{6}{c|}{$P2020$ ($p_0 = 182.5$)}                                                                                                                                                                                                                                                                                                                                                                                                                                                                                                          &                   & \multicolumn{6}{c|}{$Full \; dependence$ ($p_0 = 185$)}                                                                                                                                                                                                                                                                                                                                                                                                                                                                                                \\ \cline{2-14} 
                                   & \multicolumn{1}{c|}{Acc}                                                                & \multicolumn{1}{c|}{MCC}                                                                & \multicolumn{1}{c|}{TPR}                                                                & \multicolumn{1}{c|}{FPR}                                                                & \multicolumn{1}{c|}{AUC}                                                                & Fr Loss                                                             & \multirow{7}{*}{} & \multicolumn{1}{c|}{Acc}                                                                & \multicolumn{1}{c|}{MCC}                                                                & \multicolumn{1}{c|}{TPR}                                                                & \multicolumn{1}{c|}{FPR}                                                                & \multicolumn{1}{c|}{AUC}                                                                & Fr Loss                                                             \\ \cline{1-14} \cline{1-14} \cline{1-14} 
$\mathsf{mGHS}_{\mathsf{MPM}}$     & \multicolumn{1}{c|}{\begin{tabular}[c]{@{}c@{}}0.720\\ (0.013)\end{tabular}}            & \multicolumn{1}{c|}{\begin{tabular}[c]{@{}c@{}}0.215\\ (0.014)\end{tabular}}            & \multicolumn{1}{c|}{\begin{tabular}[c]{@{}c@{}}$\mathbf{0.808}$\\ (0.025)\end{tabular}} & \multicolumn{1}{c|}{\begin{tabular}[c]{@{}c@{}}0.284\\ (0.013)\end{tabular}}            & \multicolumn{1}{c|}{\begin{tabular}[c]{@{}c@{}}0.853\\ (0.016)\end{tabular}}            & \begin{tabular}[c]{@{}c@{}}18.878\\ (1.944)\end{tabular}            &                   & \multicolumn{1}{c|}{\begin{tabular}[c]{@{}c@{}}0.589\\ (0.030)\end{tabular}}            & \multicolumn{1}{c|}{\begin{tabular}[c]{@{}c@{}}0.107\\ (0.021)\end{tabular}}            & \multicolumn{1}{c|}{\begin{tabular}[c]{@{}c@{}}$\mathbf{0.692}$\\ (0.036)\end{tabular}} & \multicolumn{1}{c|}{\begin{tabular}[c]{@{}c@{}}0.415\\ (0.031)\end{tabular}}            & \multicolumn{1}{c|}{\begin{tabular}[c]{@{}c@{}}0.714\\ (0.035)\end{tabular}}            & \begin{tabular}[c]{@{}c@{}}17.127\\ (1.491)\end{tabular}            \\ \cline{1-7} \cline{9-14} 
$\mathsf{mGHS}_{t_\alpha}$         & \multicolumn{1}{c|}{\begin{tabular}[c]{@{}c@{}}0.948\\ (0.004)\end{tabular}}            & \multicolumn{1}{c|}{\begin{tabular}[c]{@{}c@{}}0.459\\ (0.022)\end{tabular}}            & \multicolumn{1}{c|}{\begin{tabular}[c]{@{}c@{}}0.625\\ (0.030)\end{tabular}}            & \multicolumn{1}{c|}{\begin{tabular}[c]{@{}c@{}}0.040\\ (0.005)\end{tabular}}            & \multicolumn{1}{c|}{\begin{tabular}[c]{@{}c@{}}0.853\\ (0.016)\end{tabular}}            & \begin{tabular}[c]{@{}c@{}}18.878\\ (1.944)\end{tabular}            &                   & \multicolumn{1}{c|}{\begin{tabular}[c]{@{}c@{}}0.958\\ (0.004)\end{tabular}}            & \multicolumn{1}{c|}{\begin{tabular}[c]{@{}c@{}}$\mathbf{0.285}$\\ (0.046)\end{tabular}} & \multicolumn{1}{c|}{\begin{tabular}[c]{@{}c@{}}0.230\\ (0.071)\end{tabular}}            & \multicolumn{1}{c|}{\begin{tabular}[c]{@{}c@{}}0.013\\ (0.006)\end{tabular}}            & \multicolumn{1}{c|}{\begin{tabular}[c]{@{}c@{}}0.714\\ (0.035)\end{tabular}}            & \begin{tabular}[c]{@{}c@{}}17.127\\ (1.491)\end{tabular}            \\ \cline{1-7} \cline{9-14} 
$\mathsf{GHS}_{\mathsf{MPM}}$      & \multicolumn{1}{c|}{\begin{tabular}[c]{@{}c@{}}0.710\\ (0.016)\end{tabular}}            & \multicolumn{1}{c|}{\begin{tabular}[c]{@{}c@{}}0.181\\ (0.015)\end{tabular}}            & \multicolumn{1}{c|}{\begin{tabular}[c]{@{}c@{}}0.733\\ (0.026)\end{tabular}}            & \multicolumn{1}{c|}{\begin{tabular}[c]{@{}c@{}}0.291\\ (0.017)\end{tabular}}            & \multicolumn{1}{c|}{\begin{tabular}[c]{@{}c@{}}0.800\\ (0.016)\end{tabular}}            & \begin{tabular}[c]{@{}c@{}}20.299\\ (1.650)\end{tabular}            &                   & \multicolumn{1}{c|}{\begin{tabular}[c]{@{}c@{}}0.564\\ (0.029)\end{tabular}}            & \multicolumn{1}{c|}{\begin{tabular}[c]{@{}c@{}}0.072\\ (0.022)\end{tabular}}            & \multicolumn{1}{c|}{\begin{tabular}[c]{@{}c@{}}0.626\\ (0.040)\end{tabular}}            & \multicolumn{1}{c|}{\begin{tabular}[c]{@{}c@{}}0.438\\ (0.029)\end{tabular}}            & \multicolumn{1}{c|}{\begin{tabular}[c]{@{}c@{}}0.647\\ (0.039)\end{tabular}}            & \begin{tabular}[c]{@{}c@{}}17.256\\ (1.261)\end{tabular}            \\ \cline{1-7} \cline{9-14} 
$\mathsf{fJGL}$                    & \multicolumn{1}{c|}{\begin{tabular}[c]{@{}c@{}}0.935\\ (0.010)\end{tabular}}            & \multicolumn{1}{c|}{\begin{tabular}[c]{@{}c@{}}0.393\\ (0.030)\end{tabular}}            & \multicolumn{1}{c|}{\begin{tabular}[c]{@{}c@{}}0.588\\ (0.042)\end{tabular}}            & \multicolumn{1}{c|}{\begin{tabular}[c]{@{}c@{}}0.052\\ (0.011)\end{tabular}}            & \multicolumn{1}{c|}{\begin{tabular}[c]{@{}c@{}}0.768\\ (0.018)\end{tabular}}            & \begin{tabular}[c]{@{}c@{}}18.557\\ (1.070)\end{tabular}            &                   & \multicolumn{1}{c|}{\begin{tabular}[c]{@{}c@{}}0.955\\ (0.007)\end{tabular}}            & \multicolumn{1}{c|}{\begin{tabular}[c]{@{}c@{}}0.240\\ (0.042)\end{tabular}}            & \multicolumn{1}{c|}{\begin{tabular}[c]{@{}c@{}}0.201\\ (0.079)\end{tabular}}            & \multicolumn{1}{c|}{\begin{tabular}[c]{@{}c@{}}0.016\\ (0.010)\end{tabular}}            & \multicolumn{1}{c|}{\begin{tabular}[c]{@{}c@{}}0.593\\ (0.035)\end{tabular}}            & \begin{tabular}[c]{@{}c@{}}$\mathbf{16.103}$\\ (1.027)\end{tabular} \\ \cline{1-7} \cline{9-14} 
$\mathsf{gJGL}$                    & \multicolumn{1}{c|}{\begin{tabular}[c]{@{}c@{}}0.923\\ (0.011)\end{tabular}}            & \multicolumn{1}{c|}{\begin{tabular}[c]{@{}c@{}}0.335\\ (0.024)\end{tabular}}            & \multicolumn{1}{c|}{\begin{tabular}[c]{@{}c@{}}0.540\\ (0.043)\end{tabular}}            & \multicolumn{1}{c|}{\begin{tabular}[c]{@{}c@{}}0.062\\ (0.013)\end{tabular}}            & \multicolumn{1}{c|}{\begin{tabular}[c]{@{}c@{}}0.739\\ (0.018)\end{tabular}}            & \begin{tabular}[c]{@{}c@{}}20.104\\ (1.107)\end{tabular}            &                   & \multicolumn{1}{c|}{\begin{tabular}[c]{@{}c@{}}0.955\\ (0.008)\end{tabular}}            & \multicolumn{1}{c|}{\begin{tabular}[c]{@{}c@{}}0.223\\ (0.038)\end{tabular}}            & \multicolumn{1}{c|}{\begin{tabular}[c]{@{}c@{}}0.182\\ (0.071)\end{tabular}}            & \multicolumn{1}{c|}{\begin{tabular}[c]{@{}c@{}}0.015\\ (0.010)\end{tabular}}            & \multicolumn{1}{c|}{\begin{tabular}[c]{@{}c@{}}0.583\\ (0.031)\end{tabular}}            & \begin{tabular}[c]{@{}c@{}}16.772\\ (0.924)\end{tabular}            \\ \cline{1-7} \cline{9-14} 
$\mathsf{GemBAG}_{\mathsf{MPM}}$   & \multicolumn{1}{c|}{\begin{tabular}[c]{@{}c@{}}$\mathbf{0.975}$\\ (0.002)\end{tabular}} & \multicolumn{1}{c|}{\begin{tabular}[c]{@{}c@{}}$\mathbf{0.550}$\\ (0.041)\end{tabular}} & \multicolumn{1}{c|}{\begin{tabular}[c]{@{}c@{}}0.321\\ (0.052)\end{tabular}}            & \multicolumn{1}{c|}{\begin{tabular}[c]{@{}c@{}}$\mathbf{0.000}$\\ (0.000)\end{tabular}} & \multicolumn{1}{c|}{\begin{tabular}[c]{@{}c@{}}$\mathbf{0.869}$\\ (0.015)\end{tabular}} & \begin{tabular}[c]{@{}c@{}}$\mathbf{15.676}$\\ (1.264)\end{tabular} &                   & \multicolumn{1}{c|}{\begin{tabular}[c]{@{}c@{}}$\mathbf{0.966}$\\ (0.001)\end{tabular}} & \multicolumn{1}{c|}{\begin{tabular}[c]{@{}c@{}}0.277\\ (0.038)\end{tabular}}            & \multicolumn{1}{c|}{\begin{tabular}[c]{@{}c@{}}0.084\\ (0.022)\end{tabular}}            & \multicolumn{1}{c|}{\begin{tabular}[c]{@{}c@{}}$\mathbf{0.000}$\\ (0.000)\end{tabular}} & \multicolumn{1}{c|}{\begin{tabular}[c]{@{}c@{}}$\mathbf{0.808}$\\ (0.037)\end{tabular}} & \begin{tabular}[c]{@{}c@{}}16.411\\ (1.044)\end{tabular}            \\ \hline
\end{tabular}
\end{adjustbox}
\caption{Simulation results for $n = 50$ and $p = 100$ ($50$ replicates). Methods mGHS and GHS are evaluated over $B = 10000$ post burn-in samples.}
\label{tab:simn50p100}
\end{table}

\begin{table}[H]
\begin{adjustbox}{width = \textwidth}
\begin{tabular}{|c|cccccc|c|cccccc|}
\hline
\multirow{2}{*}{$n = 100, p = 250$} & \multicolumn{6}{c|}{$Independent$ ($p_0 = 532.5$)}                                                                                                                                                                                                                                                                                                                                                                                                                                                                                                    &                   & \multicolumn{6}{c|}{$Coupled$ ($p_0 = 477.5$)}                                                                                                                                                                                                                                                                                                                                                                                                                                                                                                        \\ \cline{2-14} 
                                    & \multicolumn{1}{c|}{Acc}                                                                & \multicolumn{1}{c|}{MCC}                                                                & \multicolumn{1}{c|}{TPR}                                                                & \multicolumn{1}{c|}{FPR}                                                                & \multicolumn{1}{c|}{AUC}                                                                & Fr Loss                                                             & \multirow{7}{*}{} & \multicolumn{1}{c|}{Acc}                                                                & \multicolumn{1}{c|}{MCC}                                                                & \multicolumn{1}{c|}{TPR}                                                                & \multicolumn{1}{c|}{FPR}                                                                & \multicolumn{1}{c|}{AUC}                                                                & Fr Loss                                                             \\ \cline{1-14} \cline{1-14} \cline{1-14}
$\mathsf{mGHS}_{\mathsf{MPM}}$      & \multicolumn{1}{c|}{\begin{tabular}[c]{@{}c@{}}0.632\\ (0.012)\end{tabular}}            & \multicolumn{1}{c|}{\begin{tabular}[c]{@{}c@{}}0.115\\ (0.006)\end{tabular}}            & \multicolumn{1}{c|}{\begin{tabular}[c]{@{}c@{}}0.808\\ (0.015)\end{tabular}}            & \multicolumn{1}{c|}{\begin{tabular}[c]{@{}c@{}}0.371\\ (0.012)\end{tabular}}            & \multicolumn{1}{c|}{\begin{tabular}[c]{@{}c@{}}0.830\\ (0.012)\end{tabular}}            & \begin{tabular}[c]{@{}c@{}}34.766\\ (1.629)\end{tabular}            &                   & \multicolumn{1}{c|}{\begin{tabular}[c]{@{}c@{}}0.556\\ (0.010)\end{tabular}}            & \multicolumn{1}{c|}{\begin{tabular}[c]{@{}c@{}}0.077\\ (0.007)\end{tabular}}            & \multicolumn{1}{c|}{\begin{tabular}[c]{@{}c@{}}$\mathbf{0.761}$\\ (0.021)\end{tabular}} & \multicolumn{1}{c|}{\begin{tabular}[c]{@{}c@{}}0.447\\ (0.010)\end{tabular}}            & \multicolumn{1}{c|}{\begin{tabular}[c]{@{}c@{}}0.761\\ (0.018)\end{tabular}}            & \begin{tabular}[c]{@{}c@{}}31.934\\ (1.062)\end{tabular}            \\ \cline{1-7} \cline{9-14} 
$\mathsf{mGHS}_{t_\alpha}$          & \multicolumn{1}{c|}{\begin{tabular}[c]{@{}c@{}}0.976\\ (0.002)\end{tabular}}            & \multicolumn{1}{c|}{\begin{tabular}[c]{@{}c@{}}$\mathbf{0.420}$\\ (0.016)\end{tabular}} & \multicolumn{1}{c|}{\begin{tabular}[c]{@{}c@{}}0.524\\ (0.021)\end{tabular}}            & \multicolumn{1}{c|}{\begin{tabular}[c]{@{}c@{}}0.015\\ (0.002)\end{tabular}}            & \multicolumn{1}{c|}{\begin{tabular}[c]{@{}c@{}}0.830\\ (0.012)\end{tabular}}            & \begin{tabular}[c]{@{}c@{}}34.766\\ (1.629)\end{tabular}            &                   & \multicolumn{1}{c|}{\begin{tabular}[c]{@{}c@{}}0.983\\ (0.001)\end{tabular}}            & \multicolumn{1}{c|}{\begin{tabular}[c]{@{}c@{}}$\mathbf{0.350}$\\ (0.019)\end{tabular}} & \multicolumn{1}{c|}{\begin{tabular}[c]{@{}c@{}}0.308\\ (0.036)\end{tabular}}            & \multicolumn{1}{c|}{\begin{tabular}[c]{@{}c@{}}0.007\\ (0.002)\end{tabular}}            & \multicolumn{1}{c|}{\begin{tabular}[c]{@{}c@{}}0.761\\ (0.018)\end{tabular}}            & \begin{tabular}[c]{@{}c@{}}31.934\\ (1.062)\end{tabular}            \\ \cline{1-7} \cline{9-14} 
$\mathsf{GHS}_{\mathsf{MPM}}$       & \multicolumn{1}{c|}{\begin{tabular}[c]{@{}c@{}}0.639\\ (0.007)\end{tabular}}            & \multicolumn{1}{c|}{\begin{tabular}[c]{@{}c@{}}0.118\\ (0.005)\end{tabular}}            & \multicolumn{1}{c|}{\begin{tabular}[c]{@{}c@{}}$\mathbf{0.812}$\\ (0.015)\end{tabular}} & \multicolumn{1}{c|}{\begin{tabular}[c]{@{}c@{}}0.364\\ (0.007)\end{tabular}}            & \multicolumn{1}{c|}{\begin{tabular}[c]{@{}c@{}}$\mathbf{0.835}$\\ (0.011)\end{tabular}} & \begin{tabular}[c]{@{}c@{}}$\mathbf{34.721}$\\ (1.596)\end{tabular} &                   & \multicolumn{1}{c|}{\begin{tabular}[c]{@{}c@{}}0.551\\ (0.010)\end{tabular}}            & \multicolumn{1}{c|}{\begin{tabular}[c]{@{}c@{}}0.069\\ (0.007)\end{tabular}}            & \multicolumn{1}{c|}{\begin{tabular}[c]{@{}c@{}}0.729\\ (0.023)\end{tabular}}            & \multicolumn{1}{c|}{\begin{tabular}[c]{@{}c@{}}0.451\\ (0.010)\end{tabular}}            & \multicolumn{1}{c|}{\begin{tabular}[c]{@{}c@{}}0.732\\ (0.019)\end{tabular}}            & \begin{tabular}[c]{@{}c@{}}32.946\\ (1.024)\end{tabular}            \\ \cline{1-7} \cline{9-14} 
$\mathsf{fJGL}$                     & \multicolumn{1}{c|}{\begin{tabular}[c]{@{}c@{}}0.956\\ (0.005)\end{tabular}}            & \multicolumn{1}{c|}{\begin{tabular}[c]{@{}c@{}}0.345\\ (0.016)\end{tabular}}            & \multicolumn{1}{c|}{\begin{tabular}[c]{@{}c@{}}0.617\\ (0.025)\end{tabular}}            & \multicolumn{1}{c|}{\begin{tabular}[c]{@{}c@{}}0.038\\ (0.006)\end{tabular}}            & \multicolumn{1}{c|}{\begin{tabular}[c]{@{}c@{}}0.790\\ (0.011)\end{tabular}}            & \begin{tabular}[c]{@{}c@{}}37.527\\ (1.235)\end{tabular}            &                   & \multicolumn{1}{c|}{\begin{tabular}[c]{@{}c@{}}0.971\\ (0.004)\end{tabular}}            & \multicolumn{1}{c|}{\begin{tabular}[c]{@{}c@{}}0.307\\ (0.020)\end{tabular}}            & \multicolumn{1}{c|}{\begin{tabular}[c]{@{}c@{}}0.423\\ (0.030)\end{tabular}}            & \multicolumn{1}{c|}{\begin{tabular}[c]{@{}c@{}}0.021\\ (0.004)\end{tabular}}            & \multicolumn{1}{c|}{\begin{tabular}[c]{@{}c@{}}0.701\\ (0.014)\end{tabular}}            & \begin{tabular}[c]{@{}c@{}}31.616\\ (0.839)\end{tabular}            \\ \cline{1-7} \cline{9-14} 
$\mathsf{gJGL}$                     & \multicolumn{1}{c|}{\begin{tabular}[c]{@{}c@{}}0.956\\ (0.005)\end{tabular}}            & \multicolumn{1}{c|}{\begin{tabular}[c]{@{}c@{}}0.344\\ (0.016)\end{tabular}}            & \multicolumn{1}{c|}{\begin{tabular}[c]{@{}c@{}}0.618\\ (0.024)\end{tabular}}            & \multicolumn{1}{c|}{\begin{tabular}[c]{@{}c@{}}0.038\\ (0.006)\end{tabular}}            & \multicolumn{1}{c|}{\begin{tabular}[c]{@{}c@{}}0.790\\ (0.010)\end{tabular}}            & \begin{tabular}[c]{@{}c@{}}37.581\\ (1.169)\end{tabular}            &                   & \multicolumn{1}{c|}{\begin{tabular}[c]{@{}c@{}}0.970\\ (0.005)\end{tabular}}            & \multicolumn{1}{c|}{\begin{tabular}[c]{@{}c@{}}0.292\\ (0.017)\end{tabular}}            & \multicolumn{1}{c|}{\begin{tabular}[c]{@{}c@{}}0.407\\ (0.040)\end{tabular}}            & \multicolumn{1}{c|}{\begin{tabular}[c]{@{}c@{}}0.022\\ (0.005)\end{tabular}}            & \multicolumn{1}{c|}{\begin{tabular}[c]{@{}c@{}}0.693\\ (0.018)\end{tabular}}            & \begin{tabular}[c]{@{}c@{}}32.616\\ (0.917)\end{tabular}            \\ \cline{1-7} \cline{9-14} 
$\mathsf{GemBAG}_{\mathsf{MPM}}$    & \multicolumn{1}{c|}{\begin{tabular}[c]{@{}c@{}}$\mathbf{0.985}$\\ (0.000)\end{tabular}} & \multicolumn{1}{c|}{\begin{tabular}[c]{@{}c@{}}0.344\\ (0.018)\end{tabular}}            & \multicolumn{1}{c|}{\begin{tabular}[c]{@{}c@{}}0.147\\ (0.013)\end{tabular}}            & \multicolumn{1}{c|}{\begin{tabular}[c]{@{}c@{}}$\mathbf{0.000}$\\ (0.000)\end{tabular}} & \multicolumn{1}{c|}{\begin{tabular}[c]{@{}c@{}}0.697\\ (0.020)\end{tabular}}            & \begin{tabular}[c]{@{}c@{}}46.156\\ (1.840)\end{tabular}            &                   & \multicolumn{1}{c|}{\begin{tabular}[c]{@{}c@{}}$\mathbf{0.986}$\\ (0.000)\end{tabular}} & \multicolumn{1}{c|}{\begin{tabular}[c]{@{}c@{}}0.326\\ (0.022)\end{tabular}}            & \multicolumn{1}{c|}{\begin{tabular}[c]{@{}c@{}}0.130\\ (0.014)\end{tabular}}            & \multicolumn{1}{c|}{\begin{tabular}[c]{@{}c@{}}$\mathbf{0.000}$\\ (0.000)\end{tabular}} & \multicolumn{1}{c|}{\begin{tabular}[c]{@{}c@{}}$\mathbf{0.836}$\\ (0.010)\end{tabular}} & \begin{tabular}[c]{@{}c@{}}$\mathbf{30.824}$\\ (0.927)\end{tabular} \\ \hline \hline
\multirow{2}{*}{}                   & \multicolumn{6}{c|}{$P2020$ ($p_0 = 482.5$)}                                                                                                                                                                                                                                                                                                                                                                                                                                                                                                          &                   & \multicolumn{6}{c|}{$Full \; dependence$ ($p_0 = 485$)}                                                                                                                                                                                                                                                                                                                                                                                                                                                                                             \\ \cline{2-14} 
                                    & \multicolumn{1}{c|}{Acc}                                                                & \multicolumn{1}{c|}{MCC}                                                                & \multicolumn{1}{c|}{TPR}                                                                & \multicolumn{1}{c|}{FPR}                                                                & \multicolumn{1}{c|}{AUC}                                                                & Fr Loss                                                             & \multirow{7}{*}{} & \multicolumn{1}{c|}{Acc}                                                                & \multicolumn{1}{c|}{MCC}                                                                & \multicolumn{1}{c|}{TPR}                                                                & \multicolumn{1}{c|}{FPR}                                                                & \multicolumn{1}{c|}{AUC}                                                                & Fr Loss                                                             \\ \cline{1-14} \cline{1-14} \cline{1-14} 
$\mathsf{mGHS}_{\mathsf{MPM}}$      & \multicolumn{1}{c|}{\begin{tabular}[c]{@{}c@{}}0.654\\ (0.007)\end{tabular}}            & \multicolumn{1}{c|}{\begin{tabular}[c]{@{}c@{}}0.132\\ (0.004)\end{tabular}}            & \multicolumn{1}{c|}{\begin{tabular}[c]{@{}c@{}}$\mathbf{0.863}$\\ (0.013)\end{tabular}} & \multicolumn{1}{c|}{\begin{tabular}[c]{@{}c@{}}0.350\\ (0.007)\end{tabular}}            & \multicolumn{1}{c|}{\begin{tabular}[c]{@{}c@{}}0.885\\ (0.008)\end{tabular}}            & \begin{tabular}[c]{@{}c@{}}26.270\\ (1.321)\end{tabular}            &                   & \multicolumn{1}{c|}{\begin{tabular}[c]{@{}c@{}}0.575\\ (0.011)\end{tabular}}            & \multicolumn{1}{c|}{\begin{tabular}[c]{@{}c@{}}0.095\\ (0.006)\end{tabular}}            & \multicolumn{1}{c|}{\begin{tabular}[c]{@{}c@{}}$\mathbf{0.811}$\\ (0.017)\end{tabular}} & \multicolumn{1}{c|}{\begin{tabular}[c]{@{}c@{}}0.429\\ (0.011)\end{tabular}}            & \multicolumn{1}{c|}{\begin{tabular}[c]{@{}c@{}}0.815\\ (0.014)\end{tabular}}            & \begin{tabular}[c]{@{}c@{}}30.133\\ (1.058)\end{tabular}            \\ \cline{1-7} \cline{9-14} 
$\mathsf{mGHS}_{t_\alpha}$          & \multicolumn{1}{c|}{\begin{tabular}[c]{@{}c@{}}0.972\\ (0.002)\end{tabular}}            & \multicolumn{1}{c|}{\begin{tabular}[c]{@{}c@{}}0.460\\ (0.013)\end{tabular}}            & \multicolumn{1}{c|}{\begin{tabular}[c]{@{}c@{}}0.699\\ (0.017)\end{tabular}}            & \multicolumn{1}{c|}{\begin{tabular}[c]{@{}c@{}}0.024\\ (0.002)\end{tabular}}            & \multicolumn{1}{c|}{\begin{tabular}[c]{@{}c@{}}0.885\\ (0.008)\end{tabular}}            & \begin{tabular}[c]{@{}c@{}}26.270\\ (1.321)\end{tabular}            &                   & \multicolumn{1}{c|}{\begin{tabular}[c]{@{}c@{}}0.981\\ (0.002)\end{tabular}}            & \multicolumn{1}{c|}{\begin{tabular}[c]{@{}c@{}}0.406\\ (0.017)\end{tabular}}            & \multicolumn{1}{c|}{\begin{tabular}[c]{@{}c@{}}0.440\\ (0.034)\end{tabular}}            & \multicolumn{1}{c|}{\begin{tabular}[c]{@{}c@{}}0.011\\ (0.002)\end{tabular}}            & \multicolumn{1}{c|}{\begin{tabular}[c]{@{}c@{}}0.815\\ (0.014)\end{tabular}}            & \begin{tabular}[c]{@{}c@{}}30.133\\ (1.058)\end{tabular}            \\ \cline{1-7} \cline{9-14} 
$\mathsf{GHS}_{\mathsf{MPM}}$       & \multicolumn{1}{c|}{\begin{tabular}[c]{@{}c@{}}0.659\\ (0.007)\end{tabular}}            & \multicolumn{1}{c|}{\begin{tabular}[c]{@{}c@{}}0.123\\ (0.005)\end{tabular}}            & \multicolumn{1}{c|}{\begin{tabular}[c]{@{}c@{}}0.817\\ (0.015)\end{tabular}}            & \multicolumn{1}{c|}{\begin{tabular}[c]{@{}c@{}}0.344\\ (0.007)\end{tabular}}            & \multicolumn{1}{c|}{\begin{tabular}[c]{@{}c@{}}0.850\\ (0.010)\end{tabular}}            & \begin{tabular}[c]{@{}c@{}}29.366\\ (1.298)\end{tabular}            &                   & \multicolumn{1}{c|}{\begin{tabular}[c]{@{}c@{}}0.552\\ (0.010)\end{tabular}}            & \multicolumn{1}{c|}{\begin{tabular}[c]{@{}c@{}}0.068\\ (0.007)\end{tabular}}            & \multicolumn{1}{c|}{\begin{tabular}[c]{@{}c@{}}0.725\\ (0.022)\end{tabular}}            & \multicolumn{1}{c|}{\begin{tabular}[c]{@{}c@{}}0.451\\ (0.010)\end{tabular}}            & \multicolumn{1}{c|}{\begin{tabular}[c]{@{}c@{}}0.728\\ (0.018)\end{tabular}}            & \begin{tabular}[c]{@{}c@{}}32.782\\ (0.948)\end{tabular}            \\ \cline{1-7} \cline{9-14} 
$\mathsf{fJGL}$                     & \multicolumn{1}{c|}{\begin{tabular}[c]{@{}c@{}}0.963\\ (0.004)\end{tabular}}            & \multicolumn{1}{c|}{\begin{tabular}[c]{@{}c@{}}0.416\\ (0.018)\end{tabular}}            & \multicolumn{1}{c|}{\begin{tabular}[c]{@{}c@{}}0.717\\ (0.021)\end{tabular}}            & \multicolumn{1}{c|}{\begin{tabular}[c]{@{}c@{}}0.033\\ (0.004)\end{tabular}}            & \multicolumn{1}{c|}{\begin{tabular}[c]{@{}c@{}}0.842\\ (0.009)\end{tabular}}            & \begin{tabular}[c]{@{}c@{}}31.347\\ (1.244)\end{tabular}            &                   & \multicolumn{1}{c|}{\begin{tabular}[c]{@{}c@{}}0.972\\ (0.004)\end{tabular}}            & \multicolumn{1}{c|}{\begin{tabular}[c]{@{}c@{}}0.395\\ (0.022)\end{tabular}}            & \multicolumn{1}{c|}{\begin{tabular}[c]{@{}c@{}}0.559\\ (0.040)\end{tabular}}            & \multicolumn{1}{c|}{\begin{tabular}[c]{@{}c@{}}0.021\\ (0.004)\end{tabular}}            & \multicolumn{1}{c|}{\begin{tabular}[c]{@{}c@{}}0.769\\ (0.019)\end{tabular}}            & \begin{tabular}[c]{@{}c@{}}26.937\\ (1.079)\end{tabular}            \\ \cline{1-7} \cline{9-14} 
$\mathsf{gJGL}$                     & \multicolumn{1}{c|}{\begin{tabular}[c]{@{}c@{}}0.954\\ (0.006)\end{tabular}}            & \multicolumn{1}{c|}{\begin{tabular}[c]{@{}c@{}}0.347\\ (0.020)\end{tabular}}            & \multicolumn{1}{c|}{\begin{tabular}[c]{@{}c@{}}0.655\\ (0.023)\end{tabular}}            & \multicolumn{1}{c|}{\begin{tabular}[c]{@{}c@{}}0.041\\ (0.006)\end{tabular}}            & \multicolumn{1}{c|}{\begin{tabular}[c]{@{}c@{}}0.807\\ (0.010)\end{tabular}}            & \begin{tabular}[c]{@{}c@{}}37.849\\ (1.392)\end{tabular}            &                   & \multicolumn{1}{c|}{\begin{tabular}[c]{@{}c@{}}0.969\\ (0.005)\end{tabular}}            & \multicolumn{1}{c|}{\begin{tabular}[c]{@{}c@{}}0.291\\ (0.017)\end{tabular}}            & \multicolumn{1}{c|}{\begin{tabular}[c]{@{}c@{}}0.413\\ (0.042)\end{tabular}}            & \multicolumn{1}{c|}{\begin{tabular}[c]{@{}c@{}}0.023\\ (0.006)\end{tabular}}            & \multicolumn{1}{c|}{\begin{tabular}[c]{@{}c@{}}0.695\\ (0.019)\end{tabular}}            & \begin{tabular}[c]{@{}c@{}}32.168\\ (0.913)\end{tabular}            \\ \cline{1-7} \cline{9-14} 
$\mathsf{GemBAG}_{\mathsf{MPM}}$    & \multicolumn{1}{c|}{\begin{tabular}[c]{@{}c@{}}$\mathbf{0.992}$\\ (0.000)\end{tabular}} & \multicolumn{1}{c|}{\begin{tabular}[c]{@{}c@{}}$\mathbf{0.713}$\\ (0.010)\end{tabular}} & \multicolumn{1}{c|}{\begin{tabular}[c]{@{}c@{}}0.516\\ (0.013)\end{tabular}}            & \multicolumn{1}{c|}{\begin{tabular}[c]{@{}c@{}}$\mathbf{0.000}$\\ (0.000)\end{tabular}} & \multicolumn{1}{c|}{\begin{tabular}[c]{@{}c@{}}$\mathbf{0.893}$\\ (0.007)\end{tabular}} & \begin{tabular}[c]{@{}c@{}}$\mathbf{18.421}$\\ (0.865)\end{tabular} &                   & \multicolumn{1}{c|}{\begin{tabular}[c]{@{}c@{}}$\mathbf{0.989}$\\ (0.000)\end{tabular}} & \multicolumn{1}{c|}{\begin{tabular}[c]{@{}c@{}}$\mathbf{0.534}$\\ (0.017)\end{tabular}} & \multicolumn{1}{c|}{\begin{tabular}[c]{@{}c@{}}0.293\\ (0.017)\end{tabular}}            & \multicolumn{1}{c|}{\begin{tabular}[c]{@{}c@{}}$\mathbf{0.000}$\\ (0.000)\end{tabular}} & \multicolumn{1}{c|}{\begin{tabular}[c]{@{}c@{}}$\mathbf{0.893}$\\ (0.008)\end{tabular}} & \begin{tabular}[c]{@{}c@{}}$\mathbf{26.442}$\\ (0.976)\end{tabular} \\ \hline
\end{tabular}
\end{adjustbox}
\caption{Simulation results for $n = 100$ and $p = 250$ ($50$ replicates). Methods mGHS and GHS are evaluated over $B = 10000$ post burn-in samples.}
\label{tab:simn100p250}
\end{table}


\begin{table}[H]
\begin{adjustbox}{width = \textwidth}
\begin{tabular}{|c|cccccc|c|cccccc|}
\hline
\multirow{2}{*}{$n = 100, p = 500$} & \multicolumn{6}{c|}{$Independent$ ($p_0 = 271.25$)}                                                                                                                                                                                                                                                                                                                                                                                                                                                                                   &                   & \multicolumn{6}{c|}{$Coupled$ ($p_0 = 279.5$)}                                                                                                                                                                                                                                                                                                                                                                                                                                                                                        \\ \cline{2-14} 
                                    & \multicolumn{1}{c|}{Acc}                                                                & \multicolumn{1}{c|}{MCC}                                                                & \multicolumn{1}{c|}{TPR}                                                                & \multicolumn{1}{c|}{FPR}                                                                & \multicolumn{1}{c|}{AUC}                                                                & Fr Loss                                                             & \multirow{4}{*}{} & \multicolumn{1}{c|}{Acc}                                                                & \multicolumn{1}{c|}{MCC}                                                                & \multicolumn{1}{c|}{TPR}                                                                & \multicolumn{1}{c|}{FPR}                                                                & \multicolumn{1}{c|}{AUC}                                                                & Fr Loss                                                             \\ \cline{1-14} \cline{1-14} \cline{1-14}
$\mathsf{mGHS}_{\mathsf{MPM}}$      & \multicolumn{1}{c|}{\begin{tabular}[c]{@{}c@{}}0.518\\ (0.003)\end{tabular}}            & \multicolumn{1}{c|}{\begin{tabular}[c]{@{}c@{}}0.034\\ (0.002)\end{tabular}}            & \multicolumn{1}{c|}{\begin{tabular}[c]{@{}c@{}}$\mathbf{0.845}$\\ (0.023)\end{tabular}} & \multicolumn{1}{c|}{\begin{tabular}[c]{@{}c@{}}0.482\\ (0.003)\end{tabular}}            & \multicolumn{1}{c|}{\begin{tabular}[c]{@{}c@{}}$\mathbf{0.832}$\\ (0.018)\end{tabular}} & \begin{tabular}[c]{@{}c@{}}41.906\\ (1.633)\end{tabular}            &                   & \multicolumn{1}{c|}{\begin{tabular}[c]{@{}c@{}}0.525\\ (0.004)\end{tabular}}            & \multicolumn{1}{c|}{\begin{tabular}[c]{@{}c@{}}0.043\\ (0.002)\end{tabular}}            & \multicolumn{1}{c|}{\begin{tabular}[c]{@{}c@{}}$\mathbf{0.929}$\\ (0.014)\end{tabular}} & \multicolumn{1}{c|}{\begin{tabular}[c]{@{}c@{}}0.476\\ (0.004)\end{tabular}}            & \multicolumn{1}{c|}{\begin{tabular}[c]{@{}c@{}}0.922\\ (0.010)\end{tabular}}            & \begin{tabular}[c]{@{}c@{}}39.677\\ (1.851)\end{tabular}            \\ \cline{1-7} \cline{9-14} 
$\mathsf{mGHS}_{t_\alpha}$          & \multicolumn{1}{c|}{\begin{tabular}[c]{@{}c@{}}0.997\\ (0.000)\end{tabular}}            & \multicolumn{1}{c|}{\begin{tabular}[c]{@{}c@{}}$\mathbf{0.430}$\\ (0.021)\end{tabular}} & \multicolumn{1}{c|}{\begin{tabular}[c]{@{}c@{}}0.461\\ (0.030)\end{tabular}}            & \multicolumn{1}{c|}{\begin{tabular}[c]{@{}c@{}}0.001\\ (0.000)\end{tabular}}            & \multicolumn{1}{c|}{\begin{tabular}[c]{@{}c@{}}$\mathbf{0.832}$\\ (0.018)\end{tabular}} & \begin{tabular}[c]{@{}c@{}}41.906\\ (1.633)\end{tabular}            &                   & \multicolumn{1}{c|}{\begin{tabular}[c]{@{}c@{}}0.994\\ (0.003)\end{tabular}}            & \multicolumn{1}{c|}{\begin{tabular}[c]{@{}c@{}}0.425\\ (0.064)\end{tabular}}            & \multicolumn{1}{c|}{\begin{tabular}[c]{@{}c@{}}0.747\\ (0.022)\end{tabular}}            & \multicolumn{1}{c|}{\begin{tabular}[c]{@{}c@{}}0.006\\ (0.003)\end{tabular}}            & \multicolumn{1}{c|}{\begin{tabular}[c]{@{}c@{}}0.922\\ (0.010)\end{tabular}}            & \begin{tabular}[c]{@{}c@{}}39.677\\ (1.851)\end{tabular}            \\ \cline{1-7} \cline{9-14} 
$\mathsf{GemBAG}_{\mathsf{MPM}}$      & \multicolumn{1}{c|}{\begin{tabular}[c]{@{}c@{}}$\mathbf{0.998}$\\ (0.000)\end{tabular}} & \multicolumn{1}{c|}{\begin{tabular}[c]{@{}c@{}}0.379\\ (0.032)\end{tabular}}            & \multicolumn{1}{c|}{\begin{tabular}[c]{@{}c@{}}0.172\\ (0.023)\end{tabular}}            & \multicolumn{1}{c|}{\begin{tabular}[c]{@{}c@{}}$\mathbf{0.000}$\\ (0.000)\end{tabular}} & \multicolumn{1}{c|}{\begin{tabular}[c]{@{}c@{}}0.799\\ (0.013)\end{tabular}}            & \begin{tabular}[c]{@{}c@{}}$\mathbf{40.747}$\\ (1.495)\end{tabular} &                   & \multicolumn{1}{c|}{\begin{tabular}[c]{@{}c@{}}$\mathbf{0.999}$\\ (0.000)\end{tabular}} & \multicolumn{1}{c|}{\begin{tabular}[c]{@{}c@{}}$\mathbf{0.740}$\\ (0.015)\end{tabular}} & \multicolumn{1}{c|}{\begin{tabular}[c]{@{}c@{}}0.634\\ (0.036)\end{tabular}}            & \multicolumn{1}{c|}{\begin{tabular}[c]{@{}c@{}}$\mathbf{0.000}$\\ (0.000)\end{tabular}} & \multicolumn{1}{c|}{\begin{tabular}[c]{@{}c@{}}$\mathbf{0.962}$\\ (0.014)\end{tabular}} & \begin{tabular}[c]{@{}c@{}}$\mathbf{33.100}$\\ (3.257)\end{tabular} \\ \hline \hline
\multirow{2}{*}{}                   & \multicolumn{6}{c|}{$P2020$ ($p_0 = 270.5$)}                                                                                                                                                                                                                                                                                                                                                                                                                                                                                          &                   & \multicolumn{6}{c|}{$Full \; Dependence$ ($p_0 = 273$)}                                                                                                                                                                                                                                                                                                                                                                                                                                                                               \\ \cline{2-14} 
                                    & \multicolumn{1}{c|}{Acc}                                                                & \multicolumn{1}{c|}{MCC}                                                                & \multicolumn{1}{c|}{TPR}                                                                & \multicolumn{1}{c|}{FPR}                                                                & \multicolumn{1}{c|}{AUC}                                                                & Fr Loss                                                             & \multirow{4}{*}{} & \multicolumn{1}{c|}{Acc}                                                                & \multicolumn{1}{c|}{MCC}                                                                & \multicolumn{1}{c|}{TPR}                                                                & \multicolumn{1}{c|}{FPR}                                                                & \multicolumn{1}{c|}{AUC}                                                                & Fr Loss                                                             \\ \cline{1-14} \cline{1-14} \cline{1-14} 
$\mathsf{mGHS}_{\mathsf{MPM}}$      & \multicolumn{1}{c|}{\begin{tabular}[c]{@{}c@{}}0.522\\ (0.004)\end{tabular}}            & \multicolumn{1}{c|}{\begin{tabular}[c]{@{}c@{}}0.041\\ (0.002)\end{tabular}}            & \multicolumn{1}{c|}{\begin{tabular}[c]{@{}c@{}}$\mathbf{0.915}$\\ (0.018)\end{tabular}} & \multicolumn{1}{c|}{\begin{tabular}[c]{@{}c@{}}0.479\\ (0.003)\end{tabular}}            & \multicolumn{1}{c|}{\begin{tabular}[c]{@{}c@{}}0.909\\ (0.015)\end{tabular}}            & \begin{tabular}[c]{@{}c@{}}38.121\\ (2.107)\end{tabular}            &                   & \multicolumn{1}{c|}{\begin{tabular}[c]{@{}c@{}}0.523\\ (0.004)\end{tabular}}            & \multicolumn{1}{c|}{\begin{tabular}[c]{@{}c@{}}0.043\\ (0.002)\end{tabular}}            & \multicolumn{1}{c|}{\begin{tabular}[c]{@{}c@{}}$\mathbf{0.938}$\\ (0.016)\end{tabular}} & \multicolumn{1}{c|}{\begin{tabular}[c]{@{}c@{}}0.478\\ (0.004)\end{tabular}}            & \multicolumn{1}{c|}{\begin{tabular}[c]{@{}c@{}}0.933\\ (0.013)\end{tabular}}            & \begin{tabular}[c]{@{}c@{}}38.741\\ (2.042)\end{tabular}            \\ \cline{1-7} \cline{9-14} 
$\mathsf{mGHS}_{t_\alpha}$          & \multicolumn{1}{c|}{\begin{tabular}[c]{@{}c@{}}0.979\\ (0.013)\end{tabular}}            & \multicolumn{1}{c|}{\begin{tabular}[c]{@{}c@{}}0.274\\ (0.085)\end{tabular}}            & \multicolumn{1}{c|}{\begin{tabular}[c]{@{}c@{}}0.770\\ (0.020)\end{tabular}}            & \multicolumn{1}{c|}{\begin{tabular}[c]{@{}c@{}}0.020\\ (0.013)\end{tabular}}            & \multicolumn{1}{c|}{\begin{tabular}[c]{@{}c@{}}0.909\\ (0.015)\end{tabular}}            & \begin{tabular}[c]{@{}c@{}}38.121\\ (2.107)\end{tabular}            &                   & \multicolumn{1}{c|}{\begin{tabular}[c]{@{}c@{}}0.979\\ (0.013)\end{tabular}}            & \multicolumn{1}{c|}{\begin{tabular}[c]{@{}c@{}}0.286\\ (0.080)\end{tabular}}            & \multicolumn{1}{c|}{\begin{tabular}[c]{@{}c@{}}0.819\\ (0.019)\end{tabular}}            & \multicolumn{1}{c|}{\begin{tabular}[c]{@{}c@{}}0.020\\ (0.013)\end{tabular}}            & \multicolumn{1}{c|}{\begin{tabular}[c]{@{}c@{}}0.933\\ (0.013)\end{tabular}}            & \begin{tabular}[c]{@{}c@{}}38.741\\ (2.042)\end{tabular}            \\ \cline{1-7} \cline{9-14} 
$\mathsf{GemBAG}_{\mathsf{MPM}}$      & \multicolumn{1}{c|}{\begin{tabular}[c]{@{}c@{}}$\mathbf{0.999}$\\ (0.000)\end{tabular}} & \multicolumn{1}{c|}{\begin{tabular}[c]{@{}c@{}}$\mathbf{0.839}$\\ (0.012)\end{tabular}} & \multicolumn{1}{c|}{\begin{tabular}[c]{@{}c@{}}0.723\\ (0.030)\end{tabular}}            & \multicolumn{1}{c|}{\begin{tabular}[c]{@{}c@{}}$\mathbf{0.000}$\\ (0.000)\end{tabular}} & \multicolumn{1}{c|}{\begin{tabular}[c]{@{}c@{}}$\mathbf{0.966}$\\ (0.011)\end{tabular}} & \begin{tabular}[c]{@{}c@{}}$\mathbf{23.992}$\\ (4.506)\end{tabular} &                   & \multicolumn{1}{c|}{\begin{tabular}[c]{@{}c@{}}$\mathbf{0.999}$\\ (0.000)\end{tabular}} & \multicolumn{1}{c|}{\begin{tabular}[c]{@{}c@{}}$\mathbf{0.873}$\\ (0.014)\end{tabular}} & \multicolumn{1}{c|}{\begin{tabular}[c]{@{}c@{}}0.771\\ (0.028)\end{tabular}}            & \multicolumn{1}{c|}{\begin{tabular}[c]{@{}c@{}}$\mathbf{0.000}$\\ (0.000)\end{tabular}} & \multicolumn{1}{c|}{\begin{tabular}[c]{@{}c@{}}$\mathbf{0.979}$\\ (0.006)\end{tabular}} & \begin{tabular}[c]{@{}c@{}}$\mathbf{22.112}$\\ (3.407)\end{tabular} \\ \hline
\end{tabular}
\end{adjustbox}
\caption{Simulation results for $n = 100$ and $p = 500$ ($25$ replicates). Method mGHS is evaluated over $B = 10000$ post burn-in samples.}
\label{tab:simn100p500}
\end{table}

\section{Application to a bike-sharing dataset}
\label{sec:application}
%
We perform an analysis of the Capital Bikeshare system data\footnote{Data are available at \href{http://www.capitalbikeshare.com/system-data}{http://www.capitalbikeshare.com/system-data}}, a benchmark dataset previously analyzed in \cite{Zhu-2015} and \cite{Yang-2021}. This is the first analysis of this dataset with a full Bayesian graphical model. 
The dataset contains records of bike rentals in a bicycle sharing system with more than $500$ stations located in the Washington D.C. area, where each ride is labeled as \textit{casual} (paying for a single day) or \textit{member} (membership payment). Data from years 2016, 2017, and 2018 are used, for a total of $n = 1092$ registered days. Only the $p = 239$ most active stations are selected. Therefore, for $i = 1, \dots, 1092$ and $j = 1, \dots, 239$, let $y_{ij}^c$ and $y_{ij}^m$ be the number of registered casual and member trips initiated at station $j$ on day $i$, respectively. After correcting for the seasonal trend, each station data is marginally standardized and transformed with the Yeo-Johnson transformation \citep{Yeo-2000} to better approximate a Gaussian distribution. Finally, the data are divided by year and rider membership for a total of $K = 6$ groups. Matrices $\mathbf{Y}_k$, $k = 1, \dots, 6$ are marginally standardized such that $\boldsymbol{\mu}_k = \mathbf{0}$  and the standard deviations are equal to 1 for each group.

For each class, $80\%$ of the observations are used as training set and the remaining $20\%$ as test set. For $k = 1, \dots, 6$, let $\hat{\boldsymbol{\Omega}}_k$ be the estimated precision matrix of the $k$-th training set. Here we take the posterior mean. Following \cite{Fan-2009}, the observations of each test set is partitioned as $\mathbf{y}_i^k = \left(\mathbf{y}_{i, j_1}^k, \mathbf{y}_{i, j_2}^k\right)$, where $\mathbf{y}_{i, j_1}^k = \left(y_{i, 1}^k, \dots, y_{i, 120}^k\right)$ and $\mathbf{y}_{i, j_2}^k = \left(y_{i, 121}^k, \dots, y_{i, 239}^k\right)$, $i = 1, \dots, n_k$. The corresponding partition for $\boldsymbol{\Omega}_k$ and $\boldsymbol{\Sigma}_k$ are
\begin{equation*}
	\boldsymbol{\Omega}_k = \begin{bmatrix} \boldsymbol{\Omega}_{k_{11}} & \boldsymbol{\Omega}_{k_{12}} \\ \boldsymbol{\Omega}_{k_{21}} & \boldsymbol{\Omega}_{k_{22}} \end{bmatrix} \quad \text{and} \quad \boldsymbol{\Sigma}_k = \begin{bmatrix} \boldsymbol{\Sigma}_{k_{11}} & \boldsymbol{\Sigma}_{k_{12}} \\ \boldsymbol{\Sigma}_{k_{21}} & \boldsymbol{\Sigma}_{k_{22}} \end{bmatrix}.
\end{equation*}
The performances of the models are evaluated by predicting $\mathbf{y}_{i, j_2}^k$ based on $\mathbf{y}_{i, j_1}^k$ and $\hat{\boldsymbol{\Omega}}_k$. Under the Gaussian assumption, the best linear predictor is
\begin{equation*}
	\hat{\mathbf{y}}_{i, j_2}^k = \mathbb{E}\left(\mathbf{y}_{i, j_2}^k\mid\mathbf{y}_{i, j_1}^k\right) = \hat{\boldsymbol{\Sigma}}_{k_{21}} \hat{\boldsymbol{\Sigma}}_{k_{11}}^{-1} \mathbf{y}_{i, j_1}^k.
\end{equation*}
To assess the prediction performances of the methods we rely on the \textit{average absolute forecast error} (AAFE), defined as
\begin{equation*}
	\text{AAFE}_k = \frac{1}{119}\frac{1}{ \vert \mathbb{T}_k \vert} \sum_{i \in \mathbb{T}_k} \sum_{j = 121}^{239} \vert y_{ij}^k - \hat{y}_{ij}^k\vert,
\end{equation*}
where $\mathbb{T}_k$ denotes the test set indexes for group $k$. We denote the mean AAFE across groups as mAAFE.

The multiple Graphical Horseshoe is tested against the ordinary Graphical Horseshoe of \cite{Bahdra-2019} and the GemBAG of \cite{Yang-2021}. For the estimation of the threshold in the mGHS model we set the hyperparameter to $a = 30$ and $b = 25$, whereas in GemBAG we estimated hyperparameters $v_0$ and $v_1$ according to the BIC criterion as in Section \ref{sec:simulation}. For computational reasons, the joint Graphical LASSO of \cite{Danaher-2014} is excluded from the analysis. We checked the convergence of mGHS algorithm by estimating the potential scale reduction factor \citep[psrf,][]{Gelman-1992} of parameters $\omega_{ij}^K$, $i = 1, \dots, 239$, $j > i$, $k = 1, \dots, 6$, over $4$ replications. The distribution of the estimated psrf is shown in Figure 1 of the Supplementary Material; roughly 99$\%$ of the estimated values lie in the interval $[1.0, 1.2]$. Finally, the trace plots of the log-posterior are shown in Figure 2 of the Supplementary Material and do not suggest a lack of convergence of the chains.

\begin{figure}[h!]
\centering
\begin{subfigure}{.5\textwidth}
  \centering
  \hspace*{-0.8in}
  \includegraphics[scale = 0.5]{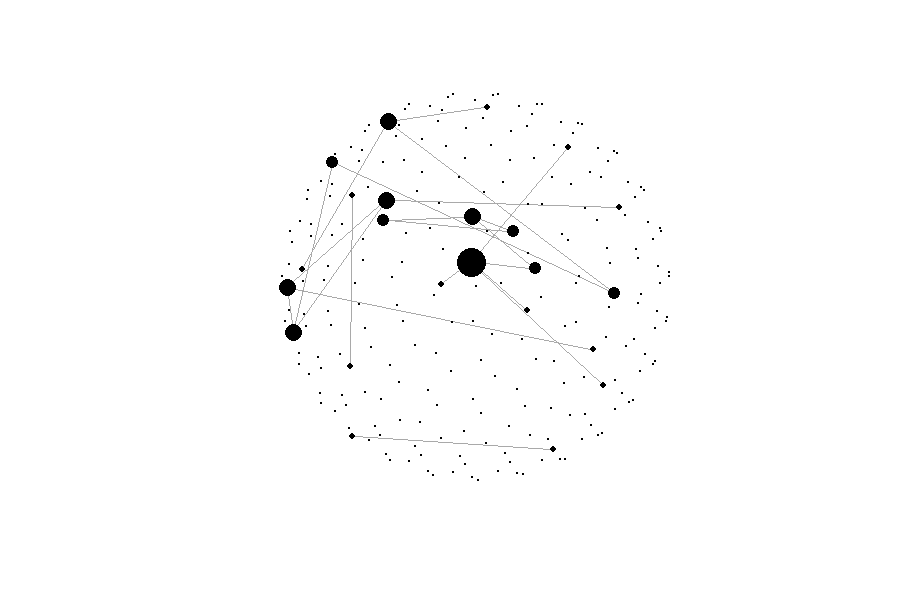}
  \caption{Casual network estimated by mGHS}
  \label{fig:casual1}
\end{subfigure}%
\begin{subfigure}{.5\textwidth}
  \centering
  \hspace*{-0.8in}
  \includegraphics[scale = 0.5]{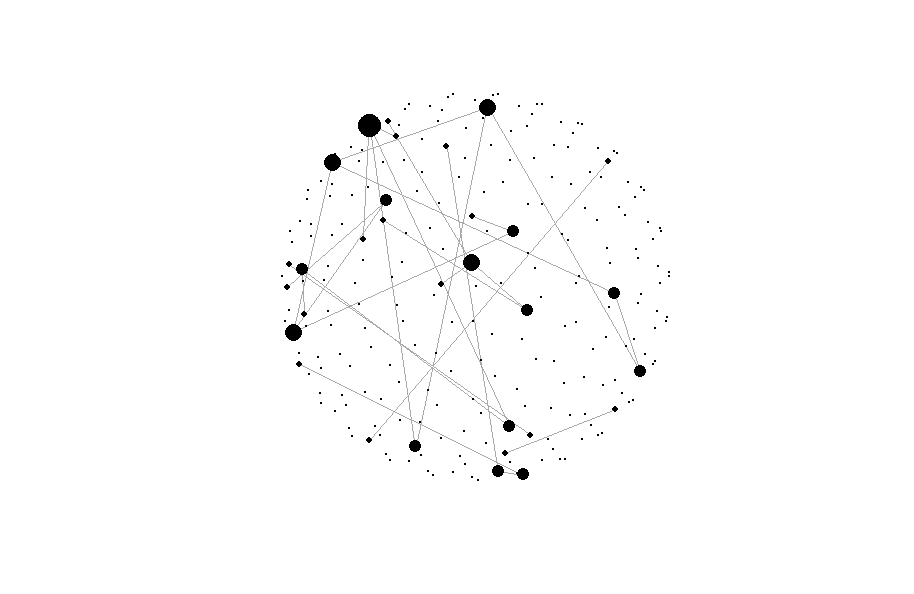}
  \caption{Member network estimated by mGHS}
  \label{fig:member1}
\end{subfigure}
\begin{subfigure}{.5\textwidth}
  \centering
  \hspace*{-0.8in}
  \includegraphics[scale = 0.5]{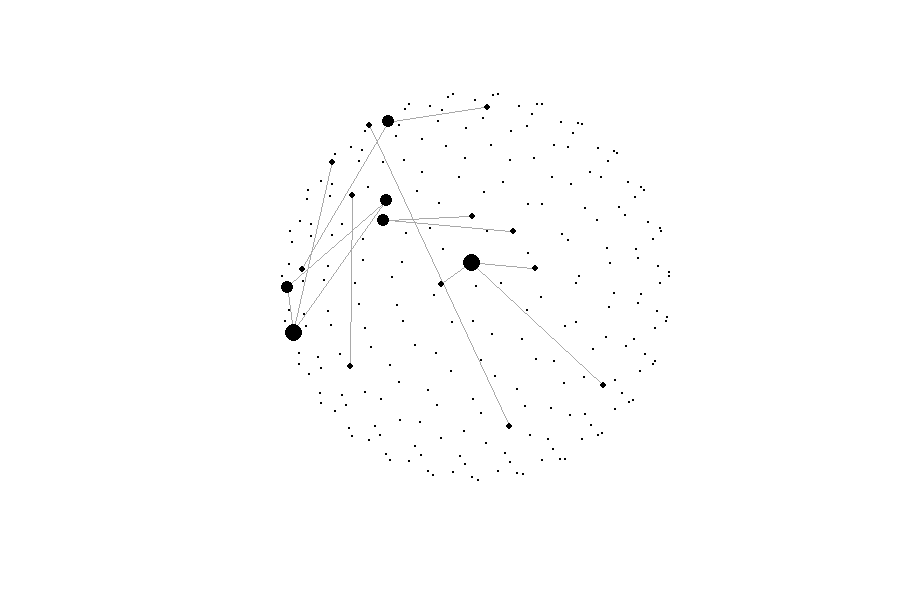}
  \caption{Casual network estimated by GHS}
  \label{fig:casual2}
\end{subfigure}%
\begin{subfigure}{.5\textwidth}
  \centering
  \hspace*{-0.8in}
  \includegraphics[scale = 0.5]{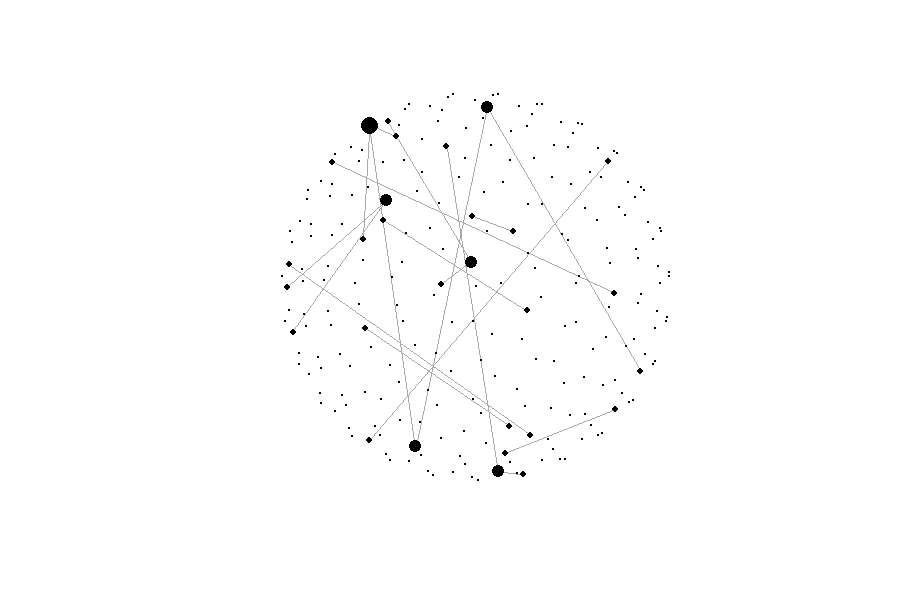}
  \caption{Member network estimated by GHS}
  \label{fig:member2}
\end{subfigure}
\caption{Intersection of the estimated networks across three years; the size of the nodes depends on the number of edges associated to the related station}
\label{fig:applicationdata}
\end{figure}

With $\text{mAAFE} = 0.596$, the best predictive model is the mGHS, whereas the ordinary GHS shows similar predictive performance ($\text{mAAFE} = 0.600$). The latter, however, provides a sparser model: regardless of the method used for selecting the edges a posteriori, the mGHS always estimates denser networks, including connections between stations that the GHS is not able to capture. Finally, the GemBAG provides at the same time the sparsest model and the worst predictive performance, with $\text{mAAFE} = 0.613$. 

\noindent To further understand how the connections between stations work among the casual and member users, we plot the estimated networks for each group for both GHS and mGHS (Figures 3 and 4 in Appendix B of Supplementary Materials), where we select those edges with a posterior inclusion probability higher than 0.9. The estimated networks for casual users are denser in both models, suggesting a higher activity of casual rides. However, the number of edges shared across the years is higher for the registered users, implying more regular activities of those who choose to pay a seasonal ticket. The intersection of the estimated networks across three years for the registered and casual users is shown in Figure \ref{fig:applicationdata} for both GHS and mGHS, where the size of the nodes depends on the number of edges associated with the related stations. The two models estimate similar networks for both types of users, however, mGHS gives more importance to the stations identified by GHS and includes some additional ones.

The hypothesis of a more regular behavior of the registered users is supported also by the estimated correlation matrix between groups, i.e. the posterior mean of $\mathbf{R}$:
\begin{equation*}
\centering
\scriptsize
\begin{blockarray}{cccccccc}
& \text{casual 2016} & \text{casual 2017} & \text{casual 2018} & \text{member 2016} & \text{member 2017} & \text{member 2018} & \\
\begin{block}{c(cccccc)c}
&  1.000 & 0.969 & 0.893 & 0.479 & 0.515 & 0.483 & \text{casual 2016} \\
&  0.969 & 1.000 & 0.958 & 0.518 & 0.562 & 0.526 & \text{casual 2017} \\
 & 0.893 & 0.958 & 1.000 & 0.461 & 0.502 & 0.475 & \text{casual 2018} \\
\hat{\mathbf{R}} = &  0.479 & 0.518 & 0.461 & 1.000 & 0.984 & 0.971 & \text{member 2016} \\
 & 0.515 & 0.562 & 0.502 & 0.984 & 1.000 & 0.980 & \text{member 2017} \\
 & 0.483 & 0.526 & 0.475 & 0.971 & 0.980 & 1.000 & \text{member 2018} \\
\end{block}
\end{blockarray}
\end{equation*}
The correlation is high across the years for both types of users. In particular, it remains close to $1$ even after two years for the rides with membership payment (correlation between 2016 and 2018 is $0.971$). On the contrary, the decrease is larger for the casual rides, with a correlation of $0.893$.


\section{Conclusion}
\label{sec:conclusion}
%

We have introduced a novel fully Bayesian method for the analysis of high-dimensional dependent precision matrices. In particular, we provided an efficient approach that works up to hundreds of variables. We empirically showed that the model is able to borrow information between groups when appropriately supported by the data. Simulation studies empirically demonstrated that the proposed approach has good performances in terms of edge selection; 
the proposed joint model performs at least as well as the separate analysis of each group with the ordinary Graphical Horseshoe \citep{Bahdra-2019}. We applied our method to a benchmark dataset with a slight improvement in prediction performance. Compared to the ordinary Graphical Horseshoe, the proposed model borrowed information across groups and selected a higher number of common edges across the years. Moreover, the estimation of correlation matrix $\mathbf{R}$ provided unique insights about the behavior of bike-sharing users.
We also proposed a new approach for posterior edge selection that accounts for posterior dependencies between parameters $\lambda_{ij, k}^2$'s. This method can be easily extended to other common frameworks, for example, variable selection in regression models. Further improvements concern the introduction of different thresholds $t_{ij}^\alpha$ behavioredge or adaptive methods to improve the proposal distribution. The proposed cut model provides only an approximation of the posterior distribution, and, in models with cuts in general, the algorithm may fail to converge to a well-defined distribution \citep{plummer-2015}. Whereas cut models can outperform fully Bayesian models in terms of performance and computational efficiency, a careful assessment of the output produced by models with cuts should be always performed.

Note that very recently \cite{Lingjaerde-2022} have proposed an approach, alternative to the one presented in this paper, for the analysis of multiple graphical models with horseshoe priors, termed the joint graphical horseshoe. The approach proposed in this paper, with respect to the joint graphical horseshoe, is characterized by a few important and unique features, since it provides full Bayesian inference, it adapts well to setting with heterogeneous levels of network similarity, it learns the level of network similarity across groups from the data, and it has been successfully applied to networks with large $p$ (up to 500 nodes).

Among possible extensions, we may consider a spike-and-slab type of prior on the off-diagonal elements of the correlation matrix $\mathbf{R}$.
This approach would not only give a deeper insight into the similarity across the groups, but it would speed the model up when the groups are not significantly related: the $\mathcal{G}_{3p}$ distribution would reduce to an Inverse-Gamma when the $k$-th row of the matrix $\mathbf{R}$ is zero, avoiding the need of the rejection sampling discussed in Section \ref{sec:G3p}.

A main challenge, and still a limitation, of the proposed approach, is the computational complexity of the algorithm since it becomes infeasible when the number of covariates $p$ is extremely large, e.g., in the thousands. 
Alternative computational approaches that could be explored include the thresholding approach of \cite{Johndrow-2020} that could be adapted to sample from multivariate Normal distributions under the Horseshoe prior, and eventually lead to a significant reduction in computational times.

The R code for mGHS model, simulations studies and application to bike-sharing dataset is available at \href{https://github.com/cbusatto/mGHS}{https://github.com/cbusatto/mGHS}.

\bigskip
\noindent {\large\bf Supplementary Materials} \\
In Supplementary Materials (file \texttt{mGHS\_suppl\_mat.pdf}) can be found:
\begin{description}
    \item[mGHS algorithm:] the pseudo-code for mGHS algorithm can be found in Appendix A.
    %
    %
    \item[Supplementary Figures:] distribution of the estimated psrf and post-burnin trace of the log-posterior distribution of the $4$ chains for the analysis of the convergence of mGHS algorithm are shown in Figure 1 and 2 of Appendix B, respectively, whereas estimated networks by both GHS and mGHS models for bike-sharing dataset are given in Figures 3 and 4 of Appendix B.
\end{description}

\bigskip
\noindent {\large\bf Acknowledgments}

\bigskip
\noindent {\large\bf Funding details}\\
Both authors were partially supported by the “Dipartimenti Eccellenti 2018-2022” ministerial funds (Italy). 

\bigskip
\noindent {\large\bf Disclosure statement} \\
The authors report there are no competing interests to declare.

\clearpage
\newpage
\bibliographystyle{apalike}
\bibliography{mGHS_biblio}
\clearpage 

\appendix

\section{Appendix A}
\label{app:a}


\subsection{Technical details of the modified rejection sampling method}
\label{app:a1}

The acceptance probability of each step of the algorithm is compute as follows:
\begin{itemize}
	\item Step 1: the probability of immediate acceptance is
	\begin{equation*}
		P\left(E_1\right) = \Phi_{0, \omega^2}\left(t_2\right) - \Phi_{0, \omega^2}\left(t_1\right),
	\end{equation*}
	where $\Phi_{\mu, \sigma^2}\left(\cdot\right)$ denotes the cumulative density function of a gaussian distribution with mean $\mu$ and variance $\sigma^2$;
	\item Step 2: the acceptance probability of Step 2 is 
	\begin{equation*}
		P\left(E_2\right) =1 - P\left(E_1\right) - P\left(E_3\right),
	\end{equation*}
	where $P\left(E_3\right)$ is the acceptance probability of Step 3.
	\item Step 3: the probability of acceptance this step is
	\begin{align*}
		P\left(E_3\right) = & \; \int_{-\infty}^{t_1} h(t) dt + \int_{t_2}^\infty h(t) dt - \int_{-\frac{\mu}{\sigma}}^{t_1} g(t) dt + \int_{t_2}^\infty g(t) dt \nonumber\\
		= & \; \int_{-\infty}^\infty h(t) dt - \int_{t_1}^{t_2} h(t) dt - \int_{-\frac{\mu}{\sigma}}^\infty g(t) dt + \int_{t_1}^{t_2} g(t) dt \nonumber\\
		= & \; \int_{t_1}^{t_2} g(t) - h(t) dt.
	\end{align*}
\end{itemize}


\subsection{Rejection sampling for sampling from the difference distribution $d(t)$}
\label{app:a2}

\begin{figure}
    \centering
    \includegraphics[scale = 0.5]{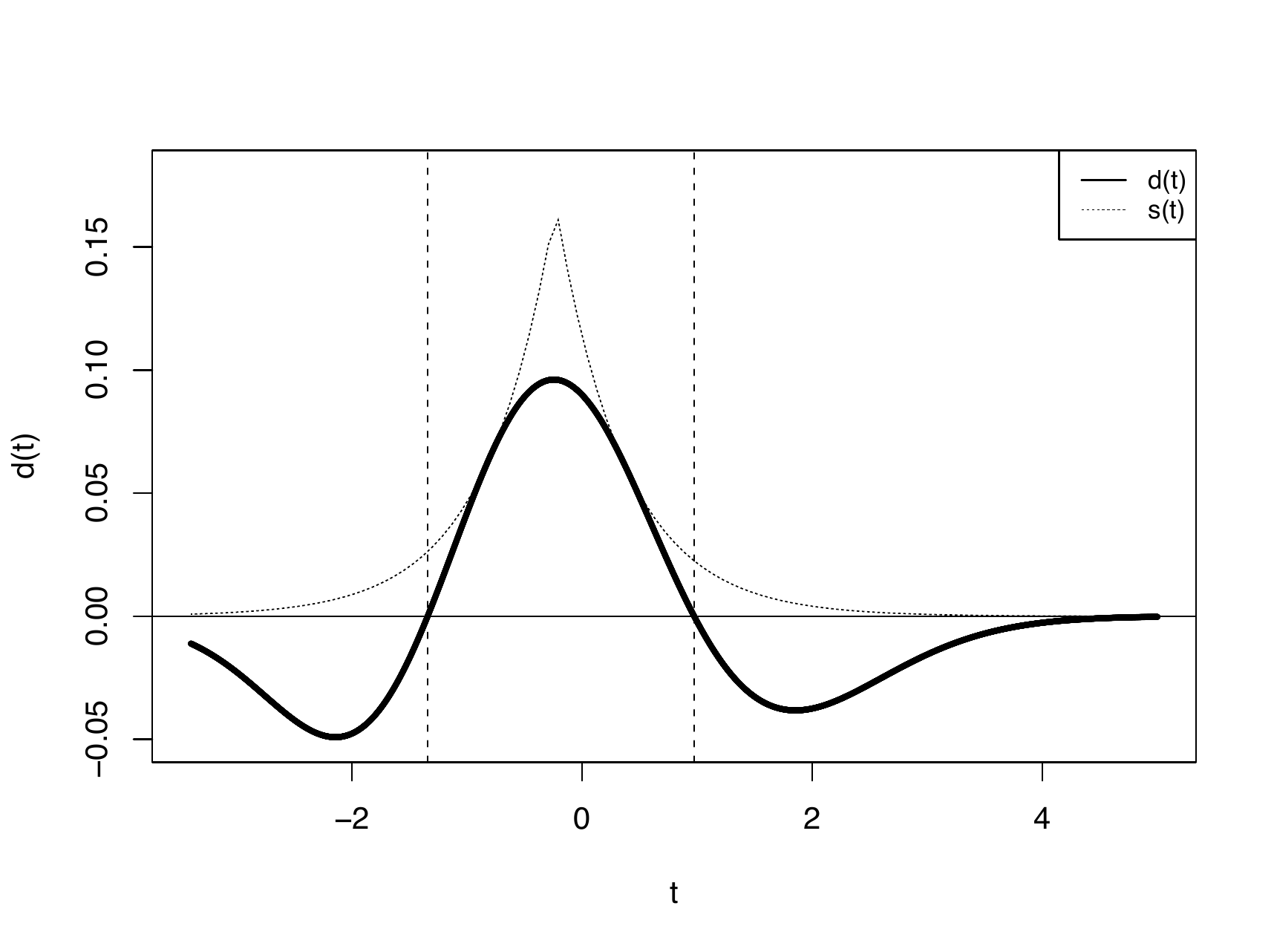}
    \caption{distributions $d(t)$ and $s(t)$; dotted lines represent $t_1$ and $t_2$.}
	\label{fig:2}
\end{figure}

Sampling from $d(t)$ in Step 3 can be achieved by means of a standard rejection sampling. Let $s(t)$ be the proposal distribution, we adapt a double-exponential (Laplace) distribution of the form
\begin{equation*}
	s(t) = \frac{c}{\sqrt{2 \pi}} e^{-\frac{\vert t - b\vert}{\delta}}, \quad -\infty < t < \infty,
\end{equation*}
in order to minimize the area between $s(t)$ and $d(t)$ \citep{Ahrens-1982, Stadlober-1982}. This happens when the hat function $s(t)$ touches $d(t)$ at two different points $L$ and $R$, with $R > L$. As explained in \cite{Dieter-1981}, if $d(t)$ is covered by a double-exponential distribution, optimal parameters $c$, $b$, and $\delta$ can be estimated within two steps: first, points $L$, $R$ and parameter $\delta$ are computed simultaneously (for instance by Newton iteration) as
 \begin{align*}
 	d'\left(L\right) = &\; \frac{1}{\delta} d\left(L\right) \\
 	d'\left(R\right) = &\; -\frac{1}{\delta} d\left(R\right) \\
 	\delta = &\; \frac{1}{2} \left(R - L\right), 
 \end{align*}
whereas parameters $c$ and $b$ are calculated as
\begin{align*}
	b = & \; \frac{1}{2} \left(L + R + \delta ln \left(\frac{d(R)}{d(L)}\right)\right) \\
	c = &\; e \sqrt{2 \pi d(R) d(L)}.
\end{align*}
Figure \ref{fig:2} shows the difference function $d(t)$ and its optimal hat function $s(t)$.

The algorithm can be further sped up by noting that the quantities $t_1$, $t_2$, $b$, $c$ and $\delta$ only depend on the ratio $\beta/\alpha$. The computation of these parameters, which involve iterative methods, can be avoided by tabulating the needed quantities for a restricted grid of the parameters $\gamma$, $\alpha$, and $\beta$.


\subsection{Proof of Proposition \ref{prep_1}}
\label{app:a3}

Recalling that $t = (x-\mu)/\sigma$, where $x > 0$, the acceptance probability of the first two steps of the algorithm can be computed as
\begin{align*}
	\mathbb{P}\left(T_{acc} \right) = & \; \mathbb{P}\left(U \le \frac{g(t)}{h(t)} \right) \nonumber \\
	= & \; \int_{-\frac{\mu}{\sigma}}^{\infty} \mathbb{P}\left(U \le \frac{g(t)}{h(t)} \big| T = t \right) h(t) dt \nonumber \\
	= & \; \int_{-\frac{\mu}{\sigma}}^{t_1} \frac{g(t)}{h(t)} h(t) dt + \int_{t_1}^{t_2} h(t) dt + \int_{t_2}^{\infty} \frac{g(t)}{h(t)} h(t) dt \nonumber \\
	= & \;  \int_{-\frac{\mu}{\sigma}}^{t_1} g(t) dt + \int_{t_1}^{t_2} h(t) dt + \int_{t_2}^{\infty}g(t) dt.
\end{align*}
Thus, the probability of rejection is $\mathbb{P}\left(T_{rej}\right) = 1 - \mathbb{P} \left(T_{acc}\right) = \int_{t_1}^{t_2}g(t) - h(t) dt$. Since the Step 3 draws a sample from $\int_{t_1}^{t_2}g(t) - h(t) dt$, the acceptance probability of the method is exactly $1$. 

To show that the distribution of accepted values follows the target density $g(t)$, the cumulative density function $\mathbb{P}\left(T \le u \big| T_{acc} \right) = \frac{\mathbb{P}\left(T \le u , T_{acc} \right)}{\mathbb{P}\left(T_{acc} \right)} = \mathbb{P}\left(T \le u , T_{acc} \right) \nonumber$ has to be equal to $F_{g(t)}(u) = \int_{-\frac{\mu}{\sigma}}^u g(t) dt$. Three different cases are studied:
\begin{itemize}
	\item \textbf{Case} $u < t_1:$
		\begin{align*}
			\mathbb{P}\left(T \le u , T_{acc} \right) = & \; \int_{-\frac{\mu}{\sigma}}^{u} \mathbb{P}\left(U \le \frac{g(t)}{h(t)} \right) h(t) dt \nonumber \\
			= & \; \int_{-\frac{\mu}{\sigma}}^{u} g(t) dt \nonumber \\
			= & \; F_{g(t)}(u);
		\end{align*}
	\item \textbf{Case} $u \in \left[ t_1, t_2\right]:$
		\begin{align*}
			\mathbb{P}\left(T \le u, T_{acc} \right) = & \; \mathbb{P}\left(T \le t_1 , T_{acc} \right) + \mathbb{P}\left(t_1 < T \le u , T_{acc} \right) \nonumber \\
			= & \; F_{g(t)}(t_1) + \int_{t_1}^{u} \mathbb{P}\left(U \le \frac{g(t)}{h(t)} \right) h(t)  dt + \int_{t_1}^u g(t) - h(t) dt \nonumber \\
			= & \; F_{g(t)}(t_1) + \int_{t_1}^{u} h(t) dt + \int_{t_1}^u g(t) - h(t) dt \nonumber \\
			= & \; F_{g(t)}(u);
		\end{align*}
	\item \textbf{Case} $t_2 < u:$
		\begin{align*}
			\mathbb{P}\left(T \le u , T_{acc} \right) = & \; \mathbb{P}\left(T \le t_2 , T_{acc} \right) + \mathbb{P}\left(t_2 < T \le u , T_{acc} \right) \nonumber \\
			= & \; F_{g(t)}(t_2) + \int_{t_2}^{u} \mathbb{P}\left(U \le \frac{g(t)}{h(t)} \right) h(t) dt \nonumber \\
			= & \; F_{g(t)}(t_2) + \int_{t_2}^{u} g(t) dt  \nonumber \\
			= & \; F_{g(t)}(u).
		\end{align*}
\end{itemize}
Therefore, the method actually samples from the target distribution.


\section{KL divergence for the $\mathcal{G}_{3p}$ distribution}
\label{app:a4}
Here the asymptotic behaviour of a $\mathcal{G}_{3p}$ distribution for limit cases $\beta/\alpha \to -\infty$, $\beta/\alpha \to \infty$ and $\gamma \to \infty$ is described. The analysis relies on the KL divergence. In the first case, $\beta/\alpha \to -\infty$ the $\mathcal{G}_{3p}$ distribution is compared to a Gamma distribution and yields a closed-form result, whereas when $\beta/\alpha \to \infty$ and $\gamma \to \infty$ the target density is approximated with a Gaussian distribution based on empirical results.
\begin{itemize}
	\item \textbf{Proof of Proposition \ref{prep_KL}}: \\
	The KL divergence between distribution $q_x \sim \mathcal{G}_{3p} \left(\gamma, \alpha, \beta\right)$ and distribution $p_x \sim \mathsf{Ga} \left(d, c\right)$ is
	\begin{equation}
		\mathrm{KL}\left(p \Vert q\right) =  \int_0^\infty p_x \log \left(\frac{p_x}{q_x}\right) dx = \; \int_0^\infty p_x\log\left(p_x\right) dx - \int_0^\infty p_x\log\left(q_x\right) dx. \label{eq:KL}
	\end{equation}
	Denoting the two integrals in \eqref{eq:KL} as $ I\left(d, c\right)  = \int_0^\infty p_x\log\left(p_x\right) dx$ and $I\left(d, c, \gamma, \alpha, \beta\right) = \int_0^\infty p_x\log\left(q_x\right) dx$, it yields
	\begin{align*}
		I\left(d, c\right) = &\; \int_0^{\infty} \log\left(\frac{c^d}{\Gamma(d)} e^{-cx}x^{d-1}\right) \frac{c^d}{\Gamma(d)} e^{-cx}x^{d-1}dx \nonumber\\
		= &\; \log\left(\frac{c^d}{\Gamma\left(d\right)}\right) - \frac{c^{d+1}}{\Gamma\left(d\right)} \int_0^\infty e^{-cx} x^d dx +  \frac{c^{d} (d-1)}{\Gamma\left(d\right)} \int_0^\infty \log(x) e^{-cx} x^{d-1} dx\nonumber \\
		= &\; \log\left(\frac{c^d}{\Gamma\left(d\right)}\right) - \frac{c^{d+1}}{\Gamma\left(d\right)} \frac{\Gamma\left(d+1\right)}{c^{d+1}} +  \frac{c^{d} (d-1)}{\Gamma\left(d\right)} \frac{\Gamma(d)}{c^d} \left(\frac{\Gamma'\left(d\right)}{\Gamma(d)} - \log{c}\right) \nonumber\\
		= &\; \log\left(\frac{c^d}{\Gamma\left(d\right)}\right) - d +  (d-1) \left(\frac{\Gamma'\left(d\right)}{\Gamma(d)} - \log{c}\right) \nonumber \\
		= &\; \log(c) + (d-1) \frac{\Gamma'\left(d\right)}{\Gamma(d)} - d - \log\left(\Gamma(d)\right)
	\end{align*}
	and 
	\begin{align*}
		I\left(d, c, \gamma, \alpha, \beta\right) = &\; \int_0^{\infty} \log\left( \frac{\left(2 \alpha^2\right)^{\frac{\gamma+1}{2}} e^{-\frac{\beta^2}{8 \alpha^2}}  }{\gamma! D_{-\gamma-1}\left(-\frac{\beta}{\alpha \sqrt{2}}\right)}   e^{-\alpha^2x^2 + \beta x}x^\gamma\right) \frac{c^d}{\Gamma(d)} e^{-cx}x^{d-1}dx \nonumber \\
		= &\; \log\left(\frac{\left(2 \alpha^2\right)^{\frac{\gamma+1}{2}} e^{-\frac{\beta^2}{8 \alpha^2}}  }{\gamma! D_{-\gamma-1}\left(-\frac{\beta}{\alpha \sqrt{2}}\right)} \right) + \frac{c^d}{\Gamma\left(d\right)} \int_0^\infty \left(-\alpha^2x^2 + \beta x\right) e^{-cx} x^{d-1} dx \nonumber\\
		& \quad \quad +  \frac{c^{d} \gamma}{\Gamma\left(d\right)} \int_0^\infty \log(x) e^{-cx} x^{d-1} dx \nonumber \\
		= &\; \log\left(\frac{\left(2 \alpha^2\right)^{\frac{\gamma+1}{2}} e^{-\frac{\beta^2}{8 \alpha^2}}  }{\gamma! D_{-\gamma-1}\left(-\frac{\beta}{\alpha \sqrt{2}}\right)} \right) -  \frac{\alpha^2 c^d \Gamma\left(d+2\right)}{c^{d+2}\Gamma\left(d\right)} + \frac{\beta c^d \Gamma\left(d+1\right)}{c^{d+1}\Gamma\left(d\right)} + \gamma \left(\frac{\Gamma'\left(d\right)}{\Gamma(d)} - \log\left(c\right)\right) \nonumber\\
		= &\; \log\left(\frac{\left(2 \alpha^2\right)^{\frac{\gamma+1}{2}} e^{-\frac{\beta^2}{8 \alpha^2}}  }{\gamma! D_{-\gamma-1}\left(-\frac{\beta}{\alpha \sqrt{2}}\right)} \right) -  \frac{\alpha^2 d (d+1)}{c^{2}} + \frac{\beta d}{c} + \gamma \left(\frac{\Gamma'\left(d\right)}{\Gamma(d)} - \log\left(c\right)\right).
	\end{align*}
	Thus, 
	\begin{align}
		I\left(d, c\right) - I\left(d, c, \gamma, \alpha, \beta\right) = &\; \log(c) + (d-1) \frac{\Gamma'\left(d\right)}{\Gamma(d)} - d - \log\left(\Gamma(d)\right)  - \log\left(\frac{\left(2 \alpha^2\right)^{\frac{\gamma+1}{2}} e^{-\frac{\beta^2}{8 \alpha^2}}  }{\gamma! D_{-\gamma-1}\left(-\frac{\beta}{\alpha \sqrt{2}}\right)} \right) \nonumber \\
		& \quad \quad +  \frac{\alpha^2 d (d+1)}{c^{2}} - \frac{\beta d}{c} - \gamma \left(\frac{\Gamma'\left(d\right)}{\Gamma(d)} - \log\left(c\right)\right) \nonumber\\
		= &\; \left(\gamma+1\right) \log\left(c\right) + \left(d-1-\gamma\right) \frac{\Gamma'\left(d\right)}{\Gamma(d)} - d\left(1 + \frac{\beta}{c} - \frac{\alpha^2 (d+1)}{c^2}\right) \nonumber\\
		& \quad \quad + \log\left(\frac{\Gamma(\gamma+1)}{\Gamma(d)}\right)  - \log\left(\frac{\left(2\alpha^2\right)^{\frac{\gamma+1}{2}} e^{-\frac{\beta^2}{8 \alpha^2}}}{D_{-\gamma-1}\left(-\frac{\beta}{\alpha \sqrt{2}}\right)}\right) \label{eq:KL2}.
	\end{align}
	Let $d = \frac{\mu^2}{\sigma^2}$ and $c = \frac{\mu}{\sigma^2}$ so that the Gamma distribution has the same mean and variance of the $\mathcal{G}_{3p}$ distribution. Exploiting the properties of the Parabolic Cylinder functions it yields
 
    \begin{align*}
	\lim_{\frac{\beta}{\alpha} \to -\infty} d = &\; \lim_{\frac{\beta}{\alpha} \to -\infty} \frac{ \left(\gamma+1\right)^2 D_{-\gamma-2}\left(-\frac{\beta}{\alpha \sqrt{2}}\right)^2}{D_{-\gamma-1}\left(-\frac{\beta}{\alpha \sqrt{2}}\right)} \left(\frac{\gamma+2}{\gamma+1} D_{-\gamma-3}\left(-\frac{\beta}{\alpha \sqrt{2}}\right) - \frac{D_{-\gamma-2}\left(-\frac{\beta}{\alpha \sqrt{2}}\right)^2}{D_{-\gamma-1}\left(-\frac{\beta}{\alpha \sqrt{2}}\right)}\right)^{-1}\nonumber\\
		= &\; \frac{1}{\lim_{\frac{\beta}{\alpha} \to -\infty} \frac{D_{-\gamma-1}\left(-\frac{\beta}{\alpha \sqrt{2}}\right)}{ \left(\gamma+1\right) D_{-\gamma-2}\left(-\frac{\beta}{\alpha \sqrt{2}}\right)^2} \left(\left(\gamma+2\right) D_{-\gamma-3}\left(-\frac{\beta}{\alpha \sqrt{2}}\right) - \left(\gamma+1\right) \frac{D_{-\gamma-2}\left(-\frac{\beta}{\alpha \sqrt{2}}\right)^2}{D_{-\gamma-1}\left(-\frac{\beta}{\alpha \sqrt{2}}\right)}\right)} \nonumber\\
		= &\; \frac{1}{\lim_{\frac{\beta}{\alpha} \to -\infty}  \frac{\left(\gamma+2\right) D_{-\gamma-3}\left(-\frac{\beta}{\alpha \sqrt{2}}\right) D_{-\gamma-1}\left(-\frac{\beta}{\alpha \sqrt{2}}\right)}{\left(\gamma+1\right) D_{-\gamma-2}\left(-\frac{\beta}{\alpha \sqrt{2}}\right)^2} - 1}\nonumber\\
		= &\; \frac{1}{\frac{\gamma+2}{\gamma+1} - 1} = \gamma+1
	\end{align*}
	and 
	\begin{align*}
		\lim_{\frac{\beta}{\alpha} \to -\infty} c = &\; \lim_{\frac{\beta}{\alpha} \to -\infty} \frac{d}{\mu} \nonumber\\
		= &\; \left(\gamma+1\right)  \lim_{\frac{\beta}{\alpha} \to -\infty} \frac{1}{\mu} \nonumber\\
		= &\; \left(\gamma+1\right)  \lim_{\frac{\beta}{\alpha} \to -\infty} \frac{\alpha \sqrt{2}}{\gamma+1} \frac{D_{-\gamma-1}\left(-\frac{\beta}{\alpha \sqrt{2}}\right)}{D_{-\gamma-2}\left(-\frac{\beta}{\alpha \sqrt{2}}\right)} \nonumber\\
		= &\;  \lim_{\frac{\beta}{\alpha} \to -\infty} \alpha\sqrt{2} \frac{D_{-\gamma-1}\left(-\frac{\beta}{\alpha \sqrt{2}}\right)}{D_{-\gamma-2}\left(-\frac{\beta}{\alpha \sqrt{2}}\right)} \quad \bigg(D_v(z) = zD_{v-1}(z) - \left(v-1\right)D_{v-2}(z)\bigg) \nonumber\\
		= &\;  \lim_{\frac{\beta}{\alpha} \to -\infty}  \alpha\sqrt{2} \left(-\frac{\beta}{\alpha \sqrt{2}} + \left(\gamma + 2\right) \frac{D_{-\gamma-3}\left(-\frac{\beta}{\alpha \sqrt{2}}\right)}{D_{-\gamma-2}\left(-\frac{\beta}{\alpha \sqrt{2}}\right)}\right) \nonumber\\
		= &\; -\beta.
	\end{align*}
	Plugging these results into \eqref{eq:KL2} yields
	\begin{align*}
		KL_{\frac{\beta}{\alpha} \to -\infty}\left(q, p\right) &= \; \left(\gamma+1\right) \log\left(-\beta\right) - \left(\gamma+1\right)\left( \frac{\alpha^2 (d+1)}{\beta^2}\right) - \log\left(\frac{\left(2\alpha^2\right)^{\frac{\gamma+1}{2}} e^{-\frac{\beta^2}{8 \alpha^2}}}{D_{-\gamma-1}\left(-\frac{\beta}{\alpha \sqrt{2}}\right)}\right) \nonumber \\
		= &\; \log\left(\left(\frac{-\beta}{\alpha \sqrt{2}}\right)^{\gamma+1} \frac{D_{-\gamma-1}\left(-\frac{\beta}{\alpha \sqrt{2}}\right)}{e^{-\frac{\beta^2}{4 \left(2\alpha^2\right)}}}\right)  - \left(\gamma+1\right)\left( \frac{\alpha^2 (d+1)}{\beta^2}\right) = 0, 
	\end{align*}
	since $\lim_{z \to \infty} \frac{ D_{-v}\left(z\right)}{z^{-v} e^{\frac{-z^2}{4}}} = 1$

	\item \textbf{Asymptotic behaviour when $\frac{\beta}{\alpha} \to +\infty$ or $\gamma \to +\infty$}:\\
	When $\frac{\beta}{\alpha} \to +\infty$ the Gamma-$3p$ is approximated with a $\mathcal{N}\left(\mu, \sigma^2\right)$ distribution, with
	\begin{align}
		\lim_{\frac{\beta}{\alpha} \to +\infty} \mu =&\; \lim_{\frac{\beta}{\alpha} \to +\infty}\frac{\gamma + 1}{\alpha \sqrt{2}} \frac{D_{-\gamma-2}\left(-\frac{\beta}{\alpha \sqrt{2}}\right)}{ D_{-\gamma-1}\left(-\frac{\beta}{\alpha \sqrt{2}}\right)} \quad \bigg( \lim_{x \to -\infty} v \frac{D_{-v-1}\left(z\right)}{ D_{-v}\left(z\right)} = - z \bigg) \nonumber\\
		= &\; \frac{\beta}{2 \alpha^2} \label{eq:e11}
	\end{align}
	and
	\begin{align}
		\lim_{\frac{\beta}{\alpha} \to +\infty} \sigma^2 = & \;\lim_{\frac{\beta}{\alpha} \to +\infty}\frac{\gamma + 1^2}{2\alpha^2} \frac{D_{-\gamma-2}\left(-\frac{\beta}{\alpha \sqrt{2}}\right)}{D_{-\gamma-1}\left(-\frac{\beta}{\alpha \sqrt{2}}\right)} \left( \frac{\gamma+2}{\gamma+1} \frac{D_{-\gamma-3}\left(-\frac{\beta}{\alpha \sqrt{2}}\right)}{D_{-\gamma-2}\left(-\frac{\beta}{\alpha \sqrt{2}}\right)} - \frac{D_{-\gamma-2}\left(-\frac{\beta}{\alpha \sqrt{2}}\right)}{D_{-\gamma-1}\left(-\frac{\beta}{\alpha \sqrt{2}}\right)}\right) \nonumber\\
		= &\;\frac{1}{2 \alpha^2}. \label{eq:e12}
	\end{align}
	Following \cite{Segura2020}, when $v \to \infty$ a sharp approximation for the ratio of Parabolic Cylinder functions is $v \frac{D_{-v-1}\left(z\right)}{ D_{-v}\left(z\right)} \approx -z + \frac{1}{2} \left(z + \sqrt{z^2 + 4v - 2}\right)$. Therefore, the mean and variance of the gaussian approximation become
	\begin{align}
		\lim_{\gamma \to +\infty} \mu = &\; \lim_{\gamma \to +\infty}\frac{\gamma + 1}{\alpha \sqrt{2}} \frac{D_{-\gamma-2}\left(-\frac{\beta}{\alpha \sqrt{2}}\right)}{ D_{-\gamma-1}\left(-\frac{\beta}{\alpha \sqrt{2}}\right)}  \nonumber\\
		= &\;\frac{\beta}{4 \alpha^2} + \frac{1}{\alpha\sqrt{8}}\sqrt{\frac{\beta^2}{2\alpha^2} + 4\gamma + 2} \label{eq:e21}
	\end{align}
	and
	\begin{align}
		\lim_{\gamma \to +\infty} \sigma^2 = &\; \lim_{\gamma \to +\infty}\frac{\gamma + 1^2}{2\alpha^2} \frac{D_{-\gamma-2}\left(-\frac{\beta}{\alpha \sqrt{2}}\right)}{D_{-\gamma-1}\left(-\frac{\beta}{\alpha \sqrt{2}}\right)} \left( \frac{\gamma+2}{\gamma+1} \frac{D_{-\gamma-3}\left(-\frac{\beta}{\alpha \sqrt{2}}\right)}{D_{-\gamma-2}\left(-\frac{\beta}{\alpha \sqrt{2}}\right)} - \frac{D_{-\gamma-2}\left(-\frac{\beta}{\alpha \sqrt{2}}\right)}{D_{-\gamma-1}\left(-\frac{\beta}{\alpha \sqrt{2}}\right)}\right) \nonumber\\
		= &\; \frac{\gamma+1}{4 \alpha^2} \left(\frac{\beta}{\alpha \sqrt{2}} + \sqrt{\frac{\beta^2}{2\alpha^2} + 4\gamma + 6}\right) - \lim_{\gamma \to +\infty} \mu^2 \nonumber \\
		= &\; \frac{\gamma+1}{4 \alpha^2} \left(\frac{\beta}{\alpha \sqrt{2}} + \sqrt{\frac{\beta^2}{2\alpha^2} + 4\gamma + 6}\right) - \left(\frac{\beta}{4 \alpha^2} + \frac{1}{\alpha\sqrt{8}}\sqrt{\frac{\beta^2}{2\alpha^2} + 4\gamma + 2}\right)^2. \label{eq:e22}
	\end{align}
	Tables \ref{tab:ba1} and \ref{tab:c1} show the KL divergence for increasing values of the ratio $\beta /\alpha$ and $\gamma$. The integral is numerically approximated with the command \textsf{KLD} from package \textsf{LaplacesDemon} for software \textsf{R}. The approximated KL divergence is evaluated over the interval $\left(\mu - 5\sigma, \mu + 5\sigma\right)$. Values of the parameters higher than those shown in the table \ref{tab:ba1} give overflow problems. The results in the tables below depend only on the values of $\gamma$ and the ratio $\beta/\alpha$, that is, for different values of $\alpha$ the KL divergence between $q \sim \mathcal{G}_{3p}\left(\gamma, \alpha, \beta\right)$ and $p \sim \mathcal{N}\left(\mu, \sigma^2\right)$ does not change. The sequence of KL divergence is always decreasing in Table \ref{tab:c1}, for both $\mathrm{KL}(q\Vert p)$ and $\mathrm{KL}(p\Vert q)$. In Table \ref{tab:ba1} the sequence is decreasing only for $\mathrm{KL}(p\Vert q)$, however the mean between the two is decreasing.

% ...

\begin{table}[H]
\begin{adjustbox}{width = \textwidth}
\begin{tabular}{|c|ccccccccccccccc|}
\hline
$\mathrm{KL}$ & \multicolumn{1}{c|}{$\frac{\beta}{\alpha} =0.002$} & \multicolumn{1}{c|}{$\frac{\beta}{\alpha} =0.2$} & \multicolumn{1}{c|}{$\frac{\beta}{\alpha} =0.5$} & \multicolumn{1}{c|}{$\frac{\beta}{\alpha} =1$} & \multicolumn{1}{c|}{$\frac{\beta}{\alpha} =3$} & \multicolumn{1}{c|}{$\frac{\beta}{\alpha} =5$} & \multicolumn{1}{c|}{$\frac{\beta}{\alpha} =8$} & \multicolumn{1}{c|}{} & \multicolumn{1}{c|}{$\frac{\beta}{\alpha} =0.002$} & \multicolumn{1}{c|}{$\frac{\beta}{\alpha} =0.2$} & \multicolumn{1}{c|}{$\frac{\beta}{\alpha} =0.5$} & \multicolumn{1}{c|}{$\frac{\beta}{\alpha} =1$} & \multicolumn{1}{c|}{$\frac{\beta}{\alpha} =3$} & \multicolumn{1}{c|}{$\frac{\beta}{\alpha} =5$} & $\frac{\beta}{\alpha} =8$ \\ \hline \hline
$\gamma = 1$   & \multicolumn{1}{c|}{0.284}                         & \multicolumn{1}{c|}{0.273}                       & \multicolumn{1}{c|}{0.257}                       & \multicolumn{1}{c|}{0.227}                     & \multicolumn{1}{c|}{0.105}                     & \multicolumn{1}{c|}{0.041}                     & \multicolumn{1}{c|}{0.016}                     & \multicolumn{1}{c|}{} & \multicolumn{1}{c|}{0.411}                         & \multicolumn{1}{c|}{0.394}                       & \multicolumn{1}{c|}{0.372}                       & \multicolumn{1}{c|}{0.329}                     & \multicolumn{1}{c|}{0.139}                     & \multicolumn{1}{c|}{0.047}                     & 0.016                     \\ \cline{1-8} \cline{10-16} 
$\gamma = 3$   & \multicolumn{1}{c|}{1.206}                         & \multicolumn{1}{c|}{1.164}                       & \multicolumn{1}{c|}{1.100}                       & \multicolumn{1}{c|}{0.983}                     & \multicolumn{1}{c|}{0.545}                     & \multicolumn{1}{c|}{0.281}                     & \multicolumn{1}{c|}{0.127}                     & \multicolumn{1}{c|}{} & \multicolumn{1}{c|}{2.365}                         & \multicolumn{1}{c|}{2.288}                       & \multicolumn{1}{c|}{2.153}                       & \multicolumn{1}{c|}{1.906}                     & \multicolumn{1}{c|}{0.886}                     & \multicolumn{1}{c|}{0.355}                     & 0.139                     \\ \cline{1-8} \cline{10-16} 
$\gamma = 5$   & \multicolumn{1}{c|}{2.185}                         & \multicolumn{1}{c|}{2.115}                       & \multicolumn{1}{c|}{2.007}                       & \multicolumn{1}{c|}{1.820}                     & \multicolumn{1}{c|}{1.104}                     & \multicolumn{1}{c|}{0.636}                     & \multicolumn{1}{c|}{0.318}                     & \multicolumn{1}{c|}{} & \multicolumn{1}{c|}{5.032}                         & \multicolumn{1}{c|}{4.858}                       & \multicolumn{1}{c|}{4.577}                       & \multicolumn{1}{c|}{4.064}                     & \multicolumn{1}{c|}{1.992}                     & \multicolumn{1}{c|}{0.871}                     & 0.366                     \\ \cline{1-8} \cline{10-16} 
$\gamma = 10$  & \multicolumn{1}{c|}{4.633}                         & \multicolumn{1}{c|}{4.505}                       & \multicolumn{1}{c|}{4.319}                       & \multicolumn{1}{c|}{3.995}                     & \multicolumn{1}{c|}{2.736}                     & \multicolumn{1}{c|}{1.810}                     & \multicolumn{1}{c|}{1.038}                     & \multicolumn{1}{c|}{} & \multicolumn{1}{c|}{13.207}                        & \multicolumn{1}{c|}{12.743}                      & \multicolumn{1}{c|}{12.057}                      & \multicolumn{1}{c|}{10.804}                    & \multicolumn{1}{c|}{5.698}                     & \multicolumn{1}{c|}{2.811}                     & 1.303                     \\ \cline{1-8} \cline{10-16} 
$\gamma = 15$  & \multicolumn{1}{c|}{6.875}                         & \multicolumn{1}{c|}{6.740}                       & \multicolumn{1}{c|}{6.516}                       & \multicolumn{1}{c|}{6.124}                     & \multicolumn{1}{c|}{4.501}                     & \multicolumn{1}{c|}{3.206}                     & \multicolumn{1}{c|}{1.990}                     & \multicolumn{1}{c|}{} & \multicolumn{1}{c|}{22.597}                        & \multicolumn{1}{c|}{21.884}                      & \multicolumn{1}{c|}{20.721}                      & \multicolumn{1}{c|}{18.653}                    & \multicolumn{1}{c|}{10.295}                    & \multicolumn{1}{c|}{5.419}                     & 2.672                     \\ \cline{1-8} \cline{10-16} 
$\gamma = 30$  & \multicolumn{1}{c|}{10.882}                        & \multicolumn{1}{c|}{10.839}                      & \multicolumn{1}{c|}{10.751}                      & \multicolumn{1}{c|}{10.537}                    & \multicolumn{1}{c|}{9.160}                     & \multicolumn{1}{c|}{7.568}                     & \multicolumn{1}{c|}{5.517}                     & \multicolumn{1}{c|}{} & \multicolumn{1}{c|}{53.998}                        & \multicolumn{1}{c|}{52.316}                      & \multicolumn{1}{c|}{49.941}                      & \multicolumn{1}{c|}{45.377}                    & \multicolumn{1}{c|}{27.042}                    & \multicolumn{1}{c|}{15.766}                    & 8.645                     \\ \cline{1-8} \cline{10-16} 
$\gamma = 50$  & \multicolumn{1}{c|}{12.710}                        & \multicolumn{1}{c|}{12.739}                      & \multicolumn{1}{c|}{12.749}                      & \multicolumn{1}{c|}{12.708}                    & \multicolumn{1}{c|}{12.090}                    & \multicolumn{1}{c|}{11.137}                    & \multicolumn{1}{c|}{9.542}                     & \multicolumn{1}{c|}{} & \multicolumn{1}{c|}{97.765}                        & \multicolumn{1}{c|}{95.186}                      & \multicolumn{1}{c|}{90.974}                      & \multicolumn{1}{c|}{83.411}                    & \multicolumn{1}{c|}{52.121}                    & \multicolumn{1}{c|}{32.489}                    & 19.310                    \\ \cline{1-8} \cline{10-16} 
$\gamma = 100$ & \multicolumn{1}{c|}{14.162}                        & \multicolumn{1}{c|}{14.225}                      & \multicolumn{1}{c|}{14.278}                      & \multicolumn{1}{c|}{14.349}                    & \multicolumn{1}{c|}{14.163}                    & \multicolumn{1}{c|}{13.755}                    & \multicolumn{1}{c|}{13.110}                    & \multicolumn{1}{c|}{} & \multicolumn{1}{c|}{208.476}                       & \multicolumn{1}{c|}{203.439}                     & \multicolumn{1}{c|}{195.323}                     & \multicolumn{1}{c|}{180.003}                   & \multicolumn{1}{c|}{117.316}                   & \multicolumn{1}{c|}{77.428}                    & 49.973                    \\ \hline
\end{tabular}
\end{adjustbox}
\caption{KL divergence when $\beta / \alpha$ increases: $\mathrm{KL}(q\Vert p)$ (left) and $\mathrm{KL}(p\Vert q)$ (right) where $q \sim \mathcal{G}_{3p}\left(\gamma, \alpha, \beta\right)$ and $p \sim \mathcal{N}\left(\mu, \sigma^2\right)$, with $\mu$ and $\sigma^2$ computed as in \eqref{eq:e11}-\eqref{eq:e12}.}
\label{tab:ba1}
\end{table}

\begin{table}[H]
\begin{adjustbox}{width = \textwidth}
\begin{tabular}{|c|ccccccccccccccc|}
\hline
$\mathrm{KL}$  & \multicolumn{1}{c|}{$\frac{\beta}{\alpha} =0.002$} & \multicolumn{1}{c|}{$\frac{\beta}{\alpha} =0.2$} & \multicolumn{1}{c|}{$\frac{\beta}{\alpha} =0.5$} & \multicolumn{1}{c|}{$\frac{\beta}{\alpha} =1$} & \multicolumn{1}{c|}{$\frac{\beta}{\alpha} =3$} & \multicolumn{1}{c|}{$\frac{\beta}{\alpha} =5$} & \multicolumn{1}{c|}{$\frac{\beta}{\alpha} =8$} & \multicolumn{1}{c|}{} & \multicolumn{1}{c|}{$\frac{\beta}{\alpha} =0.002$} & \multicolumn{1}{c|}{$\frac{\beta}{\alpha} =0.2$} & \multicolumn{1}{c|}{$\frac{\beta}{\alpha} =0.5$} & \multicolumn{1}{c|}{$\frac{\beta}{\alpha} =1$} & \multicolumn{1}{c|}{$\frac{\beta}{\alpha} =3$} & \multicolumn{1}{c|}{$\frac{\beta}{\alpha} =5$} & $\frac{\beta}{\alpha} =8$ \\ \hline \hline
$\gamma = 1$   & \multicolumn{1}{c|}{0.022}                          & \multicolumn{1}{c|}{0.021}                       & \multicolumn{1}{c|}{0.018}                       & \multicolumn{1}{c|}{0.016}                     & \multicolumn{1}{c|}{0.011}                     & \multicolumn{1}{c|}{0.007}                     & \multicolumn{1}{c|}{0.004}                     & \multicolumn{1}{c|}{} & \multicolumn{1}{c|}{0.023}                         & \multicolumn{1}{c|}{0.022}                       & \multicolumn{1}{c|}{0.020}                       & \multicolumn{1}{c|}{0.017}                     & \multicolumn{1}{c|}{0.010}                     & \multicolumn{1}{c|}{0.007}                     & 0.003                     \\ \cline{1-8} \cline{10-16} 
$\gamma = 3$   & \multicolumn{1}{c|}{0.015}                         & \multicolumn{1}{c|}{0.013}                       & \multicolumn{1}{c|}{0.012}                       & \multicolumn{1}{c|}{0.010}                     & \multicolumn{1}{c|}{0.005}                     & \multicolumn{1}{c|}{0.004}                     & \multicolumn{1}{c|}{0.002}                     & \multicolumn{1}{c|}{} & \multicolumn{1}{c|}{0.025}                         & \multicolumn{1}{c|}{0.023}                       & \multicolumn{1}{c|}{0.020}                       & \multicolumn{1}{c|}{0.015}                     & \multicolumn{1}{c|}{0.005}                     & \multicolumn{1}{c|}{0.004}                     & 0.002                     \\ \cline{1-8} \cline{10-16} 
$\gamma = 5$   & \multicolumn{1}{c|}{0.010}                         & \multicolumn{1}{c|}{0.009}                       & \multicolumn{1}{c|}{0.008}                       & \multicolumn{1}{c|}{0.006}                     & \multicolumn{1}{c|}{0.003}                     & \multicolumn{1}{c|}{0.002}                     & \multicolumn{1}{c|}{0.002}                     & \multicolumn{1}{c|}{} & \multicolumn{1}{c|}{0.016}                         & \multicolumn{1}{c|}{0.015}                       & \multicolumn{1}{c|}{0.013}                       & \multicolumn{1}{c|}{0.010}                     & \multicolumn{1}{c|}{0004}                      & \multicolumn{1}{c|}{0.002}                     & 0.002                     \\ \cline{1-8} \cline{10-16} 
$\gamma = 10$  & \multicolumn{1}{c|}{0.005}                         & \multicolumn{1}{c|}{0.004}                       & \multicolumn{1}{c|}{0.004}                       & \multicolumn{1}{c|}{0.003}                     & \multicolumn{1}{c|}{0.002}                     & \multicolumn{1}{c|}{0.001}                     & \multicolumn{1}{c|}{0.001}                     & \multicolumn{1}{c|}{} & \multicolumn{1}{c|}{0.006}                         & \multicolumn{1}{c|}{0.006}                       & \multicolumn{1}{c|}{0.005}                       & \multicolumn{1}{c|}{0.004}                     & \multicolumn{1}{c|}{0.002}                     & \multicolumn{1}{c|}{0.001}                     & 0.001                     \\ \cline{1-8} \cline{10-16} 
$\gamma = 15$  & \multicolumn{1}{c|}{0.003}                         & \multicolumn{1}{c|}{0.003}                       & \multicolumn{1}{c|}{0.003}                       & \multicolumn{1}{c|}{0.002}                     & \multicolumn{1}{c|}{0.001}                     & \multicolumn{1}{c|}{0.001}                     & \multicolumn{1}{c|}{$< 0.001$}                 & \multicolumn{1}{c|}{} & \multicolumn{1}{c|}{0.003}                         & \multicolumn{1}{c|}{0.003}                       & \multicolumn{1}{c|}{0.003}                       & \multicolumn{1}{c|}{0.003}                     & \multicolumn{1}{c|}{0.001}                     & \multicolumn{1}{c|}{0.001}                     & $< 0.001$                 \\ \cline{1-8} \cline{10-16} 
$\gamma = 30$  & \multicolumn{1}{c|}{0.001}                         & \multicolumn{1}{c|}{0.001}                       & \multicolumn{1}{c|}{0.001}                       & \multicolumn{1}{c|}{0.001}                     & \multicolumn{1}{c|}{$< 0.001$}                 & \multicolumn{1}{c|}{$< 0.001$}                 & \multicolumn{1}{c|}{$< 0.001$}                 & \multicolumn{1}{c|}{} & \multicolumn{1}{c|}{0.002}                         & \multicolumn{1}{c|}{0.001}                       & \multicolumn{1}{c|}{0.001}                       & \multicolumn{1}{c|}{0.001}                     & \multicolumn{1}{c|}{$< 0.001$}                 & \multicolumn{1}{c|}{$< 0.001$}                 & $< 0.001$                 \\ \cline{1-8} \cline{10-16} 
$\gamma = 50$  & \multicolumn{1}{c|}{$< 0.001$}                     & \multicolumn{1}{c|}{$< 0.001$}                   & \multicolumn{1}{c|}{$< 0.001$}                   & \multicolumn{1}{c|}{$< 0.001$}                 & \multicolumn{1}{c|}{$< 0.001$}                 & \multicolumn{1}{c|}{$< 0.001$}                 & \multicolumn{1}{c|}{$< 0.001$}                 & \multicolumn{1}{c|}{} & \multicolumn{1}{c|}{$< 0.001$}                     & \multicolumn{1}{c|}{$< 0.001$}                   & \multicolumn{1}{c|}{$< 0.001$}                   & \multicolumn{1}{c|}{$< 0.001$}                 & \multicolumn{1}{c|}{$< 0.001$}                 & \multicolumn{1}{c|}{$< 0.001$}                 & $< 0.001$                 \\ \cline{1-8} \cline{10-16} 
$\gamma = 100$ & \multicolumn{1}{c|}{$< 0.001$}                     & \multicolumn{1}{c|}{$< 0.001$}                   & \multicolumn{1}{c|}{$< 0.001$}                   & \multicolumn{1}{c|}{$< 0.001$}                 & \multicolumn{1}{c|}{$< 0.001$}                 & \multicolumn{1}{c|}{$< 0.001$}                 & \multicolumn{1}{c|}{$< 0.001$}                 & \multicolumn{1}{c|}{} & \multicolumn{1}{c|}{$< 0.001$}                     & \multicolumn{1}{c|}{$< 0.001$}                   & \multicolumn{1}{c|}{$< 0.001$}                   & \multicolumn{1}{c|}{$< 0.001$}                 & \multicolumn{1}{c|}{$< 0.001$}                 & \multicolumn{1}{c|}{$< 0.001$}                 & $< 0.001$                 \\ \hline
\end{tabular}
\end{adjustbox}
\caption{KL divergence when $\gamma$ increases: $\mathrm{KL}(q\Vert p)$ (left) and $\mathrm{KL}(p\Vert q)$ (right) where $q \sim \mathcal{G}_{3p}\left(\gamma, \alpha, \beta\right)$ and $p \sim \mathcal{N}\left(\mu, \sigma^2\right)$, with $\mu$ and $\sigma^2$ computed as in \eqref{eq:e21}-\eqref{eq:e22}.}
\label{tab:c1}
\end{table}
\end{itemize}


%

\end{document}


\def\spacingset#1{\renewcommand{\baselinestretch}%
{#1}\small\normalsize} \spacingset{1}

\if0\blind
{
  \title{\bf Supplementary material: \\ Inference of multiple high-dimensional networks with the Graphical Horseshoe prior}
  \author{Claudio Busatto \\
    Department of Statistics, Computer Science, Applications "G. Parenti", \\ University of Florence, Florence, Italy\\
    and \\
    Francesco Claudio Stingo \\
    Department of Statistics, Computer Science, Applications "G. Parenti", \\ University of Florence, Florence, Italy}
  \maketitle
} \fi

\if1\blind
{
  \bigskip
  \bigskip
  \bigskip
  \begin{center}
    {\LARGE\bf Multiple Graphical Horseshoe estimator for modeling correlated precision matrices}
\end{center}
  \medskip
} \fi

\spacingset{1.45}

\noindent This supplementary provides the pseudo-code for mGHS algorithm and additional graphs and tables for the analysis of the bike-sharing dataset in Section 7. In particular, Appendix \ref{app:a5} shows the estimated networks for each group across the years for both GHS and mGHS and the analysis of the convergence of mGHS algorithm.

\newpage
\appendix

\section{Pseudo-code for mGHS algorithm}
\label{app:algapp}

\begin{algorithm}
\scriptsize
	\label{alg:alg1}
	\caption{multiple Graphical Horseshoe algorithm} 
 	{\textbf{Input}: $\mathbf{S}_1, \dots, \mathbf{S}_K \in \mathbb{R}^{p \times p}$, $K, p, B, bn \in \mathbb{N}$, $\mathbf{n} \in \mathbb{N}^K$}\;
 	
  	set $\boldsymbol{\Omega}_k = \mathbf{I}_p$, $\boldsymbol{\Sigma}_k = \mathbf{I}_p$, $\boldsymbol{\Lambda}_k = \mathbf{1}_{p\times p}$, $\boldsymbol{\eta}_k = \mathbf{1}_{p\times p}$, $\boldsymbol{\tau} = \mathbf{1}_K$, $\boldsymbol{\zeta} = \mathbf{1}_K$, $\mathbf{R} = \mathbf{I}_K$ and $\boldsymbol{\mu} = \mathbf{0}_K$\;
 	
 	\For(){$b = 1 \text{ to } B$} 
 	{
 		\For(){$k = 1 \text{ to } K$} 
 		{
 			\For(){$j = 1 \text{ to } p$} 
 			{
 				compute $\mathbf{t}_{j, k} := \left\{t_{j, k}^i = \mathbf{r}_{k}^\intercal \mathbf{R}_{-k}^{-1} \boldsymbol{\Delta}_{ij, -k}^{-1} \boldsymbol{\omega}_{ij}^{ -k}, \; i = 1, \dots, p, i \ne j\right\}$\;
 				 				
 				compute $\mathbf{m}_{j, k} := \left\{m_{j, k}^i = t_{j, k}^i \sqrt{\tau_k \lambda_{ij}^k} , \; i = 1, \dots, p, i \ne j\right\}$\;
 				compute $\mathbf{D}_{j, k} := \left\{D_{j, k}^{ii} = \mu_k  \tau_k \lambda_{ij}^k , \; i = 1, \dots, p, i \ne j\right\}$\;
 				compute $\mathbf{O}_{j,k} = \boldsymbol{\Sigma}^k_{-j} - \boldsymbol{\sigma}^k_{j}\left(\boldsymbol{\sigma}^k_{j} / \sigma_{jj}^k\right)^\intercal$ and $\mathbf{W}_{j,k} = \mathbf{D}_{j, k} + s_{jj}^k \mathbf{O}_{j,k}$\;	
 				sample $\gamma_{jj}^k \sim \mathcal{IG}\left(n_k / 2 + 1, s_{jj}^k/ 2\right)$\;
                    sample $\mathbf{v}_{j,k} \sim \mathcal{N}_{p-1} \left(\mathbf{W}_{j,k}^{-1}\left(\mathbf{D}_{j, k}^{-1} \mathbf{m}_{j, k} - \mathbf{s}_{-j}^k\right), \mathbf{W}_{j,k}^{-1}\right)$\; 
 				 				
 				sample $\mathbf{e}_{j, k} := \left\{e_{j, k}^i \sim \mathcal{IG}\left(1, 1 + 1/l_{j,k}^i\right), \; i = 1, \dots, p, i \ne j\right\}$\;
 				sample $\mathbf{l}_{j, k} := \left\{l_{j, k}^i \sim \mathcal{G}_{3p}\left(1, \alpha_{\lambda_{ij, k}}, \beta_{\lambda_{ij, k}}\right)^{-2}, \; i = 1, \dots, p, i \ne j \right\}$, where $\alpha_{\lambda_{ij,k}} = \left(\frac{v_i^2}{2 \tau_k \mu_k} + \frac{1}{\eta_{ij}^k}\right)^{1/2}$ and $\beta_{\lambda_{ij, k}} = \frac{v_i}{ \sqrt{\tau_k} \mu_k} t_{j, k}^i $\; 
 				
 				set $\boldsymbol{\Sigma}_{-j}^k = \mathbf{O}_{j,k} + \mathbf{O}_{j,k} \mathbf{v}_{j,k}\mathbf{v}_{j,k}^\intercal \mathbf{O}_{j,k} / \gamma_{jj}^k$, $\boldsymbol{\sigma}_j^k =  -\mathbf{O}_{j,k}\mathbf{v}_{j,k} / \gamma_{jj}^k$, $\sigma_{jj}^k = 1 / \gamma_{jj}^k$, $\boldsymbol{\omega}_{j}^k = \mathbf{v}_{j,k}$, $\omega_{jj}^k = \gamma_{jj}^k + \mathbf{v}_{j,k}^\intercal\mathbf{O}_{j,k}\mathbf{v}_{j,k}$, $\boldsymbol{\lambda}_j^k = \mathbf{l}_{j,k}$ and $\boldsymbol{\eta}_{j}^k = \mathbf{e}_{j,k}$\;
 			}
 		}
 		 		
 		compute $\mathbf{T}_k := \left\{T_k^{ij} = \mathbf{r}_{k}^\intercal \mathbf{R}_{-k}^{-1} \boldsymbol{\Delta}_{ij, -k}^{-1} \boldsymbol{\omega}_{ij}^{ -k}, \; k = 1, \dots, K, j = 2, \dots, p, i < j\right\}$\;
 		sample $\boldsymbol{\tau} := \left\{\tau_k \sim \mathcal{G}_{3p}\left(p(p-1)/2, \alpha_{\tau_k}, \beta_{\tau_k}\right)^{-2},\;  k = 1, \dots, K\right\}$, where $\alpha_{\tau_k} = \left(\frac{1}{\zeta_k} + \sum_{i<j}\frac{\left(\omega_{ij}^k\right)^2}{ 2 \lambda^k_{ij}  \mu_k}\right)^{1/2}$ and $\beta_{\tau_k} = \sum_{i<j} \frac{\omega_{ij}^k}{ \sqrt{\lambda_{ij}^k} \mu_k}  \mathbf{T}_k^{ij}$ \;
 		sample $\boldsymbol{\zeta} := \left\{\zeta_k \sim \mathcal{IG}\left(1, 1 + 1/\tau_k\right), \; k = 1, \dots, K \right\}$\; 

 		sample $\mathbf{R}_\star = \text{diag}\left(\boldsymbol{\Psi}\right)^{-1/2} \boldsymbol{\Psi} \text{diag}\left(\boldsymbol{\Psi}\right)^{-1/2}$, where $\boldsymbol{\Psi}  \sim \mathcal{IW}\left(p(p-1)/2, \mathbf{H}\right)$ and $\mathbf{H} := \left\{H_{ij} = \sum_{k = 1}^K \left(\sum_{i<j}\omega_{ij}^k\right)^{-1/2}\frac{\omega_{ij}^k}{\sqrt{\tau_k\lambda_{ij}^k}}\right\} $\;
 		\lIf(){$\mathcal{U}(0, 1) < e^{(K+1) / 2 \left(\log\vert\mathbf{R}_\star\vert - \log\vert\mathbf{R}\vert\right)}$}
 		{
 			set $\mathbf{R} = \mathbf{R}_\star$ and compute $\boldsymbol{\mu} := \left\{\mu_k = 1 - \mathbf{r}_{k}^\intercal \mathbf{R}_{-k}^{-1}\mathbf{r}_{k}, \; k = 1, \dots, K \right\}$
		}
 	}
 	\textbf{return} $\boldsymbol{\Omega}_1 \dots, \boldsymbol{\Omega}_K , \boldsymbol{\Lambda}_1, \dots, \boldsymbol{\Lambda}_K ,\boldsymbol{\tau}$ and $\mathbf{R}$\;
\end{algorithm}
\newpage

\section{Bike-sharing dataset: convergence analysis and estimated graph for each group}
\label{app:a5}

\begin{figure}[h!]
	\centering
	\includegraphics[scale = 0.45]{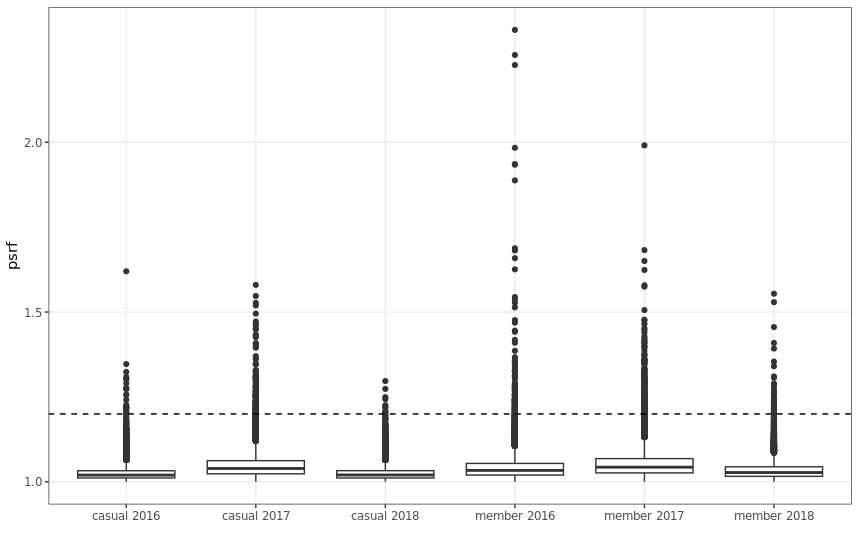}
	\caption{Distribution of the estimated potential scale reduction factors computed over a post-burning period of $5000$ updates for parameters $\omega_{ij}^k$ across $4$ replications. Optimal values of the index should lie under the dotted line ($y = 1.2$).}
	\label{fig:psrf}
\end{figure}

\begin{figure}
\centering
\begin{subfigure}{.5\textwidth}
  \centering
  \includegraphics[width = \textwidth]{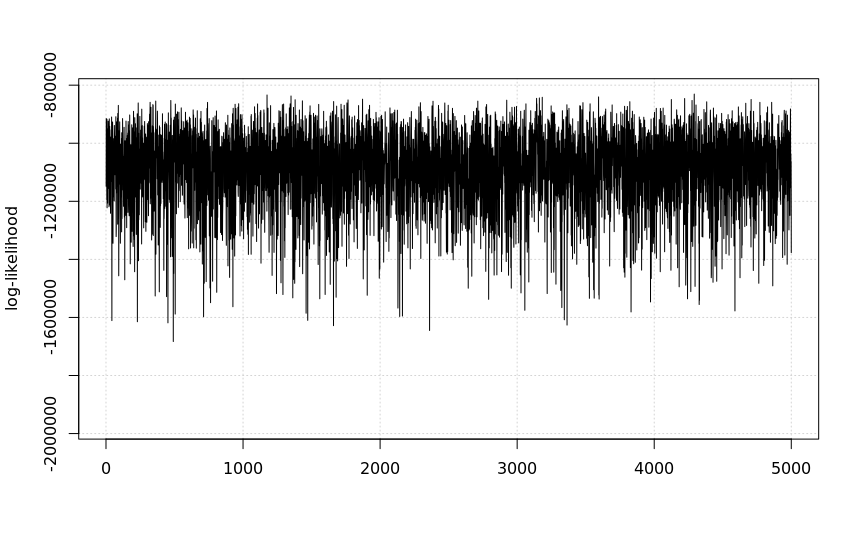}
\end{subfigure}%
\begin{subfigure}{.5\textwidth}
  \centering
  \includegraphics[width = \textwidth]{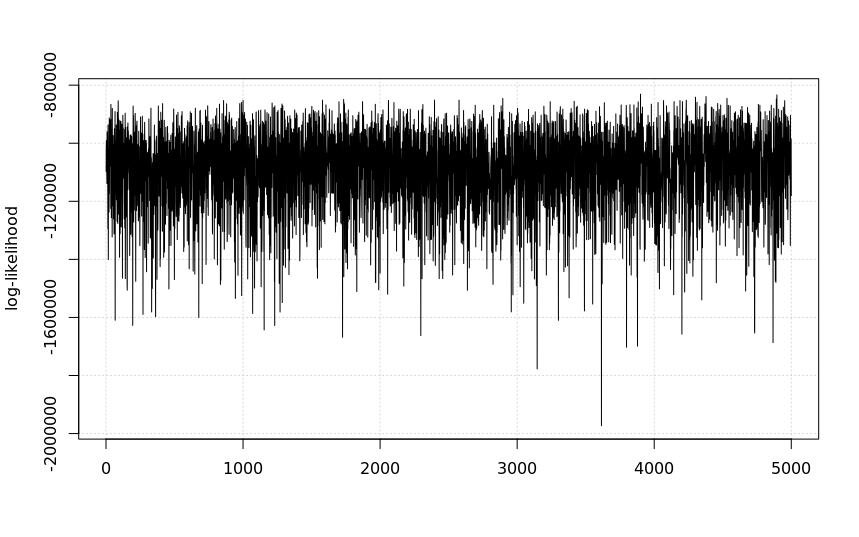}
\end{subfigure}
\begin{subfigure}{.5\textwidth}
  \centering
  \includegraphics[width = \textwidth]{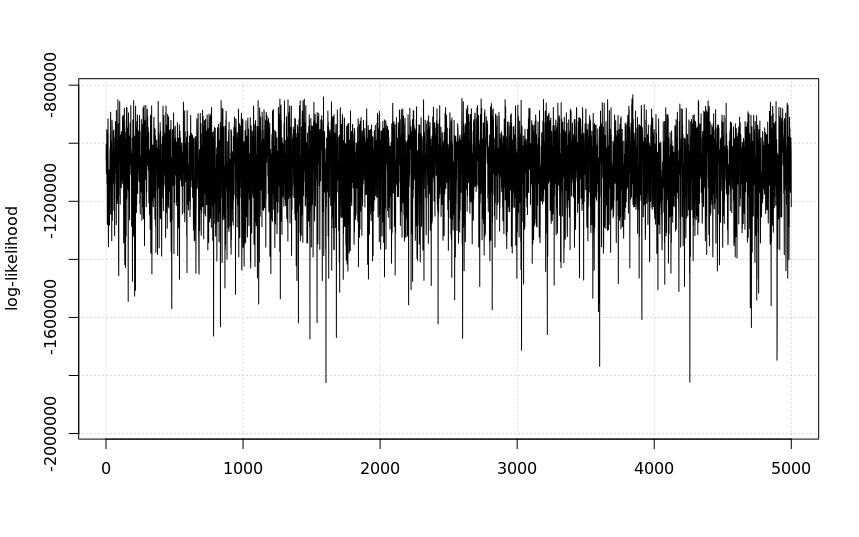}
\end{subfigure}%
\begin{subfigure}{.5\textwidth}
  \centering
  \includegraphics[width = \textwidth]{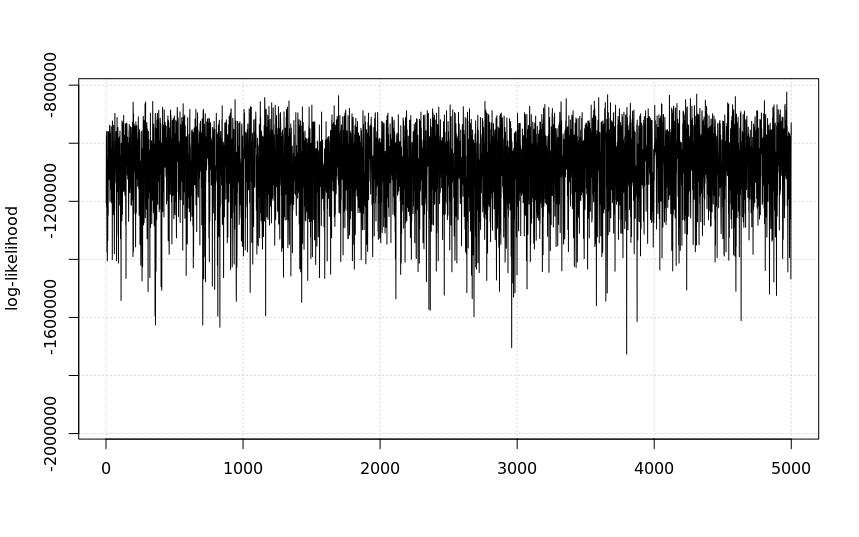}
\end{subfigure}	\caption{Post-burnin race of the joint posterior distribution of the $4$ chains.}
	\label{fig:logl}
\end{figure}

\begin{sidewaysfigure}
    \centering
    \includegraphics[scale = 0.95]{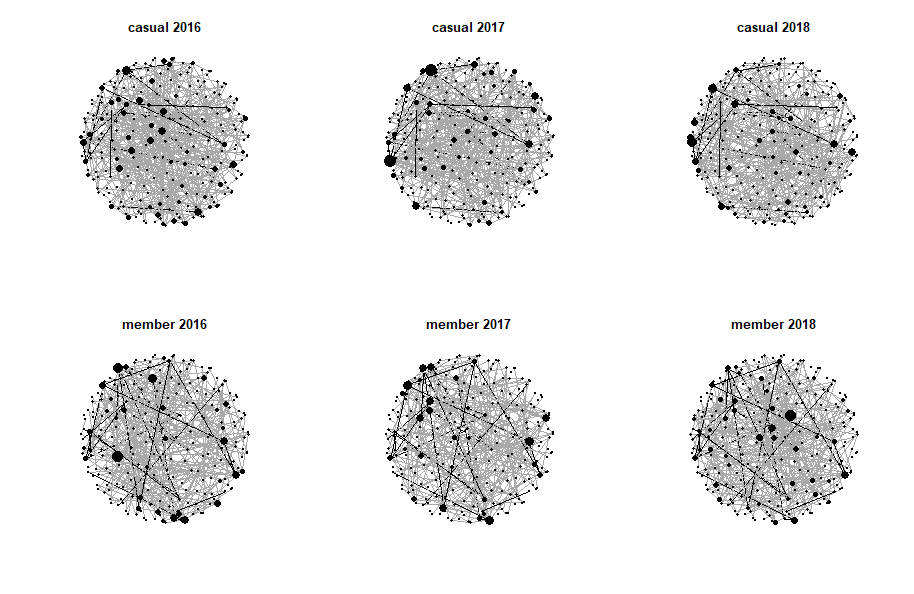}
    \caption{Estimated graph by mGHS for each group;  Black edges denote those edges included in all three years for both member and casual users and the size of the nodes depends on the number of edges associated to the related station.}
    \label{fig:mGHS_res}
\end{sidewaysfigure}

\begin{sidewaysfigure}
    \centering
    \includegraphics[scale = 0.95]{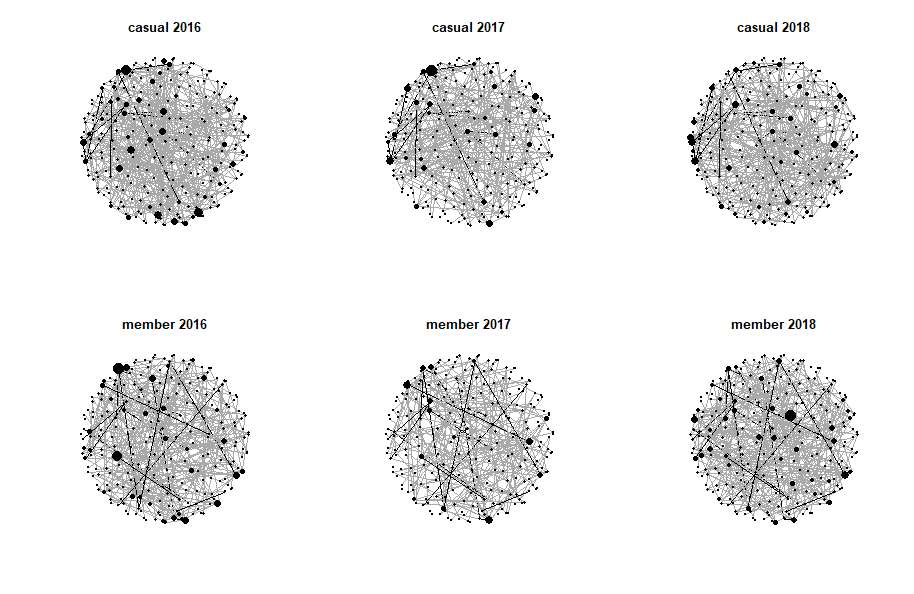}
    \caption{Estimated graph by GHS for each group; black edges denote those edges included in all three years for both member and casual users and the size of the nodes depends on the number of edges associated to the related station.}
    \label{fig:GHS_res}
\end{sidewaysfigure}

Here we provide the analysis of the convergence of mGHS algorithm. We run $M = 4$ chains by starting from different starting points for a post-burnin period of $5000$ iterations. The convergence of the algorithm is studied in Figure \ref{fig:psrf} by estimating the potential scale reduction factor for parameters $\omega_{ij}^k$, $i = 1, \dots, 239$, $j > i$, $k = 1, \dots, 6$, whereas the post-burnin trace of the joint posterior distribution of the chains is shown in Figure \ref{fig:logl}. In Figure \ref{fig:psrf}, the percentage of values that lies in the interval $[1.0, 1,2]$ is greater than $99\%$ for each group, with the only exception of the group members in 2017 for which this percentage is $98\%$. The estimated bike-sharing networks for each group with both mGHS and GHS are shown in Figure \ref{fig:mGHS_res} and \ref{fig:GHS_res}, respectively, where black edges denote those edges included in all three years for both member and casual users.
 

\vspace{0.4cm}

